\documentclass[%
reprint,
superscriptaddress,
nofootinbib,
 amsmath,amssymb,
 aps,
 prd,
floatfix,
]{revtex4-2}

\usepackage{graphicx,amssymb,amsmath,amsthm,amsfonts,epsfig,epsf,nccmath}
\usepackage[linktocpage]{hyperref}
\usepackage[usenames]{color}
\usepackage{epstopdf}
\usepackage{bm}
\usepackage{dcolumn}
\usepackage[latin1]{inputenc}
\usepackage{latexsym}
\usepackage{rotating}
\usepackage{hyperref}
\usepackage{color}
\usepackage{longtable}

\usepackage{enumerate}
\usepackage{tensor,multirow}
\usepackage{url}
\usepackage{tikz,tikz-3dplot}
\usepackage[normalem]{ulem}
\usepackage{booktabs}

\usepackage{slashed}

\newcommand{\ssqav}{\langle S^{2} \rangle_{\psi,0}}

\begin{document}

\begingroup
\renewcommand{\theHequation}{erratum.\arabic{equation}}
\renewcommand{\theHtable}{erratum.\arabic{table}}

\title{Erratum: Parity Violation in Spin-Precessing Binaries: Gravitational Waves from the Inspiral of Black Holes in Dynamical Chern-Simons Gravity [Phys. Rev. D \textbf{106}, 064009 (2022)]}

\author{Nicholas Loutrel}
\affiliation{Physical Review Letters, American Physical Society, Ridge, New York 11367, USA}

\author{Nicol\'as Yunes}
\affiliation{Illinois Center for Advanced Studies of the Universe, Department of Physics, University of Illinois at Urbana-Champaign, Urbana, Illinois 61801, USA}

\date{July 30, 2026}

\let\erratumlabel\label
\renewcommand{\label}[1]{\erratumlabel{erratum:#1}}
\maketitle
\let\label\erratumlabel

\allowdisplaybreaks[4]
\onecolumngrid

We have been asked to reproduce a subset of the calculations in this paper, and in doing so, we identified a few typos that we correct and report here. We provide below the necessary corrections. The orbit-averaged vector precession equations (8)--(12), the exact GR cubic, and the three-sector structure of the dCS carrier phase are all unchanged. The spin-spin and dipole coefficients are corrected below. Unless stated otherwise, the formulas below correspond to binaries with unequal masses, with mass ratio $0 < q < 1$, at linear order in $\bar{\zeta}_2$.

\section*{Effective spin, conservative dynamics, and roots}

The polynomial in Eq.~(A2) is missing the following, global, multiplicative factor: 
\begin{equation}
f_5(q)=\frac{q^2}{2}\left(1+q+\frac{350}{603}q^2-\frac{350}{603}q^3-q^4-q^5\right),
\label{eq:E1}
\end{equation}
but the definition of $f_3$ is unchanged. With Eq.~(\ref{eq:E1}), Eqs.~(18) and (64) retain their printed form. This normalization is the spin-quadratic expansion of the exactly-conserved dCS effective spin. This is strictly a printing typo that did not propagate to our numerical or analytical calculations. 

We identified transcription errors in Eqs.~(A7)--(A22) of the original paper. Equations~(A9), (A13), and (A19), defining $\delta D_2$, $\delta C_{2,4}$, and $\delta D_{2,4}$, respectively, are unchanged. The remaining equations should read:
\begin{align}
    \delta C_{2} &= \frac{(J-L)^{2} (J+L)^{2}}{1344 M^{2} L^{3} (1-v\chi_{\rm eff})^{2}}\left[\delta C_{2,0} + \delta C_{2,2} J^{2} + \delta C_{2,4} J^{4} \right]
    \\
    \delta C_{3} &= \frac{(J^{2} - L^{2})}{1344 L^{3} M^{2} (1+q)^{2}(1-v\chi_{\rm eff})^{2}} \left[\delta C_{3,0} + \delta C_{3,2} J^{2} + \delta C_{3,4} J^{4} + \delta C_{3,6} J^{6} \right]
    \\
    \delta D_{3} &= -\frac{(J^{2} - L^{2})}{1344 L^{3} M^{2} q (1+q)^{3} (1-v\chi_{\rm eff})^{2}}\left[\delta D_{3,0} + \delta D_{3,2} J^{2} + \delta D_{3,4} J^{4} + \delta D_{3,6} J^{6} \right]
\end{align}
with 
\begin{align}
    \delta C_{2,0} &= \frac{L^{3} M^{2} q}{(1+q)^{2}} \left(1809 + 1109 q + 1147 q^2 + 3150 q^3 + 1147 q^4 + 1109 q^5 + 1809 q^6 \right)
    \\
    \delta C_{2,2} &= -\frac{L M^{2} q}{(1+q)^{2}} \left(1809 + 2412 q + 1147 q^2 + 2800 q^3 + 1147 q^4 + 2412 q^5 + 1809 q^6 \right)
    \\
    \delta C_{3,0} &= -\frac{L^{4} M^{4} q}{(1+q)^{2}} \left( 7236 + 16884 q + 11457 q^2 + 19212 q^3 + 31150 q^4 + 20612 q^5 + 12857 q^6 + 15484 q^7 + 7236 q^8\right)
    \\
    \delta C_{3,2} &= \frac{L^{2} M^{4} q}{(1+q)^{2}} \left( 7236 + 19296 q + 19275 q^2 + 27224 q^3 + 29050 q^4 + 24424 q^5 + 20675 q^6 + 19296 q^7 + 7236 q^8\right)
    \\
    \delta C_{3,4} &= -3 L M^{2} q v \left(201 + 350 q^{2} +201 q^{4}\right) \left(4 + 5q + 4 q^{2}\right)
    \\
    \delta C_{3,6} &= -3 v^{2} q (1 + q)^{2} \left(201 + 350 q^{2} + 201 q^{4}\right)
    \\
    \delta D_{2,0} &= \frac{L^{3} M^{2} q}{(1+q)} \bigg[q S_{1}^{2} \left(1809 + 1712 q + 2450 q^2 + 3500 q^3 + 1497 q^4 + 2412 q^5 + 2412 q^6 \right) 
    \nonumber \\
    &- S_{2}^{2} \left(2412 + 2412 q + 1497 q^2 + 3500 q^3 + 2450 q^4 + 1712 q^5 + 1809 q^6 \right)\bigg]
    \\
    \delta D_{2,2} &= \frac{L M^{2} q}{(1+q)} \bigg[-q S_{1}^{2} \left(1809 + 3015 q + 1750 q^2 + 3850 q^3 + 2197 q^4 + 3015 q^5 + 2412 q^6\right) 
    \nonumber \\
    &+ S_{2}^{2} \left(2412 + 3015 q + 2197 q^2 + 3850 q^3 + 1750 q^4 + 3015 q^5 + 1809 q^6 \right)\bigg]
    \\
    \delta D_{3,0} &= \frac{L^{4}M^{4}q(1-q)}{(1+q)}\bigg[-q S_{1}^{2} \left(7236 + 16884 q + 11845 q^2 + 18200 q^3 + 25550 q^4 + 16024 q^5 + 11457 q^6 + 12060 q^7 + 4824 q^8 \right) 
    \nonumber \\
    &+ S_{2}^{2} \left(4824 + 12060 q + 11457 q^2 + 17424 q^3 + 22750 q^4 + 18200 q^5 + 
    14645 q^6 + 15484 q^7 + 7236 q^8 \right) \bigg]
    \\
    \delta D_{3,2} &= \frac{L^{2} M^{4} q(1-q)}{(1+q)} \bigg[q S_{1}^{2} \left(7236 + 19296 q + 14451 q^2 + 22400 q^3 + 24850 q^4 + 18436 q^5 + 14063 q^6 + 12060 q^7 + 4824 q^8\ \right) 
    \nonumber \\
    &- S_{2}^{2} \left(4824 + 12060 q + 12663 q^2 + 21236 q^3 + 24850 q^4 + 19600 q^5 + 15851 q^6 + 19296 q^7 + 7236 q^8 \right)\bigg]
    \\
    \delta D_{3,4} &= -3 L M^{2} q^{2} (1-q^{2}) v \left(201 + 350 q^{2} + 201 q^{4} \right) \left[(4 + q) S_{1}^{2} - (1 + 4 q) S_{2}^{2} \right]
    \\
    \delta D_{3,6} &= 3 q (1 + q)^3 (-201 + 201 q - 350 q^2 + 350 q^3 - 201 q^4 + 201 q^5) (q S_1^2 - S_2^2) v^2
\end{align}
\twocolumngrid

In Eq.~(53), the leading term $s_3^{(0)}u^{-2}$ should read $-s_3^{(0)}u^{-2}$. Correspondingly, Eq.~(57) should be replaced by 
\begin{equation}
m=\frac{s_+^{(0)}-s_-^{(0)}}{s_3^{(0)}}u^2\left[1+\frac{\bar{\zeta}_2}{3}\sqrt{\frac{2}{3y_0}}\frac{\delta y_0M^4(1-q)^2\eta^2}{q[s_+^{(0)}-s_-^{(0)}]}\right]+O(u^3).
\label{eq:E12}
\end{equation}
These were transcription errors only. The correct signs were used in the underlying calculations, and none of the other results in the paper are affected.

In Appendix B, the factor of $(1+q)^5$ in the denominator of Eq.~(B3) should read $(1+q)^4$. Also, Eqs.~(B7)--(B9) should read 
\begin{align}
\sigma_{\pm,0}^{(0)}&=Z_0-\frac{2c_1^2\eta}{\Delta_m^2}\pm\Delta_m^2 Y_0,
\label{eq:E16}\\
\sigma_{\pm,1}^{(0)}&=\frac{2c_1\eta}{\Delta_m^2},
\label{eq:E17}\\
\sigma_{\pm,2}^{(0)}&=-\frac{2\eta^2}{\Delta_m^2}.
\label{eq:E18}
\end{align}
Equation~(B8) is unchanged, but Eq.~(B29) requires $\pm2(1-q)^2Y_0$ in place of $\pm(1-q)^2Y_0$.
In Eq.~(B34), $U_{1}^{\pm}$ and $U_{2}^{\pm}$ should read
$V_{1}^{\pm}$ and $V_{2}^{\pm}$, respectively, so that
\begin{equation}
    V_{s}^{\pm} = q^{1/2} V_{1}^{\pm} + q^{-1/2} V_{2}^{\pm}\,.
\end{equation}
These were transcription errors in the abbreviated coefficient
catalogue. The underlying calculations used the uncompressed root
expressions, and no other results are affected.

\section*{Carrier phase and Table~I}

Equation~(C7) has a transcription typo in the original paper because there is a missing plus sign between the dCS correction and the ${\cal{O}}(u)$. Equation~(C7) should read
\begin{align}
\left\langle \vec{S}_{1}\cdot\vec{S}_{2}\right\rangle_{\psi}
&=
\frac{1}{4}
\left(s_{+}^{(0)}+s_{-}^{(0)}-2S_{1}^{2}-2S_{2}^{2}\right)
\nonumber \\
&+\frac{\bar{\zeta}_{2}}{4}
\left(\delta s_{+}^{(0)}+\delta s_{-}^{(0)}\right)
+{\cal O}(u).
\end{align}
In addition, the right-hand side of Eq.~(C13) has an error and it must also contain
\begin{equation}
\frac{247}{192\eta M^4}
\left(\delta s_{+}^{(0)}+\delta s_{-}^{(0)}\right).
\end{equation}

\begin{widetext}
The corrected Eq.~(C17) is
\begin{align}
\langle\delta C\rangle_\psi&=\frac{5\chi_1^2(1+q)^2}{344064q^3}(17815+20311q^3)
+\frac{5\chi_2^2(1+q)^2}{344064q^2}(20311+17815q^3)
\nonumber\\
&-\frac{5(1+q)^6}{688128M^4(1-q)^2q^4}\Bigl[18c_1^2(6537+11410q^2+6537q^4)
+17815(1-q)^2q(s_+^{(0)}+s_-^{(0)})\Bigr]
\nonumber\\
&+\frac{15c_1\chi_c(1+q)^6}{57344M^2(1-q)^2q^4}(6537-6537q+12242q^2-6537q^3+6537q^4)
\nonumber\\
&-\frac{15\chi_c^2(1+q)^4}{114688(1-q)^2q^4}(6537+11410q^3+6537q^6).
\label{eq:E23}
\end{align}
\end{widetext}
In Eqs.~(C19) and (C20), the first terms under the square roots are $(1-q^2)^2S_1^2$ and $(1-q^2)^2S_2^2$, respectively, which were transcription errors.  

The corrected output columns of Table~I are given below in the original row order; its input columns are unchanged. The entries use the finite-coupling initialization below Eq.~(68), evaluating the coefficients at the shifted integration constants, with orbital $F=5$\,Hz and hence $2F=10$\,Hz in $u_{10\,{\rm Hz}}$, and are rounded to approximately two significant digits. Strict coefficient-by-coefficient linearization gives slightly different rounding for a few cancellation-sensitive entries; the values above follow the finite-coupling initialization stated in the original paper. The statement following Eq.~(C17) that the dipole contribution can be two to three orders of magnitude below both spin contributions is withdrawn. 

\begin{table}[t]
\caption{Corrected last three columns of Table~I, in the original
row order, with $g(q)=q^4(1-q)^2/(1+q)^4$.
The tabulated input parameters are unchanged.
We use $GM_\odot/c^3=4.9254909476\times10^{-6}\,\mathrm{s}$.
Entries are rounded to two significant digits.}
\label{tab:I}
\begin{ruledtabular}
\begin{tabular}{cccc}
Row & $\langle\delta\beta_3\rangle_\psi^{\rm dCS}$
& $\langle\delta\sigma_4\rangle_\psi^{\rm dCS}$
& $g(q)\langle\delta C\rangle_\psi$ \\
\hline
1 & $+7.0\times10^{-5}$ & $-2.3\times10^{-3}$ & $-1.5\times10^{-5}$ \\
2 & $+3.0\times10^{-3}$ & $-1.5\times10^{-4}$ & $-7.5\times10^{-5}$ \\
3 & $+3.3\times10^{-3}$ & $-1.0\times10^{-4}$ & $-5.3\times10^{-5}$ \\
4 & $+3.0\times10^{-3}$ & $-1.5\times10^{-4}$ & $-7.5\times10^{-5}$ \\
5 & $+2.9\times10^{-2}$ & $+8.6\times10^{-3}$ & $-1.7\times10^{-2}$ \\
6 & $+1.8\times10^{-2}$ & $-1.5\times10^{-2}$ & $+1.4\times10^{-4}$ \\
\end{tabular}
\end{ruledtabular}
\end{table}

\section*{Effective spin and native precession angles}

In Eq.~(D1), $f_5-qf_3$ must be replaced by $f_5-q^2f_3$. 
Equations~(D2) and~(D3) retain their algebraic form, but all three $\Sigma_n$ use the corrected $f_5$ in Eq.~(\ref{eq:E1}). In Eq.~(79), the factor $1/(2J_0)$ multiplying $\Omega_{z,6}^{(-1)}(\delta s_+^{(0)}+\delta s_-^{(0)})$ must be $1/(4J_0)$. 
Equation~(D23) becomes then 
\begin{equation}
\delta\phi_{-3}=-\frac{\delta s_+^{(0)}+\delta s_-^{(0)}}{4}\left(\frac{J_0}{j_0}+\frac{j_1}{j_0^{3/2}}\tau_1\right).
\label{eq:E24}
\end{equation}
For the secular precession coefficients, it is convenient to use
$c_\pm=\sqrt{s_\pm^{(0)}-c_1^2}>0$ and $E_c\equiv M^2\chi_c$, and define  
\begin{align}
N&=c_1^2(1+q)^2-(1-q^2)(S_1^2-S_2^2)-2c_1qE_c,
\nonumber\\
P&=c_+^2\delta s_-^{(0)}+c_-^2\delta s_+^{(0)}.
\label{eq:E25}
\end{align}
The leading coefficients needed by Eqs.~(111) and~(113) are then
\begin{align}
\Omega_{z,6}^{(-1)}&=\frac{2+3/\eta}{4M^3}+\frac{3N}{4M^3 q\,c_+c_-},
\label{eq:E26}\\
\delta\Omega_{z,6}^{(-1)}&=-\frac{25(1-q)^2q^2c_1[c_1(1+q)-qE_c]}{32(1+q)M^3c_+c_-}
-\frac{3NP}{8M^3q(c_+c_-)^3}.
\label{eq:E27}
\end{align}

In Eq.~(81), the overall factor
$\Omega_{\rm osc}J_0u^6$ should be
$\Omega_{\rm osc}J_0u^6/\dot{\psi}$, with $\dot{\psi}$
given by Eq.~(33), and the final remainder should be
${\cal O}(J_0u^7/\dot{\psi})$. In addition,
$\tan(2\psi)$ in the numerator of the argument of
$\tan^{-1}$ should be $\sin(2\psi)$.
Using the same positive-real convention for $c_\pm$ introduced above,
the right-hand sides of Eqs.~(D46)--(D48) must also be multiplied by
$2q$. In addition, in Eq.~(D48), 
$c_{+}^{2}\delta s_{-}^{(0)}-c_{-}^{2}s_{+}^{(0)}$
should be
$c_{+}^{2}\delta s_{-}^{(0)}-c_{-}^{2}\delta s_{+}^{(0)}$.
These transcription errors did not affect Fig.~2, for which the
oscillatory correction $\phi_{z,0}$ was explicitly omitted, or
the IMRPhenomPv3 deformations in Eqs.~(111)--(113).
In Eq.~(108), the dipole-radiation term is proportional to
$\tilde{u}^{4}$, rather than $\tilde{u}^{2}$.

Equation~(F5) should read
\begin{equation}
\cos\theta_L={\cal C}_{L,0}
+{\cal C}_{L,2}e^{2i\psi}
+{\cal C}_{L,-2}e^{-2i\psi},
\end{equation}
and Eq.~(F6) should read
\begin{equation}
{\cal C}_{L,0}
=\frac{2J^2+2L^2-S_{+}^{2}-S_{-}^{2}}{4JL}.
\end{equation}
Equation~(F7) is correct, with
${\cal C}_{L,-2}={\cal C}_{L,2}$.

Equations~(89)--(91) contain sign errors in the rotations used to
construct the waveform. They should read
\begin{align}
\hat{n} &=
\left[
\cos\theta_L\cos\phi_C\cos\phi_z-\sin\phi_C\sin\phi_z,\,
\right.
\nonumber \\
& \left.
\cos\theta_L\cos\phi_C\sin\phi_z+\sin\phi_C\cos\phi_z,\,
-\sin\theta_L\cos\phi_C
\right],\\
\hat{\lambda} &=
\left[
-\cos\theta_L\sin\phi_C\cos\phi_z-\cos\phi_C\sin\phi_z,\,
\right. 
\nonumber \\
& \left.
\cos\phi_C\cos\phi_z-\cos\theta_L\sin\phi_C\sin\phi_z,\,
\sin\theta_L\sin\phi_C
\right],\\
\hat{L} &=
\left[
\sin\theta_L\cos\phi_z,\,
\sin\theta_L\sin\phi_z,\,
\cos\theta_L
\right].
\end{align}
Keeping the Thomas-phase definition in Eq.~(84), the carrier phase defined after Eq.~(88) should read $\phi_C=\Phi-\phi_T$. Correspondingly, the Wigner matrices in the waveform decomposition should have arguments $(-\phi_T,-\theta_L,\phi_z)$ instead of
$(\phi_T,\theta_L,-\phi_z)$. These corrections leave the precession solutions unchanged. The coefficient functions in Appendix~F can retain their
positive-$\theta_L$ definitions, with the factor $(-1)^{m-m'}$ included when expressing the corrected Wigner matrices in terms of those coefficients.


\clearpage
\endgroup
\makeatletter
\frontmatter@init
\def\addcontents@abstract{%
  \phantomsection
  \expandafter\edef\csname Parent0\endcsname
    {section*.\number\numexpr\Hy@linkcounter+1\relax}%
  \expandafter\@argswap@val\expandafter{\abstractname}{\addcontentsline{toc}{abstract}}%
}
\makeatother

\setcounter{equation}{0}
\setcounter{table}{0}
\setcounter{figure}{0}
\setcounter{footnote}{0}

\title{Parity Violation in Spin-Precessing Binaries: \\ Gravitational Waves from the Inspiral of Black Holes \\in Dynamical Chern-Simons Gravity}


\author{Nicholas Loutrel}
 \email{nicholas.loutrel@uniroma1.it}
 \affiliation{Dipartimento di Fisica, ``Sapienza'' Universit\`a di Roma \& Sezione INFN Roma1,\\ Piazzale Aldo Moro 5, 
00185, Roma, Italy}

\author{Nicol\'as Yunes}
\affiliation{
 Illinois Center for Advanced Studies of the Universe, Department of Physics\\ University of Illinois at Urbana-Champaign, Champaign, Illinois, USA
}%

\begin{abstract}
Spin precession in compact binaries is intricately tuned to the multipole structure of the underlying bodies. 
For black holes, violations of the no-hair theorems induced by modifications to general relativity correct the precession dynamics, which in turn imprints onto the amplitude and phase modulations of the gravitational waves emitted by the binary. 
Recently, the spin precession equations were derived up to second order in spin for dynamical Chern-Simons gravity, a parity violating modified theory of gravity. 
We here solve these equations and construct, for the first time, analytic expressions for the time- and frequency-domain gravitational waves emitted in the quasi-circular inspiral of spin-precessing black hole binaries in a modified theory of gravity using the post-Newtonian approximation.
Working within the small coupling approximation and using multiple scale analysis, we show that the corrections to the nutation phase enter at relative 1PN order, and the corrections to the precession angle and Thomas phase enter at relative 0PN order. 
Making use of the stationary phase approximation and shifted uniform asymptotics, we find that the Fourier phase of the waveform is characterized by three modifications, two due to the back-reaction of the precession dynamics onto the spin-orbit and spin-spin couplings that enter at 1.5PN and 2PN orders, and a 2PN modification due to the emission of dipole radiation. 
We also find that back-reaction of the precession dynamics forces the dCS corrections to the Fourier amplitude to enter at 0PN order, as opposed to 2PN order, as expected for spin-aligned binaries. 
Our work lays the first foundational stones to build an inspiral-merger-ringdown phenomenological model for spin-precessing binaries in a modified theory of gravity. 
\end{abstract}

\maketitle


\section{\label{sec:intro}Introduction}

The detection of gravitational waves (GWs) has opened the door to probing the fundamental gravitational interaction in the dynamical, strong field regime, wherein spacetime is highly curved and rapidly evolving on observation scales~\cite{LIGOScientific:2021sio}. Significant works has been done on relating constraints on the PN expansion of the GW phase of quasi-circular binaries to those of specific theories that modify general relativity (GR), with the parameterized post-Einsteinian (ppE) framework being a robust tool for doing so~\cite{Yunes:2009ke,Perkins:2022fhr}. Yet, there is still much to be learned from such tests with future observations as we delve deeper into the parameter space of coupling constants controlling the strength of non-GR effects~\cite{Berti:2022wzk}.

At present, degeneracies among parameters allow some theories to evade constraints from GW observations~\cite{Nair:2019,Perkins:2021mhb}. One such theory is dynamical Chern-Simons (dCS) gravity~\cite{Jackiw:2003pm,Alexander:2009tp}, which modifies GR by coupling a psuedo-scalar field to the parity odd quadratic curvature invariant called the Pontryagin density. The theory has its roots in the compactification of ten dimensional heterotic string theory~\cite{ALVAREZGAUME1984269,Green:1987mn,Green:1987sp,Polchinski:1998rr}, in the extension of the Holst action in loop quantum gravity when the Barbero-Imirzi paramter is promoted to a dynamical scalar field~\cite{Ashtekar:1989,Taveras:2008yf,Mercuri:2009vk,Mercuri:2009zi,Mercuri:2009zt}, and in effective field theories of inflation~\cite{Weinberg:2008hq}. As an exact theory, dCS does not appear to have a well-posed initial value problem~\cite{Delsate:2014hba}. One must therefore treat it as an effective field theory, whereby the dimensional coupling constant $\xi^{1/4}$ of the theory is assumed to be small compared to the relevant scales of the astrophysical systems under consideration. In this weak or small coupling approximation, the theory is well-posed, as proven in~\cite{Delsate:2014hba}. 

As a fundamental theory, dCS gravity is said to be parity violating even though parity is not explicitly broken at the level of the action. The theory is parity violating in the sense that modifications from GR due to the dCS coupling only appear in astrophysical systems with broken parity symmetry, such as systems that possess a preferred axis of symmetry. As an example, non-spinning black holes (BHs) described by the static Schwarzschild metric of GR are still solutions to the dCS field equations, but axial perturbations of said black holes are modified from those in GR~\cite{Cardoso:2009pk}; moreover, both axial and polar perturbations of spinning black holes are corrected in this theory with isospectrality clearly broken~\cite{Wagle:2021tam}. These intrinsic parity violations are extremely important when considering the early evolution of the universe in this theory. Tensor perturbations in the early universe possess handedness in dCS gravity, with left- and right-handed modes coupling differently to antimatter and matter, respectively. During the period of inflation, the left-handed modes decay and carry the density of antimatter in the universe with them, thus providing a natural mechanism to solve the baryogenesis problem~\cite{Garcia-Bellido:2003wva,Alexander:2004xd,Alexander:2006a}.

More generally, spinning BHs are modified due to the dCS coupling, developing scalar hair and modifying its higher $l$ multipole moments~\cite{Yunes:2009hc,Yagi:2012ya,Maselli:2017kic,McNees:2015srl,Cano:2019ore}. The former induces a spin dependent dipole moment on the BH, while the latter creates violations of the no hair theorems of GR~\cite{Israel:1967,Israel:1968,Carter:1971,Hawking:1972,Robinson:1975}. Currently, the solution of the dCS field equations describing stationary spinning black holes is known to arbitrary order in a small spin expansion~\cite{Cano:2019ore}. Binary black holes (BBHs) have been considered within this theory in the limit of spin alignment, where it was found that the GW phase is modified at relative 2PN order due to the emission of dipole radiation~\cite{Yagi:2012vf}. Said modification is known to be degenerate with the spins of the BHs, and as a result, constraints on the dCS coupling constant has proven difficult so far with current GW observations~\cite{Nair:2019,Perkins:2021mhb}. Up until recently, the best constraints on dCS gravity came from the Gravity Probe B and LAGEOS missions, which only provided extremely weak constraints on the theory~\cite{Ali-Haimoud:2011zme,Nakamura_2019}. More recently, the combined observations of the binary neutron star merger GW170817 and NICER obersvations of PSR J0030+0451 allowed for an improvement of seven orders of magnitude on the Solar System constraints, constituting the first constraints on the theory that probe the effective field theory regime~\cite{Silva:2020acr}.

Yet, the standard paradigm of non-spinning or spin-aligned quasi-circular binaries are not the only sources of relevance to GW detectors. When in binary systems, misalignment between the BHs' spins and orbital angular momentum generically induces precession of the orbital plane~\cite{Thorne:1985}. This so called spin precession induces modulation of the GW amplitude and phase as the binary coalesces~\cite{Apostolatos:1994}, an observable feature that is known to break degeneracies among the binary's physical parameters when performing parameter estimation~\cite{Chatziioannou:2014coa}. During the inspiral phase of a binary coalescence, the orbital velocity is typically small compared to the speed of light, and one may study the behavior of the binary using the post-Newtonian (PN) formalism~\cite{PoissonWill,blanchet-lrr}. With PN theory, the spin precession equations up to 2PN order contain the spin-orbit, spin-spin, and monopole-quadrupole interactions, the latter of these being dependent on the structure of the component objects of the binary~\cite{Steinhoff:2011sya,Laarakkers:1997hb}. 

For BBHs, the PN spin precession equations have been well studied. Developments in recent years have included the realization that the spin precession equations possess enough constants of motion for the system to be integrable~\cite{Racine:2008qv,Kesden:2015}. Analytic solutions to the precession equations were derived in~\cite{Kesden:2015, Gerosa:2015hba, Gerosa:2015tea, Chatziioannou:2017tdw}, with radiation reaction incorporated through the use of multiple scale analysis. The end result of said study was the development of the first analytic Fourier-domain waveform model for the inspiral of spin precessing binaries, which have been utilized to develop full inspiral-merger-rindgown (IMR) waveform models for such systems~\cite{Khan:2018fmp,Khan:2019kot,Khan:2019,Khan:2020}.

The spin precession equations in dCS gravity were computed using effective field theory methods in~\cite{Loutrel:2018ydv}. Both the spin-spin and monopole-quadrupole interactions are modified, due to the dipole-dipole interaction and the modified BH quadrupole moment, respectively. In~\cite{Loutrel:2018rxs}, a simple match calculation revealed that stringent constraints that probe the effective field theory limit should be possible on dCS gravity with spin precessing binaries. However, this calculation does not take into account the possibility of covariances among the physical parameters of the binary, which could weaken constraints.

Toward this end, we here consider the construction of analytic time-domain and Fourier-domain gravitaitonal waves emitted in the quasi-circular inspiral of spin precessing black hole binaries in dCS gravity. Much like the case in GR, the precession equations in dCS gravity have a sufficient number of constants of motion for the problem to be integrable, with specifically only the mass weighted effective spin being modified~\cite{Loutrel:2018ydv}. Utilizing this fact, we here show that one can construct a co-precessing frame in the equivalent way to the analogous BBH system in GR. Generically, the problem is too difficult to be solved to all orders is spin in a closed-form manner due to the complexity of the dCS corrections. However, working within the weak coupling approximation, we perform a small spin expansion of all dCS corrections, while allowing the GR results to be the exact solutions of~\cite{Chatziioannou:2017tdw}. In our investigation, we find that simultaneously taking the equal mass limit and the limit $\xi \rightarrow 0$ results in non-uniform behavior due to the fact that nutation of the angular momenta vanishes for equal mass binaries in GR, but not in dCS gravity. To handle this, we propose a new expansion paramter $\xi/(1-q)^{2}$ with $q$ the binary's mass ratio, which properly reduces the problem to GR in the limit of this parameter taken to zero and properly handles the non-uniform expansion about GR.

With the co-precessing frame defined, the problem reduces down to solving for two quantities that describe nutation and precession, namely the total spin magnitude $S^{2}$ and the precession phase $\phi_{z}$. The differential equation for $S^{2}$ takes the exact same form as in GR, specifically a cubic polynomial whose solution can be written in terms of Jacobi elliptic functions. The dCS modifications to this quantity mainly appear through velocity independent shifts to the constants of the solution. On the other hand, $\phi_{z}$ develops new secular and oscillatory behavior. We introduce radiation reaction into the problem through multiple scale analysis. The main result of doing so is that the phase of $S^{2}$ obtains a dCS corrections at relative 1PN order, and both the precession phase and Thomas phase are corrected at Newtonian order. Much like in GR, while the nutation phase and Thomas phase can be written in typical PN-style expansions, the precession phase cannot and must instead by expanded in \textit{functions} of the orbital velocity, rather than a power series in the orbital velocity.

To complete the solution, we compute the corrections to the orbital phase. Rather than obtaining a single deviation to the GR orbital phase, we obtain three: two at 1.5PN and 2PN order due to the back reaction of the precession dynamics on the GR spin-orbit and spin-spin couplings respectively, and a third one at 2PN order due to the emission of dipole radiation in dCS gravity. This is in contrast to spin-aligned binaries where the leading-order correction is the latter of these. The way to understand this is as follows. For spin-aligned binaries, the angle between the orbital and spin angular momenta is zero, and it remains zero in both GR and dCS gravity due to the nature of the precession equations. Thus, in the spin-aligned case, the leading order dCS correction is due to dipole emission, which enters at 2PN order in the orbital phase, as shown in~\cite{Yagi:2011xp,Yagi:2012vf}. Now, consider binaries with misaligned momenta, starting with a system with some known misalignment angle $\gamma$. The GR spin orbit coupling in the phase is the lowest PN order term that depends on this angle, with the dCS dipole radiation entring at 0.5PN order higher than this. Now evolve the spins according to the precession equations in either GR or dCS gravity. At some later time, the misalignment angle is no longer the same in the two different theories, and thus the GR spin-orbit and spin-spin couplings in the phase acquire dCS modifications.

With the solution to the precession equations in hand, we compute the analytic waveforms in the Fourier domain. To do so, we make use of the shifted uniform asymptotics (SUA) developed in~\cite{Klein:2014bua} to handle catastrophes that appear when applying the stationary phase approximation (SPA) to precessing waveforms. All throughout we validate the accuracy of our analytic results against numerical evolutions of the dCS-modified PN evolution equations. Our analytic solution of the time evolution of the direction of the total angular momentum is accurate to better than $10^{-5}$ and $10^{-2}$ for all slowly-spinning and for all arbitrary spin systems we considered respectively. 
The nutation and precession phases are accurate to $\lesssim 1$ radian and $\lesssim 10$ radians, respectively for all spinning systems we considered and for signals in the frequency band of ground based detectors.
This indicates that our analytic waveforms are accurate enough to begin the construction of inspiral-merger-ringdown (IMR) phenomenological waveforms~\cite{Khan:2018fmp}. 

Moreover, we show that the Fourier phase and amplitude of the waveform can differ from those in GR for the same binary system by $\sim 10$ radians and $\sim1\%$, respectively for both nearly equal mass and highly spinning systems, for conservative values of the dCS coupling parameter $\alpha_{4} < 1$ km. This suggests that the precessing waveform models we develop here could be critical in placing constraints on dCS gravity with GWs. Given this, we provide a detailed prescription for the construction of an IMRPhenomPv3 model in dCS gravity. Such a model is the first to describe GWs emitted by spin-precessing binaries in a modified theory, which should enhance our ability to detect or constrain non-GR effects with GW data.  

The rest of this paper presents the details of the results summarized above. Section~\ref{sec:form} introduces dCS gravity, introduces notation and presents the basics of precessing systems. Section~\ref{sec:prec} describes the analytic solution to the precession equations, with a review first of the solutions within GR. Section~\ref{sec:rr} introduces radiation-reaction and shows how this effect changes the evolution of the momenta. Section~\ref{sec:gws} presents the gravitational waveform in the time domain and in the Fourier domain, using the SUA to solve the generalized Fourier integral. Section~\ref{sec:disc} concludes and points to future work. The appendices present further details of the calculation that are too lengthy for the main body of this paper. For the rest of this paper, we follow the notation of~\cite{Misner:1973prb}, and in particular, use geometric units in which $G =   c = 1$. 

\section{\label{sec:form}Spin-precessing black hole binaries in dCS gravity}

We here provide a basic overview of the formalism and notation that we use, as well as a brief review of certain aspects of dCS gravity that are of relevance to the problem at hand. We enumerate all of the equations that are necessary for the analysis of the spin precession equations and the construction of the Fourier domain waveform in dCS gravity.

\subsection{\label{sec:not} Notation}

We consider the case of a binary system in a quasi-circular orbit, composed of two black holes with masses $m_{A}$ and spin angular momenta $\vec{S}_{A}$ with $A=1$ or $2$ that is not necessarily aligned with the orbital angular momentum $\vec{L}$. Henceforth, we employ the following notation
\begin{itemize}
\setlength\parskip{0.0cm}
\setlength\parsep{0.0cm}
    \item The magnitude of the spin angular momentum is given by $S_{A} = (\vec{S}_{A} \cdot \vec{S}_{A})^{1/2}$, where $\cdot$ corresponds to the usual Euclidean dot product between two (three-)vectors. The Kerr parameter of each black hole is $a_{A} = S_{A}/m_{A}$ and the dimensionless spin is $\chi_{A} = a_{A}/m_{A} = S_{A}/m_{A}^{2}$. 
    \item The total mass of the binary is given by $M = m_{1} + m_{2}$, while the mass ratio is $q = m_{2}/m_{1}$. In our analysis, we take $m_{2} \le m_{1}$ and thus $q \le 1$. The symmetric mass ratio is given by $\eta = q/(1+q)^{2}$, and the reduced mass is $\mu = \eta M$. 
    \item The four-position of each particle is given by $z^{\mu}_{A}(\tau)$ with $\tau$ the proper time on the particle's worldline. The four-velocity is $u^{\mu}_{A} = dz^{\mu}_{A}/d\tau$. The relative spatial separation of the binary is given by $\vec{r}$, with relative three-velocity $\vec{v} = \dot{\vec{r}}$, and relative three-acceleration $\vec{a} = \dot{\vec{v}}$, where the overdot corresponds to differentiation with respect to coordinate time $t$. The relative unit vector is $\vec{n} = \vec{r}/r$, with $r = (\vec{r} \cdot \vec{r})^{1/2}$. We define the dimensionless parameter $u = (2 \pi M F)^{1/3}$, with $F$ the orbital frequency. 
    \item The magnitude of the orbital angular momentum is given by $L = (\vec{L} \cdot \vec{L})^{1/2}$, and the unit angular momentum vector that is orthogonal to the orbital plane is $\hat{L}$. The total angular momentum is then given by
\begin{equation}
    \label{eq:J-def}
    \vec{J} = L \hat{L} + \vec{S}_{1} + \vec{S}_{2}\,.
\end{equation}
    \item Latin indices $i,j,k,..$ span constant time space-like hypersurfaces, while Greek indices $\mu,\nu,\rho,...$ span the full four-dimensional spacetime. The former are raised and lowered with the Euclidean three metric $\delta_{ij}$ (also called the Kronecker delta), while the latter are raised and lowered with the 4-dimensional spacetime metric $g_{\mu\nu}$. Multi-index quantities are written as $x_{i_{1} i_{2} ... i_{N}} = x_{i_{1}} x_{i_{2}} ... x_{i_{N}}$. 
    \item Angled brackets $<>$ on indices corresponds to the symmetric trace-free (STF) part of a tensor. For example, for tensors with spacetime indices
\begin{equation}
    T_{<\mu \nu>} = T_{(\mu \nu)} - \frac{1}{4} {T^{\rho}}_{\rho} g_{\mu \nu}
\end{equation}
    where $T_{\mu \nu}$ is an arbitrary 2-tensor, and () corresponds to the symmetrization of the indices. When performing a PN expansion, the STF projection of tensors with only spatial indices becomes
\begin{equation}
    T_{<ij>} = T_{(ij)} - \frac{1}{3} T \delta_{ij}
\end{equation}
where $T = T_{ij} \delta^{ij}$ is the trace of $T_{ij}$.
\end{itemize}

\subsection{\label{sec:main-eq} Conservative dynamics in dCS gravity}

DCS gravity is considered to be a parity violating theory of gravity, in that it modifies the action of general relativity through the coupling of a psuedo-scalar field $\vartheta$ to the parity odd quadratic curvature invariant called the Pontryagin density ${^{\star}}RR$~\cite{Jackiw:2003pm,Alexander:2009tp}. The action is 
\begin{align}
    \label{eq:action}
    S = \int d^{4}x \sqrt{-g} &\Big[\kappa R + \alpha_{4} \vartheta {^{\star}}RR 
    \nonumber \\
    &+ \frac{1}{2} \nabla^{\mu} \vartheta \nabla_{\mu} \vartheta + V(\vartheta) + {\cal{L}}_{\rm mat}\Big]\,,
\end{align}
where the first term in the brackets above is the standard Einstein-Hilbert action of GR with $\kappa = (16\pi)^{-1}$, the second term in the dCS coupling with dimension-full constant $\alpha_{4}$\footnote{Some articles in the literature use a different coupling parameter in the action. The mapping between these different notations is $\alpha_{4} = -\alpha/4$.}, the third and fourth terms are the action of the psuedo-scalar, and the last term is the Lagrangian density of any matter sources. Most studies of dCS gravity take the potential of the scalar field to be zero, i.e. $V(\vartheta) = 0$~\cite{Yagi:2011xp,Alexander:2021ssr}. The theory can be considered an effective field theory extension of GR, and as such, the coupling parameter $\alpha_{4}$ must be small. Specifically, for astrophysical systems, the small dimensionless coupling is usually $\zeta = \xi/M^{4}$, where $\xi = \alpha_{4}^{2}/\kappa$ and $M$ is the mass scale of the system under consideration in geometrized units~\cite{Yunes:2009hc,Alexander:2021ssr}. For the purposes of this paper, the mass scale is the total mass of the system.

In~\cite{Loutrel:2018ydv}, the effective matter Lagrangian describing black holes in dCS gravity was developed. Specifically, this reference found that
\begin{equation}
    {\cal{L}}_{\rm mat} = \sum_{A} \int \frac{d\tau}{\sqrt{-g}} \delta^{4}[x^{\mu} - z_{A}^{\mu}(\tau)] L_{{\rm eff},A}
\end{equation}
where
\begin{align}
    \label{eq:L-eff}
    L_{{\rm eff},A} &= p^{\mu}_{A} u_{\mu}^{A} + \frac{1}{2} S^{\mu \nu}_{A} \Omega_{\mu \nu} 
    \nonumber \\
    &+ 10\pi \frac{\alpha_{4}}{m_{A}^{2}} {^{\star}}S^{\mu \nu}_{A} u_{\mu}^{A} \nabla_{\nu} \vartheta - \frac{1}{6} J^{\mu \nu \rho \sigma}_{A} R_{\mu \nu \rho \sigma}
\end{align}
with $p^{\mu}$ the particle's four-momentum, $S^{\mu \nu}$ the spin tensor, and $J^{\mu \nu \rho \sigma}$ the canonical mass quadrupole tensor. Simultaneous variation of Eqs.~\eqref{eq:action} and~\eqref{eq:L-eff} allowing for proper asymptotic matching to known black hole solutions in dCS gravity~\cite{Yunes:2009hc,Yagi:2012ya}, and provides the necessary relativistic equations of motion of the binary. In a post-Newtonian (PN) expansion, and working in the center of mass frame, the equations of motion reduce to
\begin{equation}
    \label{eq:acc}
    \vec{a} = -\frac{M}{r^{2}}\vec{n} + \vec{a_{\rm 1PN}} + \vec{a_{\rm spin}} + \vec{a}_{{\rm quad}} + \vec{a}_{\rm scalar}\,,
\end{equation}
where the first term is the Newtonian gravitational force, $\vec{a}_{\rm 1PN}$ is the 1PN correction to the two-body interaction given in Eq.~(10.1) in~\cite{PoissonWill}, $\vec{a}_{\rm spin}$ is the correction due to spin-orbit and spin-spin coupling in GR which are given in Eqs.~(10.161)-(10.162) in~\cite{PoissonWill}, $\vec{a}_{\rm quad}$ is the quadrupole-monopole interaction which contains both GR and dCS effects, and $\vec{a}_{\rm scalar}$ is the correction due to the scalar-dipole coupling of the black holes. These equations are supplemented by a set of evolution equations for the spin of each body, specifically
\begin{align}
    \label{eq:prec}
    \dot{\vec{S}}_{A} &= \vec{\Omega} \times \vec{S}_{A}
\end{align}
with
\begin{align}
    \vec{\Omega} &= \vec{\Omega}_{\rm SO} + \vec{\Omega}_{\rm SS} + \vec{\Omega}_{\rm QM}\,,
    \\
    \vec{\Omega}_{\rm SO} &= \frac{\eta}{M} v^{5} \left(2 + \frac{3}{2} \frac{m_{B}}{m_{A}}\right) \hat{L}\,,
    \\
    \vec{\Omega}_{\rm SS} &= \frac{1}{2M^{3}} v^{6} \left(1 + \frac{25}{16} \frac{\xi}{m_{A}^{2} m_{B}^{2}}\right) \left[ \vec{S}_{B} - 3 \left(\hat{L} \cdot \vec{S}_{B}\right) \hat{L}\right]\,,
    \\
    \vec{\Omega}_{\rm QM} &= - \frac{3}{2M^{3}} \frac{m_{B}}{m_{A}} v^{6} \left(1 - \frac{201}{112} \frac{\xi}{m_{A}^{4}}\right) \left(\hat{L} \cdot \vec{S}_{A}\right) \hat{L}\,,
\end{align}
where each term represents the spin-orbit (SO), spin-spin (SS), and quadrupole-monopole (QM) couplings, respectively. Note that in this theory, the total angular momentum $\vec{J}$ is conserved, and thus the precession equation for the orbital angular momentum can be found from Eq.~\eqref{eq:J-def}. The necessary equations to characterize generic spinning binaries in dCS gravity have now been enumerated.

\subsection{\label{sec:qc} Quasi-circular binaries in dCS gravity}

The case of quasi-circular binaries was partially solved in~\cite{Yagi:2012vf}, which found the solution of the relative equations of motion in Eq.~\eqref{eq:acc}. The main results of that study were the modifications to the conservative dynamics encoded in Kepler's third law
\begin{equation}
    \label{eq:kep-3}
    r_{12} = \frac{M}{u^{2}}\left(1 + \delta C_{r} u^{4}\right)\,,
\end{equation}
and the binding energy of the binary
\begin{equation}
    E = - \frac{\mu}{2} u^{2} \left(1 + \delta C_{E} u^{4}\right)\,,
\end{equation}
where recall that $u = (2 \pi M F)^{1/3}$.
In the above, $[\delta C_{r}, \delta C_{E}]$ are linear in the coupling parameter $\zeta = \xi/M^{4}$, and depend on the spins of each body. Explicit forms for these expressions can be found in Eqs.~(6) \& (7), respectively, in~\cite{Yagi:2012vf}. The effect of dissipation due to radiation reaction on the binary was also considered in~\cite{Yagi:2012vf}, with the evolution equation for the orbital frequency being\footnote{There is an overall factor of 16 that appears in our expression and does not appear in Eq.~(9) of~\cite{Yagi:2012vf}. The reason for this is the different choice in the coupling parameter, namely $\alpha_{4}$ instead of $\alpha/4$. This means that $\zeta \rightarrow 16\zeta$. For convenience, we have further factored out the $\zeta$ dependence from $\delta C$.}
\begin{equation}
    \label{eq:dudt}
    \dot{u} = \dot{u}_{\rm GR} \left(1 + 16 \zeta \delta C u^{4}\right)
\end{equation}
where $\dot{u}_{\rm GR}$ is the leading PN order contribution due to quadrupole radiation
\begin{equation}
    \dot{u}_{\rm GR} = \frac{a_{0}}{3M} u^{9} \left[1 + \sum_{n=2} \left( a_{n} + 3 b_{n} \ln u\right) u^{n}\right] \,,
\end{equation}
with $(a_{n}, b_{n})$ given in Appendix A of~\cite{Chatziioannou:2013dza}, and
\begin{align}
    \label{eq:dC-dipole}
    \delta C &= \frac{101555}{344064} \frac{M^{2}}{m_{1}^{2}} \chi_{1}^{2} \left[1 - \frac{58833}{20311} \left(\hat{S}_{1} \cdot \hat{L}\right)^{2}\right] 
    \nonumber \\
    &- \frac{12725}{49152} \frac{\chi_{1} \chi_{2}}{\eta} \left[\left(\hat{S}_{1}\cdot\hat{S}_{2}\right) - \frac{1467}{509} \left(\hat{S}_{1}\cdot\hat{L}\right)\left(\hat{S}_{2}\cdot\hat{L}\right)\right] 
    \nonumber \\
    &+ (1\leftrightarrow2)\,.
\end{align}

\subsection{\label{sec:co-prec} Co-precessing frame}

One of the biggest developments in the study of spin precessing binaries within GR was the realization that there were a sufficient number of constants of motion for the system to be integrable, and allow one to develop a co-precessing reference frame. In~\cite{Loutrel:2018ydv}, analysis of the precession equations revealed the same number of constants of motion in dCS gravity, with only one of them being modified. The constants of motion are the spin magnitudes $S_{1,2}$, the magnitude of the orbital angular momentum $L$ and total angular momentum $J$, the direction of the total angular momentum $\hat{J}$, and the mass-weighted effective spin
\begin{align}
    \label{eq:chi-eff}
    \chi_{\rm eff} M^{2} &= \left(1 + q^{-1}\right) \left(\vec{S}_{2} + q \vec{S}_{1}\right)\cdot\hat{L} 
    \nonumber \\
    &- \frac{25}{48} \zeta_{2} q^{2} (1 - q) \left(\vec{S}_{1} \cdot \hat{L}\right)
    \nonumber \\
    &- \frac{201}{112} \frac{\zeta_{2}}{L} \frac{f_{3}(q)}{1+q} \left(\vec{S}_{1} \cdot \hat{L}\right)\left(\vec{S}_{2}\cdot\hat{L}\right)
    \nonumber \\
    &- \frac{201}{112} \frac{\zeta_{2}}{L} \frac{f_{5}(q)}{q(1+q)} \left(\vec{S}_{1}\cdot\hat{L}\right)^{2}\,.
\end{align}
where $\zeta_{2} = \xi/m_{2}^{4}$, and $[f_{3},f_{5}]$ are polynomials in $q$, which are given in Appendix~\ref{app:exp}. Note that $L$ and $J$ are only conserved in the absence of radiation reaction, but there evolution is slow compared to the orbital and precession timescales. In addition, the effective spin is modified from general relativity due to the dipole-dipole and quadrupole-monopole interaction.

The basic picture of the co-precessing frame is given in Fig.~1 of~\cite{Chatziioannou:2017tdw}. The frame is chosen such that $\vec{J}$ is aligned with the $z$-axis, $\vec{L}$ and $\vec{S} = \vec{S}_{1} + \vec{S}_{2}$ both lie in the $xz$-plane, with the angle between the $z$-axis given by $\theta_{L}$. A simple calculation gives
\begin{equation}
\label{eq:thetaL}
\cos \theta_{L} = \frac{J^{2} + L^{2} - S^{2}}{2 J L}\,.
\end{equation}
Thus, the angular momenta vectors in the co-precessing frame are given by
\begin{align}
\vec{J} &= [0, 0, J]\,,
\\
{\vec{L}} &= L [\sin\theta_{L}, 0, \cos\theta_{L}] 
\nonumber \\
&= \left[\frac{A_{1} A_{2}}{2 J}, 0, \frac{J^{2} + L^{2} - S^{2}}{2 J}\right]\,,
\\
{\vec{S}} &= [-L \sin\theta_{L}, 0, J - L\cos\theta_{L}] 
\nonumber \\
&= \left[-\frac{A_{1} A_{2}}{2J}, 0, \frac{J^{2} - L^{2} + S^{2}}{2 J}\right]\,,
\end{align}
where 
\begin{align}
\label{eq:A1-A2}
A_{1} &= \left[J^{2} - (L - S)^{2}\right]^{1/2}\,, \;\; A_{2} = \left[(L + S)^{2} - J^{2}\right]^{1/2}\,.
\end{align}

We must still specify the orientation of $(\vec{S}_{1}, \vec{S}_{2})$ in this frame. To start, it is useful to introduce a new co-precessing frame where $\vec{S}'$ is aligned with the $z'$-axis. This can be achieved by a simple Euler rotation of the original coordinate system about the $y$-axis, with the angle $\theta_{S}$ specified by
\begin{equation}
\cos \theta_{S} = \frac{J^{2} - L^{2} + S^{2}}{2 J S}\,.
\end{equation}
In the primed frame, the vectors $(\vec{S}_{1}', \vec{S}_{2}')$ are determined by the angles $(\theta',\phi')$ in the $x'y'$-plane, and satisfy $\vec{S}' = \vec{S}_{1}' + \vec{S}_{2}'$. More specifically,
\begin{equation}
\label{eq:S1-prime}
\vec{S}_{1}' = \left[\frac{A_{3} A_{4}}{2 S} \cos \phi', \frac{A_{3} A_{4}}{2 S} \sin\phi', \frac{S^{2} + S_{1}^{2} - S_{2}^{2}}{2 S}\right]
\end{equation}
with
\begin{equation}
\label{eq:A3-A4}
A_{3} = \left[S^{2} - (S_{1} - S_{2})^{2}\right]^{1/2}\,, \;\; A_{4} = \left[(S_{1} + S_{2})^{2} - S^{2}\right]^{1/2}\,.
\end{equation}
The components of the spin vectors in the un-primed frame are then found by performing another Euler rotation about the y-axis.

We are still left with specifying the angle $\phi'$. The only remaining constant of motion that has not been used is the effective mass weighted spin. To obtain $\phi'$ in terms of the constants of motion, one must insert the expressions for $(\hat{L}, \vec{S}_{1}, \vec{S}_{2})$ into Eq.~\eqref{eq:chi-eff} and solve for $\cos \phi'$. We do not do so here because there are some subtle difference between GR and dCS gravity that must be carefully taken into account. We detail this in the next section.

\section{\label{sec:prec}Analytic Solutions to the Precession Equations without Radiation Reaction}

The binary problem with spin precession exhibits a separation of scales that can be exploited to make the analytic solution of the binary's dynamics tractable. Specifically, the binary components evolve on the orbital timescale $T_{\rm orb} \sim v^{-3}$, the orbital angular momentum evolves due to the spin-coupling on the timescale $T_{\rm prec} \sim v^{-6}$, and the binary's orbit inspirals due to radiation reaction on the timescale $T_{\rm rr} \sim v^{-10}$, and where $v$ is the orbital velocity of the binary. When considering the inspiral phase of the binary's coalescence due to GW emission, the orbital velocity is typically small compared to the speed of light, and the separation of the orbital, precession, and radiation reaction timescales holds. As a result, we may apply multiple scale analysis to solve the problem at hand. The first step in this approach is to consider the unperturbed problem, specifically the spin dynamics of the binary in the absence of radiation reaction.

\subsection{Solutions in General Relativity: A Review}
\label{prec-gr}

Before considering the full problem in dCS gravity, we provide a review of how the precession solutions are constructed within GR, which was first done in~\cite{Chatziioannou:2017tdw}. The spin precession equations can be found by taking the limit $\xi\rightarrow0$ in Eq.~\eqref{eq:prec}, which yields
\begin{equation}
\dot{\vec{S}}_{1} = \left[\frac{\eta}{M} v^{5} \left(2 + \frac{3}{2} q\right) - \frac{3}{2} \frac{v^{6}}{M^{3}} \left(\vec{S}_{2} + q \vec{S}_{1}\right) \cdot \hat{L}\right] \hat{L} \times \vec{S}_{1}\,.
\end{equation}
The equation for $\vec{S}_{2}$ can be found by particle exchange (ie.~taking $1 \leftrightarrow 2$) in the above equation, and the equation for $\hat{L}$ is found from the conservation of $\vec{J}$. Following the discussion in Sec.~\ref{sec:co-prec}, the only unspecified quantity in the primed frame is the angle $\phi'$ appearing in Eq.~\eqref{eq:S1-prime}. This quantity is not a free parameter, but instead, it is fixed by the constants of motion. More specifically, $\phi'$ is determined by $\chi_{\rm eff}$ in Eq.~\eqref{eq:chi-eff}, which is invariant of the choice of frame since it is a scalar quantity. To complete the setup detailed in Sec.~\ref{sec:co-prec}, we solve for $\cos\phi'$ after taking the limit $\zeta_{2}\rightarrow0$ in Eq.~\eqref{eq:chi-eff} to obtain 
\begin{align}
\label{eq:cos-phip-gr}
\cos \phi' &= -\left[A_{1} A_{2} A_{3} A_{4} (1 - q)\right]^{-1} \Big\{\left[L^{2} + S^{2} - J^{2}\right] 
\nonumber \\
&\times \left[(1+q) S^{2} - (1 - q) (S_{1}^{2} - S_{2}^{2})\right] 
+ \frac{4 L q S^{2} \chi_{\rm eff} M^{2}}{1 + q}\Big\}\,.
\end{align}
The spin vectors are now fully specified in GR.

\subsubsection{Nutation and the Spin Magnitude $S(t)$}

The evolution of the angular momenta in the co-precessing frame is determined by one time dependent quantity, specifically the total spin magnitude $S(t)$. Physically, $S(t)$ describes nutation, the upwards and downwards bobbing motion of the precessing spin vectors. Before we seek the analytic solutions for $S(t)$, it is useful to consider the equal mass case since it has important repercussions in dCS gravity. Specializing to the equal mass case, the effective mass weighted spin and precession equation for $\vec{S}$ become
\begin{align}
\chi_{\rm eff} M^{2} &= 2 \left(\hat{L} \cdot \vec{S}\right)\,,
\\
\dot{\vec{S}} &= \left[\frac{7}{8} \frac{v^{5}}{M} - \frac{3}{2} \frac{v^{6}}{M^{3}} \left(\hat{L} \cdot \vec{S}\right)\right] \hat{L} \times \vec{S}\,,
\end{align}
respectively. From these, it can be shown that $dS^{2}/dt = 2 \vec{S} \cdot (d\vec{S}/dt) = 0$, and thus, $S$ is a constant in the equal mass case. This implies that in GR and to leading PN order, spin-precessing binaries do not experience nutation when the component masses are equal.

For $q\neq1$, the equation for $S(t)$ may be derived from the expressions in the preceding section in the following manner. The total spin $\vec{S}$ evolves according to $d\vec{S}/dt = d\vec{S}_{1}/dt + d\vec{S}_{2}/dt$, and thus the evolution of the total spin magnitude is $dS^{2}/dt = 2\vec{S} \cdot (d\vec{S}/dt)$. Performing this analysis, the evolution equation for the total spin magnitude becomes 
\begin{equation}
\label{eq:dSdt-gr}
\left(\frac{d S^{2}}{dt}\right)^{2} = - A^{2} \left(S^{6} + B S^{4} + C_{0} S^{2} + D_{0}\right)\,,
\end{equation}
where $[A,B,C_{0},D_{0}]$ are given in Appenix~\ref{app:exp}. These coefficients are funtions of the velocity, and thus vary on the longer radiation reaction timescale, but are constant on the precession timescale. Equation~\eqref{eq:dSdt-gr} can be solved analytically to obtain
\begin{equation}
\label{eq:Soft-gr}
S^{2}(t) = S_{+}^{2} + (S_{-}^{2} - S_{+}^{2}) \text{sn}^{2}(\psi,m)\,,
\end{equation}
where $\text{sn}(\cdot,\cdot)$ is the Jacobi sine (elliptic) function, $m = (S_{+}^{2} - S_{-}^{2})/(S_{+}^{2} - S_{3}^{2})$, and $(S_{+}, S_{-}, S_{3})$ are the roots of Eq.~\eqref{eq:dSdt-gr}, found through the method in Appendix~\ref{app:exp}, and
\begin{equation}
    \label{eq:dpsidt}
    \frac{d\psi}{dt} = \frac{A}{2} \left(S_{+}^{2} - S_{3}^{2}\right)^{1/2}\,.
\end{equation}
In the absence of radiation reaction, the right-hand side of the above equation is constant, and this can be trivially solved. However, radiation reaction alters the behavior of $\psi$, and we detail this in Sec.~\ref{sec:prec-rr}. This completes our derivation of $S(t)$ in the absence of radiation reaction.

\subsubsection{Non-precessing Frame and the Precession Angle $\phi_{z}(t)$}
\label{prec-phiz}

Having obtained the solution for the angular momenta in the co-precessing frame, we now seek the solutions in the physical, non-precessing frame. To describe the precession of $(\hat{L}, \vec{S}_{1}, \vec{S}_{2})$ around $\hat{J}$, we rotate the solutions in the co-precessing frame around $\hat{J}$ by a time-varying angle $\phi_{z}(t)$, called the precession angle. To obtain the evolution equation for $\phi_{z}(t)$, we combine these expressions with the precession equation $\dot{\hat{L}}$. The form of the vectors $[\hat{L},\vec{S}_{1},\vec{S}_{2}]$ in the non-precessing frame is given in Appendix~\ref{app:exp}. In the absence of radiation reaction,  Eq.~\eqref{eq:L-no-prec} only depends on time through $S^{2}$ and $\phi_{z}$, and thus we have
\begin{equation}
    \label{eq:L-partial}
    \frac{d\hat{L}}{dt} = \frac{dS^{2}}{dt} \frac{\partial \hat{L}}{\partial S^{2}} + \frac{d\phi_{z}}{dt} \frac{\partial \hat{L}}{\partial \phi_{z}}\,.
\end{equation}
It is straightforward to check that, from Eq.~\eqref{eq:L-no-prec}, $\partial \hat{L}/\partial S^{2} \perp \partial \hat{L}/\partial \phi_{z}$. Thus, to obtain the evolution equation for $\phi_{z}$, we set the left-hand side of Eq.~\eqref{eq:L-partial} equal to the precession equation for $\hat{L}$, perform the dot product with $\partial \hat{L}/\partial \phi_{z}$ to remove the dependence on the evolution of $S^{2}$, and solve for $d\phi_{z}/dt$. 
The evolution equation of $\phi_z$ takes the general form
\begin{equation}
\label{eq:dphizdt-full}
\frac{d\phi_{z}}{dt} = \frac{Q_{0} + Q_{2} S^{2} + Q_{4} S^{4}}{P_{0} + P_{2} S^{2} + P_{4} S^{4}}\,,
\end{equation}
where the constant coefficients $(P_{n}, Q_{n})$ are known functions of $(J,L,S_{1}, S_{2}, \chi_{\rm eff}, q, v)$. Inserting Eq.~\eqref{eq:Soft-gr} into the above expression, we obtain
\begin{equation}
\label{eq:dphizdt-GR}
\frac{d\phi_{z}}{dt} = J \left[\frac{b_{0} + b_{2} \text{sn}^{2}(\psi, m) + b_{4} \text{sn}^{4}(\psi,m)}{d_{0} + d_{2} \text{sn}^{2}(\psi,m) + d_{4} \text{sn}^{4}(\psi, m)}\right]\,,
\end{equation}
where the $b$ and $d$ coefficients are listed in Appendix~\ref{app:exp}. By combining this with Eq.~\eqref{eq:dpsidt}, this equation can be solved analytically to obtain
\begin{align}
\label{eq:phiz-GR}
\phi_{z}(t) &= \frac{\bar{A}_{\phi}}{\dot{\psi}} \psi + \frac{\bar{C}_{+}}{\dot{\psi}} \Pi[\bar{n}_{+}, \text{am}(\psi, m), m] 
\nonumber \\
&+ \frac{\bar{C}_{-}}{\dot{\psi}} \Pi[\bar{n}_{-}, \text{am}(\psi,m), m]
\end{align}
where $\Pi$ is the Jacobi elliptic integral of the third kind, and the constants $(\bar{A}_{\phi}, \bar{C}_{\pm}, \bar{n}_{\pm})$ are listed in Appendix~\ref{app:exp}. This solution does not take the same form as that found in~\cite{Chatziioannou:2017tdw}, but we have verified that the solution and equation for $d\phi_{z}/dt$ therein are equivalent to Eqs.~\eqref{eq:dphizdt-GR} and~\eqref{eq:phiz-GR} above. This is also why we have placed bars over the coefficients in the above solution, specifically to differentiate them from the coefficients in~\cite{Chatziioannou:2017tdw}.

\subsection{Solutions in dCS Gravity}

Now that we have reviewed the analytic solution in GR, we consider the problem in dCS gravity. The spin precession equations are now the full expression given in Eq.~\eqref{eq:prec}. One subtle difference between the GR case is that the precession equations for dCS gravity are only known to second order in a small spin expansion. As a result, we leave the GR sector of the solution to all orders in spin while expanding terms coupled to $\zeta_{2}$ to second order in small spin.

\subsubsection{Nutation and the Differences to General Relativity}

The geometric setup described in Sec.~\ref{sec:co-prec} is still valid for the problem in dCS gravity. However, there are some critical differences which make the solution to the precession equations more complicated in dCS gravity. The main difference arises in the effect of nutation. If we once again consider the equal mass case, but in dCS gravity this time, we find that $dS^{2}/dt \ne 0$, while recall that in GR, $(dS^{2}/dt)_{\rm GR} = 0$. Thus, nutation is always present in the dCS precession problem at leading PN order regardless of the mass ratio of the binary.

Consider now the mapping between the angle $\phi'$ and $\chi_{\rm eff}$. From Eq.~\eqref{eq:chi-eff}, the mapping between these quantities becomes
\begin{equation}
\label{eq:cos-phip-master}
\chi_{\rm eff} M^{2} = \zeta_{2} \alpha \cos^{2}\phi' + \left(\beta_{0} + \zeta_{2} \delta \beta\right) \cos \phi' + \gamma_{0} + \zeta_{2} \delta \gamma
\end{equation}
where $(\alpha, \beta_{0}, \delta \beta, \gamma_{0}, \delta \gamma)$ are known functions of the constants of motion. In the limit $\zeta_{2} \rightarrow 0$, this equation reduces to the GR case. Therefore, this equation can be solved perturbatively in $\zeta_{2}$ to obtain
\begin{align}
\label{eq:cos-phip-1}
\cos\phi' &= (\cos\phi')_{\rm GR} + \zeta_{2} \left[-\frac{\delta \gamma}{\beta_{0}} + \frac{\delta \beta}{\beta_{0}^{2}}\left(\gamma_{0} - \chi_{\rm eff}M^{2}\right) 
\right.
\nonumber \\
&\left.
- \frac{\alpha}{\beta_{0}^{3}}\left(\gamma_{0} - \chi_{\rm eff} M^{2}\right)^{2}\right] + {\cal{O}}(\zeta_{2}^{2})
\end{align}
where one can show that $(\cos\phi')_{\rm GR} = (\chi_{\rm eff} M^{2} - \gamma_{0})/\beta_{0}$, which can be derived by manipulating Eq.~\eqref{eq:cos-phip-gr}. 
However, we had to make an assumption about the mass ratio to obtain the above expression, specifically $q\ne1$. This is due to the fact that $\beta_{0} \sim 1-q$, and therefore, $\beta_{0} = 0$ when $q=1$. Returning to Eq.~\eqref{eq:cos-phip-master} and solving when $q=1$, we obtain
\begin{equation}
\label{eq:cphi-temp}
\cos\phi' = \zeta_{2}^{-1/2} \left(\frac{\chi_{\rm eff} M^{2} - \gamma_{0}}{\alpha}\right)^{1/2} - \frac{\delta \beta}{2 \alpha} + {\cal{O}}(\zeta_{2}^{1/2})
\end{equation}
which does not reduce properly to the GR limit when $\zeta_{2} \rightarrow 0$. 

How can it be that the precession angle takes two drastically different functional forms in the equal and non-equal mass cases? And what's worse, how can one of these two expressions diverge in the GR limit? All of this indicates that the small coupling expansion $\zeta_{2} \ll 1$ is \textit{non-uniform} in the mass ratio $q$. What we mean by non-uniformity here is that the perturbative solution for $\cos\phi'$ in Eq.~\eqref{eq:cos-phip-1} becomes non-perturbative for some set of values of the mass ratio, i.e.~when $\zeta_2/\beta_0^3 \sim \zeta_2/(1-q)^3= {\cal{O}}(1)$. The cause of this non-uniformity is the fact that nutation is still present in the equal mass case in dCS gravity, while it is absent in GR. 

Non-uniformity is a common feature of multi-variable asymptotic expansions, i.e.~asymptotic expansions of more than one variable. One solution to this problem, and the one we adopt here, is to change variables to render the expansions uniform. 
%
In particular, we define a new parameter 
\begin{align}
\bar{\zeta}_{2} = \frac{\zeta_{2}}{(1-q)^{2}}\,,    
\end{align} 
which can be used instead of $\zeta_2$ when perturbatively solving the precession equations in such a way so as to render the expansions uniform in $q$. We focus on solutions to the precession problem to linear order in $\bar{\zeta}_{2}$, since then the solutions converge to GR in the limit $\bar{\zeta}_{2} \rightarrow 0$.

\subsubsection{Solutions to ${\cal{O}}(\bar{\zeta}_2)$}

The goal now is to analytically determine the time evolution of $S(t)$. We follow the same procedure for obtaining $S(t)$ in GR, specifically to obtain an equation for $dS^{2}/dt$. Following this procedure and after linearzing in $\bar{\zeta}_{2}$, we obtain
\begin{align}
\label{eq:S-eq}
\left(\frac{dS^{2}}{dt}\right)^{2} &= - A^{2} \left(S^{6} + B S^{4} + C_{0} S^{2} + D_{0}\right) 
\nonumber \\
&+ \bar{\zeta}_{2} {\cal{F}}\left(J, L, \chi_{\rm eff}, S_{1}, S_{2}, m_{1}, m_{2}; S\right)\,,
\end{align}
where ${\cal{F}}$ is a complicated functions of $S$. The coefficients $(A, B, C_{0}, D_{0})$ are still given by Eqs.~\eqref{eq:A-coeff}-\eqref{eq:D0-coeff}.

This equation is likely impossible to solve analytically for arbitrary $S$ due to the complexity of ${\cal{F}}$. However, the dCS corrections to the precession equations are only valid to second order in the spins of the compact objects. We thus expand ${\cal{F}}$ about $(S_{1}, S_{2}, S, \chi_{\rm eff})$ all simultaneously small compared to $M^{2}$. To do this, we define an order keeping parameter $\epsilon \sim (S_{1}, S_{2}, S, \chi_{\rm eff})$, and expand about $\epsilon$.  The end result of this expansion is
\begin{align}
\label{eq:Seq-dcs}
\left(\frac{dS^{2}}{dt}\right)^{2} &= - A^{2} \left[S^{6} + B S^{4} + \left(C_{0} + \bar{\zeta}_{2} \delta C\right) S^{2} 
\right.
\nonumber \\
&\left.
+ \left(D_{0}+ \bar{\zeta}_{2} \delta D\right)\right] + {\cal{O}}(\epsilon^{4})\,,
\end{align}
with
\begin{align}
\label{eq:dC}
\delta C &= \epsilon^{2} \; \delta C_{2}(J, L) + \epsilon^{3} \; \chi_{\rm eff} \; \delta C_{3}(J, L)\,,
\\
\label{eq:dD}
\delta D &= \epsilon^{2} \; \delta D_{2}(J, L, S_{1}, S_{2}) + \epsilon^{3} \; \chi_{\rm eff} \; \delta D_{3}(J, L, S_{1}, S_{2})\,,
\end{align}
%
where $(\delta C,\delta D)$ are given in Appendix~\ref{app:exp}, and we have explicitly written out the dependence on the angular momenta. The coefficients $(\delta D_{2}, \delta D_{3})$ depend on quadratic combinations of $S_{1}$ and $S_{2}$, while $(\delta C_{2}, \delta C_{3})$ are independent of the spin magnitudes of the BHs. We have here stopped the expansion at ${\cal{O}}(\epsilon^{3})$, even though the original precession equations are only accurate to ${\cal{O}}(\epsilon^{2})$. We expect this to be acceptable based on the known expression for the scalar dipole moment to all orders in spin. When re-expanded in small spins, the scalar dipole moment is $\mu_{1,2} \sim S_{1,2} + {\cal{O}}(S_{1,2}^{3})$. In the precession equations, the scalar dipole moment enters through the dipole-dipole interaction, which scales as $\vec{\mu}_{1} \times \vec{\mu}_{2}$. Thus, the next order terms in the dipole-dipole interaction in a small spin expansion  scale as $S_{1}^{3} S_{2} + (1 \leftrightarrow 2)$, which is ${\cal{O}}(\epsilon^{4})$. The same arguments applies to the quadrupole-monopole interaction.

Schematically, Eq.~\eqref{eq:Seq-dcs} takes the same form as the GR equation, and thus, its solution is given by Eq.~\eqref{eq:Soft-gr}. The difference between the GR and dCS solutions is contained in the constants $(S_{+}, S_{-}, S_{3}, \dot{\psi}, m)$, due to the fact that the coefficients of the cubic polynomial in Eq.~\eqref{eq:Seq-dcs} acquire dCS modifications. Formally, one should re-expand all of these quantities in $\bar{\zeta}_{2} \ll 1$. However, when we include radiation reaction, we have to perform averages of the precessing solution. These averages are simpler to take with the un-expanded solutions for $S(t)$ and $\phi_{z}(t)$.

Now consider the evolution of the precession angle $\phi_{z}$. Following the procedure described in Sec.~\ref{prec-phiz}, the evolution equations becomes
\begin{align}
\label{eq:dphizdt-dcs}
\frac{d\phi_{z}}{dt} &= \frac{Q_{0} + Q_{2} S^{2} + Q_{4} S^{4}}{P_{0} + P_{2} S^{2} + P_{4} S^{4}} 
\nonumber \\
&+ \bar{\zeta}_{2} {\cal{G}}\left(J, L, \chi_{\rm eff}, S_{1}, S_{2}, m_{1}, m_{2}; S\right)\,.
\end{align}
After expanding about small spins, and inserting the expression for $S(t)$, we obtain
\begin{align}
\label{eq:dphiz-dcs}
\frac{d\phi_{z}}{dt} &= J \left[\frac{b_{0} + b_{2} \text{sn}^{2}(\psi,m) + b_{4} \text{sn}^{4}(\psi,m)}{d_{0} + d_{2} \text{sn}^{2}(\psi,m) + d_{4} \text{sn}^{4}(\psi,m)}\right] 
\nonumber \\
&+ \bar{\zeta}_{2} J \left[\delta A'_{\phi} + \delta B'_{\phi} \text{sn}^{2}(\psi,m)\right]\,,
\end{align}
where $(\delta A'_{\phi}, \delta B'_{\phi})$ are constants, and the $b$ and $c$ coefficients are still given by the expressions in Appendix~\ref{app:exp}, but recall that $(S_{+}, S_{-}, S_{3})$ are different from the GR values. 
We write $\phi_{z} = \phi_{z}^{\rm GR}(t) + \bar{\zeta}_{2} \delta \phi_{z}'(t)$, where $\phi_{z}^{\rm GR}(t)$ is given in Eq.~\eqref{eq:phiz-GR}, and solve for $\delta \phi_{z}(t)$ to obtain
\begin{equation}
\label{eq:dphiz-prime}
J^{-1} \delta \phi_{z}'(t) = \left(\delta A_{\phi}' + \frac{\delta B_{\phi}'}{m}\right) \frac{\psi}{\dot{\psi}} - \frac{\delta B_{\phi}'}{m \dot{\psi}} \text{E}[\text{am}(\psi,m),m]\,.
\end{equation}
This completes the solution to the dCS precession equations in the absence of radiation reaction.

\section{\label{sec:rr} Radiation Reaction}
We now focus on including radiation reaction into the precessional dynamics. In dCS gravity, the flux of energy and angular momentum are modified due to scalar radiation, which results in Eq.~\eqref{eq:dudt}. The dCS coefficient $\delta C$ encodes corrections from both scalar radiation and modification of Kepler's third law. The latter of these implies that $L = (\eta M^{2}/u) (1 + 2 \delta C_{r} u^{4})$, so these dCS corrections enter at relative 2PN order in the dynamics of the binary. Meanwhile, the corrections to the precession equations Eq.~\eqref{eq:prec} enter at relative 0.5PN order. This is an important point that simplifies significantly many of the arguments and calculations in this section.

Radiation reaction does not conserve the direction of the total angular momentum $\vec{J}$. However,~\cite{Chatziioannou:2017tdw} showed that the direction of $\vec{J}$ is approximately conserved over a precession cycle when including radiation reaction. Oscillations induced in the $x-$ and $y-$components of $\hat{J}$ were shown to be suppressed by two orders of magnitude relative to $\hat{J}_{z}$, with the amplitude of the oscillations scaling as $v^{3}$. In a PN expansion, one can then treat $\hat{J}$ as fixed, with radiation reaction only changing the magnitude of the total angular momentum $J$. Since the dCS modifications to the angular momentum flux enter at 2PN order, they are suppressed by $v^{4}$ relative to GR effects, and the arguments presented in~\cite{Chatziioannou:2017tdw} still hold in dCS gravity.

Note, however, that nutational resonances where the ratio of the precessional and nutation frequencies becomes an integer ratio can cause a non-negligible secular change in the direction of $\vec{J}$~\cite{Zhao:2017tro}. In this work, we assume that the binary does not experience any nutational resonances during its coalescence. The impact of dCS modifications on such resonances will be considered in future work.

\subsection{Constants of Precessional Motion}
\label{sec:const}

After verifying that the direction of $\vec{J}$ remains approximately fixed under radiation reaction, we must consider what quantities are still constant in the dCS case. To leading PN order and ignoring horizon absorption, the masses, $m_{1}$ and $m_{2}$, and magnitudes of the spin vectors, $S_{1}$ and $S_{2}$, are constant. Radiation reaction changes the magnitude of the orbital angular momentum $L$, specifically $dL/dt = - \hat{L} \cdot \vec{{\cal{G}}}$, while spin-precession changes its direction $\hat{L}$. The magnitude of the total angular momentum evolves according to $dJ/dt = - \hat{J} \cdot \vec{{\cal{G}}}$. Combining these, we may write
\begin{equation}
\frac{dJ}{dL} = \cos \theta_{L} = \frac{J^{2} + L^{2} - S^{2}}{2 J L}\,,
\end{equation}
where only $S^{2}$ evolves on the precession timescale. To obtain the evolution of $J$ on the radiation reaction timescale, we may take a precession average to obtain
\begin{equation}
\Big\langle \frac{dJ}{dL} \Big\rangle_{\psi} = \frac{J^{2} + L^{2} - \langle S^{2} \rangle_{\psi}}{2JL}\,,
\end{equation}
where the average is performed with respect to the phase of $S^{2}$, specifically $\psi$. This equation can be solved exactly to obtain,
\begin{equation}
\label{eq:J-of-L}
J^{2} = L^{2} + 2 c_{1} L - L \int \frac{\langle S^{2} \rangle_{\psi}}{L^{2}} dL\,,
\end{equation}
where $c_{1}$ is an integration constant.

Reference~\cite{Chatziioannou:2017tdw} showed that the average $\langle S^{2} \rangle_{\psi}$ is a constant at leading PN order, and can be pulled out of the above integral. The average $\langle S^{2} \rangle_{\psi}$ varies on the radiation-reaction timescale at 0.5PN order. However, this constitutes a higher PN order correction to Eq.~\eqref{eq:J-of-L}, and can be neglected without introducing larger errors in the analytic approximations. In dCS, we now show that this holds true. We begin by computing the PN expansion of the roots $(S_{+}^{2}, S_{-}^{2}, S_{3}^{2})$, whose expressions are given by the procedure detailed in Appendix~\ref{app:spn}. 
Using Eq.~\eqref{eq:J-of-L}, the PN expansion of these quantities take the form
\begin{align}
\label{eq:Spm-pn}
S_{\pm}^{2}(u) &= \sum_{n=0} s_{\pm}^{(n)} u^{n} + \bar{\zeta}_{2} \left[\delta s_{\pm}^{(0)} + {\cal{O}}(u)\right]\,,
\\
\label{eq:S3-pn}
S_{3}^{2}(u) &= \sum_{n=0} s_{3}^{(n)} u^{n-2} + \bar{\zeta}_{2} \left[\delta s_{3}^{(2)} + {\cal{O}}(u)\right]\,,
\end{align}
where recall that $u = (2 \pi M F)^{1/3} = {\cal{O}}(v)$ and $v$ is the orbital velocity. 
Here $[s_{\pm}^{(n)},s_{3}^{(n)}]$ are the coefficients of the GR expansion, and are manifestily independent of $u$ (or $v$), thus making them constants on the radiation reaction timescale. These coefficients are given up to 2PN order in Appendix~\ref{app:spn}. The dCS corrections to these are
\allowdisplaybreaks[4]
\begin{align}
\delta s_{\pm}^{(0)} &= \pm \frac{M^{4}(1-q)^{2} \delta y_{0} \eta^{2}}{3q\sqrt{6y_{0}}} 
\nonumber \\
&+ \frac{25 c_{1} q^{3}}{24(1+q)^{2}} \left[c_{1} (1+q)^{2} - M^{2} q (3+q) \chi_{\rm c}\right]\,,
\\
\delta s_{3}^{(2)} &= - \frac{25}{12} \frac{c_{1} q^{3} \left[c_{1} (1+q)^{2} - M^{2} q (3+q) \chi_{\rm c}\right]}{(1+q)^{2}} 
\nonumber \\
&+ \delta s_{+}^{(0)} + \delta s_{-}^{(0)} - 2 M^{4} \eta \chi_{e,1}
\end{align}
where $\delta y_{0}$ is given in Appendix~\ref{app:spn}, and $\chi_{e,1}$ will be presented in Eq.~\eqref{eq:chi1}.

The average $\langle S^{2} \rangle_{\psi}$ can be computed exactly from Eq.~\eqref{eq:Soft-gr}, specifically
\begin{equation}
\label{eq:S-avg}
\langle S^{2} \rangle_{\psi} = \frac{1}{m} \left[ \left(m - 1\right) S_{+}^{2} + S_{-}^{2} + \frac{E(m)}{K(m)} \left(S_{+}^{2} - S_{-}^{2}\right)\right]\,,
\end{equation}
where $K$ and $E$ are the complete elliptic integrals of the first and second kind, respectively. Using Eqs.~\eqref{eq:Spm-pn}-\eqref{eq:S3-pn}, we may PN expand the expression for $m$ to obtain
\begin{align}
\label{eq:m-pn}
m &= \left(\frac{s_{-}^{(0)} - s_{+}^{(0)}}{s_{3}^{(0)}}\right) u^{2} \left[1 - \frac{\bar{\zeta}_{2}}{3}\sqrt{\frac{2}{3y_{0}}} \frac{\delta y_{0} M^{4} (1-q)^{2} \eta^{2}}{q \left(s_{+}^{(0)} - s_{-}^{(0)}\right)}\right]
\nonumber \\
&+ {\cal{O}}(u^{3})\,.
\end{align}
Inserting all of this into Eq.~\eqref{eq:S-avg} and PN expanding, we finally have
\begin{equation}
\label{eq:Savg-pn}
\langle S^{2} \rangle_{\psi} = \frac{1}{2} \left(s_{+}^{(0)} + s_{-}^{(0)}\right) + \frac{1}{2} \left(\delta s_{+}^{(0)} + \delta s_{-}^{(0)}\right) \bar{\zeta}_{2} + {\cal{O}}(u)\,.
\end{equation}
The last step is to use Eq.~\eqref{eq:Savg-pn} to evaluate Eq.~\eqref{eq:J-of-L}. After performing the necessary integration, we obtain
\begin{equation}
\label{eq:J-avg}
J^{2} = L^{2} + 2 c_{1} L + \langle S^{2} \rangle_{\psi,0} + \frac{1}{2} \left(\delta s_{+}^{(0)} + \delta s_{-}^{(0)}\right) \bar{\zeta}_{2} \,,
\end{equation}
where $\langle S^{2} \rangle_{\psi,0} = (1/2) (s_{+}^{(0)} + s_{-}^{(0)})$. The time evolution of $J$ is now purely determined by $L$, which evolves according to Eq.~\eqref{eq:dudt}. Note that the dCS modification to $L$ enters at 2PN order, while the correction to $J$ above enters at leading PN order because $\delta s_{\pm}^{(0)}$ is independent of $v$. Thus, we may replace $L$ with its GR expression in terms of $u$ without significant loss of accuracy.

The other two constants of precession that evolve under radiation reaction are $L$ and $\chi_{\rm eff}$, the latter of these due to the fact that it depends on $L$. This is a subtle difference to the case of GR, where $\chi_{\rm eff}$ is constant under radiation reaction up to 2.5PN order~\cite{Racine:2008qv}. Rather than work in terms of $L$ as the variable that changes on the radiation reaction timescale, it is simpler to work with $u$ to avoid additional steps when computing the Fourier domain waveform. The evolution equation for $u$ is given in Eq.~\eqref{eq:dudt}, where the dCS correction is a function of $[\hat{L}, \vec{S}_{1}, \vec{S}_{2}]$ through $\delta C$. The quantity $\delta C$ is a constant for spin-aligned binaries and does not require any special treatment when computing the Fourier domain gravitational waveform. However, for precessing binaries, $\delta C$ (and all similar coefficients) is oscillatory on the precession timescale, which enters the orbital phase of the binary. These oscillations can introduce  mathematical catastrophes when computing the Fourier domain waveform using the SPA, especially if the binary is strongly precessing. The method to avoid this catastrophes is to separate out the oscillatory terms from the phase, re-write these as corrections to the GW amplitude using a Bessel decomposition, and then re-group these terms into a new, secularly evolving phase. This method was developed in~\cite{Klein:2014bua} and is called the \textit{shifted uniform asymptotic} (SUA) method. In order to properly separate out the oscillatory effects, we must consider a multiple scale analysis (MSA) of all relevant quantities.

We begin by defining two timescales, $t_{\rm pr}$ and $t_{\rm rr}$ describing the precession and radiation reaction timescales, respectively. These are related via $t_{\rm pr} = \epsilon \; t_{rr}$ with $\epsilon$ a small parameter. It then follows that $d/dt = \partial/\partial t_{\rm pr} + \epsilon \; \partial /\partial t_{\rm rr}$. The relevant equation is $\dot{u} = \epsilon \; {\cal{U}}$, where ${\cal{U}}$ is given by the right-hand-side of Eq.~\eqref{eq:dudt} and $\epsilon$ is an order keeping parameter. We write $u(t_{\rm pr}, t_{\rm rr}) = u_{0}(t_{\rm pr}, t_{\rm rr}) + \epsilon \; u_{1}(t_{\rm pr}, t_{\rm rr}) + {\cal{O}}(\epsilon^{2})$, and work perturbatively in $\epsilon$. At leading order in $\epsilon$, we obtain the equation
\begin{equation}
\frac{\partial u_{0}}{\partial t_{\rm pr}} = 0\,,
\end{equation}
which is the statement that $u$ is unchanged during a precession cycle and in the absence of radiation reaction. At first order, we have
\begin{equation}
\label{eq:msa-1}
\frac{\partial u_{1}}{\partial t_{\rm pr}} + \frac{\partial u_{0}}{\partial t_{\rm rr}} = {\cal{U}}(t_{\rm pr}, t_{\rm rr})\,.
\end{equation}
To solve this, we exploit the fact that $u$ is oscillatory on the precession timescale, such that $\langle \partial u_{n}/\partial t_{\rm pr} \rangle_{\psi} = 0$ for all $n$. Taking the average, we obtain
\begin{align}
\label{eq:u0-avg}
\Big \langle \frac{du_{0}}{dt_{\rm rr}} \Big \rangle_{\psi} &= \langle {\cal{U}}\rangle_{\psi}(t_{\rm rr})
\nonumber \\
&= \frac{a_{0}}{3M} u_{0}^{9} \Bigg[1 + \sum_{n=2} \left(\langle a_{n} \rangle_{\psi} + 3 \langle b_{n} \rangle_{\psi} \ln u\right) u^{n}_{0} 
\nonumber \\
&+ 16 \frac{q^{4}(1-q)^{2}}{(1+q)^{4}} \bar{\zeta}_{2} \langle \delta C \rangle_{\psi} u_{0}^{4}\Bigg]\,.
\end{align}
To complete the solution, we insert this back into Eq.~\eqref{eq:msa-1} to obtain
\begin{align}
\label{eq:u1-osc}
u_{1}(\psi, t_{\rm rr}) &= u_{1,{\rm sec}}(t_{\rm rr}) + \int \frac{d\psi}{\dot{\psi}} \left[ {\cal{U}}(\psi, t_{\rm rr}) - \langle {\cal{U}}\rangle_{\psi} (t_{\rm rr}) \right]\,,
\end{align}
where we have performed a change of variable in the integrand from $t_{\rm pr}$ to $\psi$. The above expression fixes $u_{1}$ up to a secular term $u_{1,{\rm sec}}$ that only varies on the radiation-reaction timescale. This quantity can be obtained by going to higher order in the MSA. As we will show, $u_{1}$ is not necessary when comparing to numerical evolutions of the dynamics, and we need only consider the secularly evolving $u_{0}$.

Now consider the effective mass-weighted spin and how it evolves under radiation reaction in dCS gravity. Due to its dependence on $u$, the evolution equation becomes
\begin{align}
\label{eq:dchi-eff-dL}
\frac{d\chi_{\rm eff}}{du} &= \lambda \bar{\zeta}_{2} \left[f_{5}(q) \left(\vec{S}_{1} \cdot \hat{L}\right)^{2} + q f_{3}(q) \left(\vec{S}_{1} \cdot \hat{L}\right) \left(\vec{S}_{2} \cdot \hat{L}\right)\right]\,,
\\
\lambda &= -\frac{201}{112} \frac{(1-q)^{2}(1+q)}{q^{2} M^{4}}\,.
\end{align}
To leading order in MSA, the secular evolution is governed by the average of Eq.~\eqref{eq:dchi-eff-dL} over $\psi$. The necessary precession averages are performed in Appendix~\ref{app:avg}; after applying said results, we find
\begin{equation}
    \Big\langle \frac{d\chi_{\rm eff}}{du} \Big\rangle_{\psi} = \lambda \bar{\zeta}_{2} \left[\Sigma_{0} + \Sigma_{1} \chi_{\rm eff} + \Sigma_{2} \chi_{\rm eff}^{2}\right]
\end{equation}
where the $\Sigma_{n}$ are constants on the radiation reaction timescale and are given in Appendix~\ref{app:rr-coeffs}. Solving this equation in the limit $\bar{\zeta}_{2} \ll 1$ gives
\begin{align}
    \label{eq:chieff-of-t}
    \chi_{\rm eff} &= \chi_{c} + \chi_{e,1} \bar{\zeta}_{2} u\,,
    \\
    \label{eq:chi1}
    \chi_{e,1} &= \lambda \left(\Sigma_{0} + \Sigma_{1} \chi_{c} + \Sigma_{2} \chi_{c}^{2}\right)
\end{align}
where $\chi_{c}$ is an integration constant, and plays the role of the standard $\chi_{\rm eff}$ of GR. This completes the discussion of the evolution of the constants of precession under radiation reactions. We detail how to handle the evolution of $u_{0}$ in Sec.~\ref{sec:orb-rr}.

To highlight the accuracy of the approximations used herein, we compare the analytic expressions for $J$ in Eq.~\eqref{eq:J-of-L} and $\chi_{\rm eff}$ in Eq.~\eqref{eq:chieff-of-t} to numerical evolutions of these quantities under radiation reaction. To obtain the numerical evolutions, we numerically integrate the precession equations in Eq.~\eqref{eq:prec} for several binaries with masses and spins provided in Table~\ref{tab1}. We start the numerical integration at an orbital frequency of 5 Hz, which fixes the initial value of $L$. The values of $\theta_{L}$ are given in Table~\ref{tab1}. We choose $J = L + (1/2)(\chi_{1} m_{1}^{2} + \chi_{2} m_{2}^{2})$, and $\phi'=0$ for initial conditions, which fixes the initial orientations of all of the angular momentum vectors and the initial value of $\chi_{\rm eff}$. The dimensionless dCS coupling parameter $\bar{\zeta}_{2}$ for each system is provided in Table~\ref{tab1}. The numerical integrations are performed in \texttt{Mathematica} with the \texttt{NDSolve} module using the \texttt{ImplicitRungeKutta} method. On the other hand, the analytic solutions depend on integration constants $[c_{1},\chi_{c},\psi_{c},\phi_{z,c}]$. To fix these we require the initial values of these quantities to be the same as those for the numerical evolutions.

\begin{figure*}
    \centering
    \includegraphics[scale=0.37, trim = 2cm 0.5cm 2cm 0.5cm, clip]{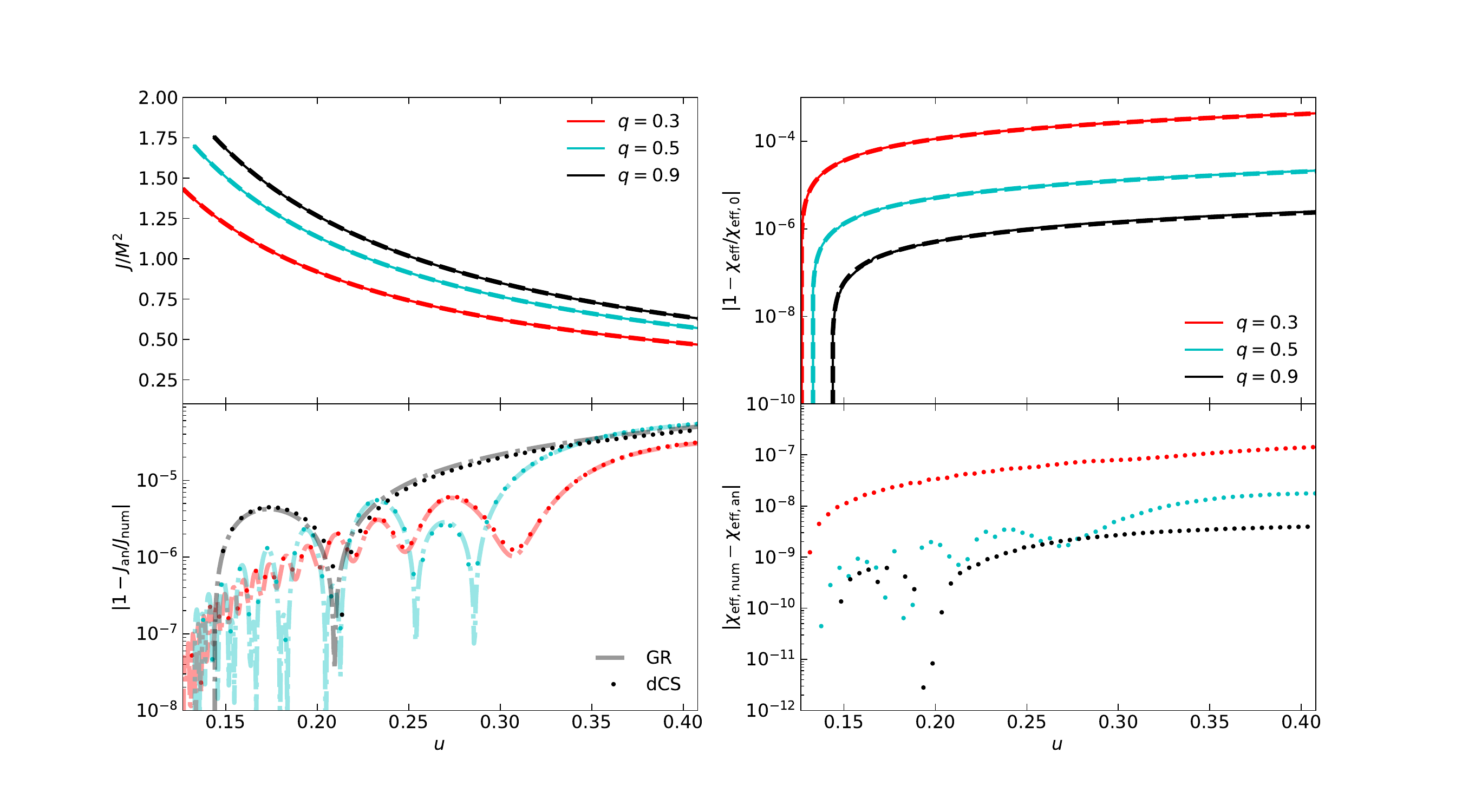}
    \includegraphics[scale=0.37, trim = 2cm 0.5cm 2cm 0.5cm, clip]{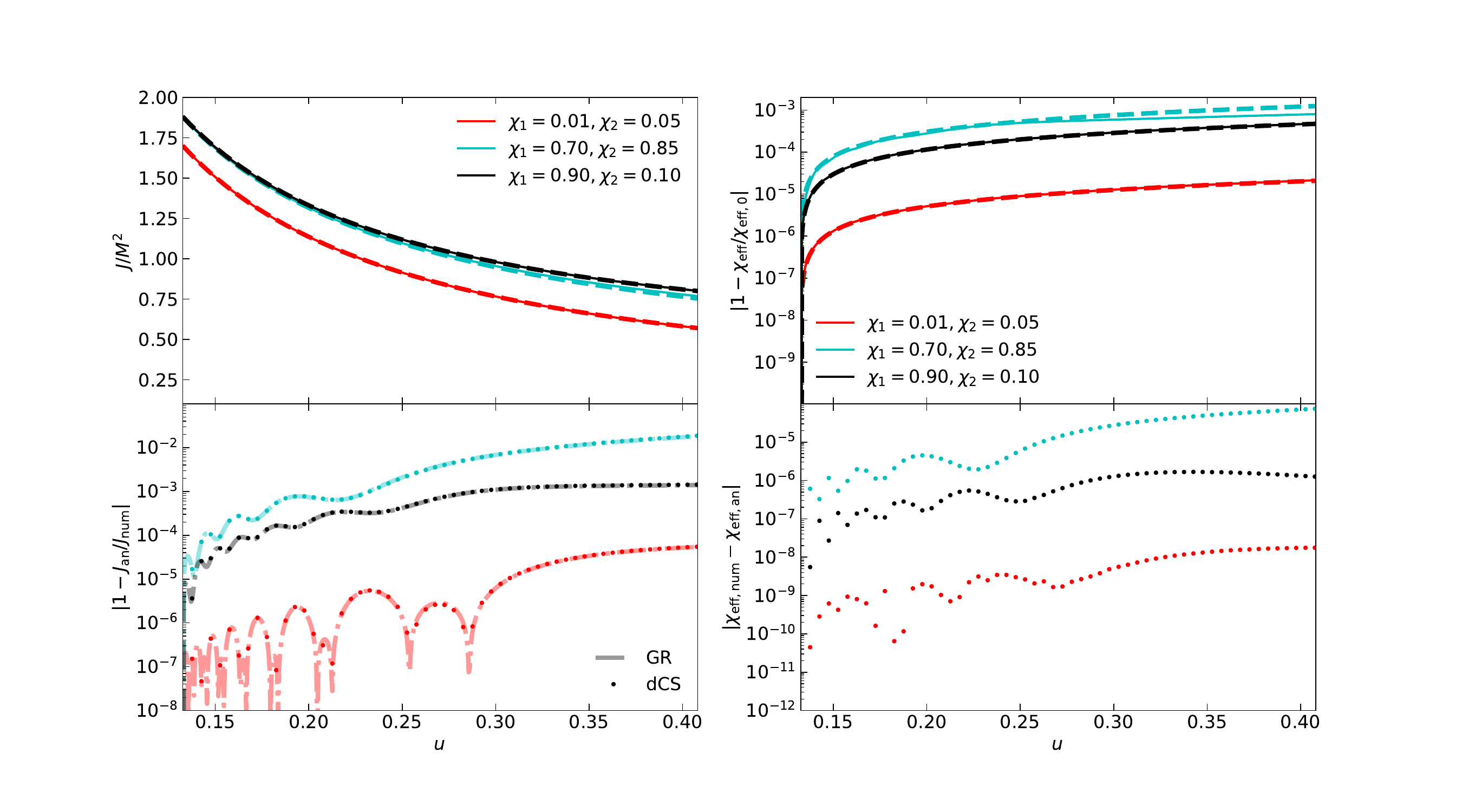}
    \caption{Left: Comparison of the analytic solution for $J(u)$ in Eq.~\eqref{eq:J-of-L} (dashed lines) to numerical evolutions of the coupled system of Eq.~\eqref{eq:prec} \&~\eqref{eq:dudt} (solid lines). The bottom panel provides the absolute error between the two solutions (dashed lines). We provide the error in the GR limit ($\bar{\zeta}_{2}\rightarrow0$) for reference (solid lines). Right: Same as the left but for $\chi_{\rm eff}$, whose analytic approximation is given in Eq.~\eqref{eq:chieff-of-t}. We plot the combination $|1-\chi_{\rm eff}/\chi_{\rm eff,0}|$ to better display the evolution, where $\chi_{\rm eff,0}$ is the initial value of the evolution. We do not provide the error in the GR limit in the bottom panel since their is no GR analog in this case.
    } 
    \label{fig:J-chi}
\end{figure*}

The comparison between the numerical evolution (solid lines) of $(J,\chi_{\rm eff})$ and their analytic approximations (dashed lines) is shown in the top panels of Fig.~\ref{fig:J-chi}, with the bottom panels showing the relative fractional error explicitly. For $J$, we provide the relative fractional error in the GR limit (dot-dashed lines) as well as in  the dCS (dotted) case described above. From this, we see that the error does not change significantly when adding the dCS corrections, indicating that the dominant uncertainty is largely controlled by the PN sequence in GR. We do not show this for $\chi_{\rm eff}$ since this is a constant in GR at the PN order we are working. Observe also that the magnitude of the uncertainty is below $10^{-4}$ and $10^{-7}$ for $J$ and $\chi_{\rm eff}$ respectively for slowly-spinning systems. The error increases for more rapidly spinning system, but it is always below a few percent and below $10^{-4}$ for $J$ and $\chi_{\rm eff}$ respectively.
The error in $J$, even in the GR case, increases with increasing spin, specifically with increasing $\chi_{1} + \chi_{2}$. This is a result of truncating the average $\langle S^{2}\rangle_{\psi}$ to leading PN order, since higher PN order terms scale as higher powers of the spins. One could improve this error by including these terms in the PN computations herein if greater accuracy is desired. 

\begin{table*}[t]
\centering
\begin{tabular}{c c c c c c c c c c c}
    \hline
    \hline
    $m_{1}[M_{\odot}]$ & $m_{2}[M_{\odot}]$ & $\chi_{1}$ & $\chi_{2}$ & $u_{\rm 10 Hz}$ & $\theta_{L}$ & $\chi_{\rm eff}$ & $\bar{\zeta}_{2}$ & $\langle \delta \beta_{3}\rangle_{\psi}^{\rm dCS}$ & $\langle \delta \sigma_{4} \rangle_{\psi}^{\rm dCS}$ & $g(q)\langle \delta C \rangle_{\psi}$\\
    \hline
    10 & 9 & 0.10 & 0.05 & 0.14 &$\pi/300$ & $3.5\times10^{-2}$ & $2.1\times10^{-2}$ & $7.0\times10^{-5}$ & $-6.3\times10^{-6}$ & $-1.5\times10^{-5}$\\
    10 & 5 & 0.10 & 0.05 & 0.13 &$\pi/150$ & $2.9\times10^{-2}$ & $3.5\times10^{-2}$ & $3.0\times10^{-3}$ & $-2.2\times10^{-4}$ & $-7.5\times10^{-5}$\\
    10 & 3 & 0.10 & 0.05 & 0.13 & $\pi/90$ & $3.1\times10^{-2}$ & $1.4\times10^{-1}$ & $3.3\times10^{-3}$ & $-2.5\times10^{-4}$ & $-5.4\times10^{-5}$\\
    \hline
    \hline
    10 & 5 & 0.10 & 0.05 & 0.13 & $\pi/150$ & $2.9\times10^{-2}$ & $3.5\times10^{-2}$ & $3.0\times10^{-3}$ & $-2.2\times10^{-4}$ & $-7.5\times10^{-5}$\\
    10 & 5 & 0.70 & 0.85 & 0.13 & $\pi/30$ & $1.7\times10^{-1}$ & $3.5\times10^{-2}$ & $2.9\times10^{-2}$ & $-1.2\times10^{-2}$ & $-1.2\times10^{-4}$\\
    10 & 5 & 0.90 & 0.10 & 0.13 & $\pi/16$ & $2.4\times10^{-1}$ & $3.5\times10^{-2}$ & $1.8\times10^{-2}$ & $-1.1\times10^{-2}$ & $1.1\times10^{-4}$\\
    \hline
    \hline
\end{tabular}
\caption{Parameters of the binary systems considered in Fig.~\ref{fig:J-chi}-\ref{fig:amp}. The first four columns provide the masses $m_{1}$ and $m_{2}$ and dimensionless spin parameters $\chi_{1}$ and $\chi_{2}$. The systems appearing in the upper half of the table have varying mass ratio with fixed spin magnitudes, and the corresponding systems are plotted in the top panels of each figure. The systems in the bottom half of the table have varying spin magnitudes with fixed mass ratio, with the corresponding plots being the bottom panels of each figure. The fifth column provides the value of the PN expansion parameter $u$ when $2F = 10$ Hz (recall that the orbital velocity is $v_{\rm orb} \sim u$). The last three columns provide the values of the dCS spin-orbit correction, spin-spin correction, and dipole radiation term appear in Eq.~\eqref{eq:spa}, with $g(q) = q^{4}(1-q)^{2}/(1+q)^{4}$.}
\label{tab1}
\end{table*}
\subsection{Precession Phases}
\label{sec:prec-rr}

We now consider the evolution of precession quantities on the radiation reaction timescale. To begin, the time evolution of the total spin magnitude $S(t)$ is still governed by Eq.~\eqref{eq:Seq-dcs}, but now, the coefficients $[A, B, C_{0}, D_{0}, \delta C, \delta D]$ are functions that change on the radiation reaction timescale. To leading order in multiple scale analysis, the solution is still given by Eq.~\eqref{eq:Soft-gr}, but with
\begin{equation}
\label{eq:dpsidt-msa}
\frac{d\psi}{dt_{\rm rr}} = \frac{A(t_{\rm rr})}{2} \sqrt{S_{+}^{2}(t_{\rm rr}) - S_{3}^{2}(t_{\rm rr})}\,.
\end{equation}
This equation can be integrated directly in a PN expansion by combining it with Eq.~\eqref{eq:u0-avg}. Caution must be taken since $A$ depends on $\chi_{\rm eff}$, and its time dependence cannot be neglected. We treat $\chi_{\rm eff}$ through Eq.~\eqref{eq:chieff-of-t}, and PN expand the above expression. After integrating, we obtain
\begin{align}
    \label{eq:psi-pn}
\psi &= \psi_{c} - \frac{5}{128} \frac{(1-q^{2})}{q} u^{-3} \Bigg(1 + \sum_{n\neq3} \psi_{n} u^{n} 
\nonumber \\
&+ \psi_{3} \ln u + \bar{\zeta}_{2} \delta \psi_{2} u^{2} \Bigg)\,,
\end{align}
where $\psi_{c}$ is an integration constant, the $\psi_{n}$ are the PN coefficients within GR and are given in Appendix~\ref{app:rr-coeffs}, and $\delta\psi_{2}$ is the leading PN order dCS correction, which is actually a 1PN deviation from GR and is given in Eq.~\eqref{eq:dpsi2} of Appendix~\ref{app:rr-coeffs}. This may seem rather strange considering that the corrections to the spin precession equations in Eq.~\eqref{eq:prec} enter at 0.5PN order. However, this deviation arises due to the shifts in $S_{+}^{2}$ and $S_{3}^{2}$ due to dCS corrections, which actually enter at 1PN order as can be seen from Eq.~\eqref{eq:S3-pn}.

For the precession phase, we write $\phi_{z} = \epsilon^{-1} \phi_{z,-1}(t_{\rm pr}, t_{\rm rr}) + \phi_{z,0}(t_{\rm pr}, t_{\rm rr}) + {\cal{O}}(\epsilon)$. To first order in the MSA, we have
\begin{align}
\frac{\partial \phi_{z,-1}}{\partial t_{\rm pr}} &= 0\,,
\\
\label{eq:msa1}
\frac{\partial \phi_{z,0}}{\partial t_{\rm pr}} + \frac{\partial \phi_{z,-1}}{\partial t_{\rm rr}} &= \Omega_{z}\,,
\end{align}
where $\Omega_{z}$ is given by the right-hand side of Eq.~\eqref{eq:dphiz-dcs}. The first of these implies $\phi_{z,-1} = \phi_{z,-1}(t_{\rm rr})$. To solve the second equation, we apply the same procedure of Sec.~\ref{sec:const}. Averaging this equation over $t_{\rm pr}$, we obtain
\begin{align}
\label{eq:dphiz-1}
\frac{d \phi_{z,-1}}{d t_{\rm rr}} &= \langle \Omega_{z} \rangle _{\psi}
\nonumber \\
&= J \Bigg\langle \frac{b_{0} + b_{2} \text{sn}^{2}(\psi,m) + b_{4} \text{sn}^{4}(\psi,m)}{d_{0} + d_{2} \text{sn}^{2}(\psi,m) + d_{4} \text{sn}^{4}(\psi,m)} \Bigg\rangle_{\psi} 
\nonumber \\
&+ \bar{\zeta}_{2} J \langle \delta A_{\phi} + \delta B_{\phi} \text{sn}^{2}(\psi, m)\rangle_{\psi}\,.
\end{align}
As pointed out in~\cite{Chatziioannou:2017tdw}, there is no closed form expression for the average of the first term above for arbitrary values of $m$. However, $m\sim u^{2}$ and is thus small in the inspiral phase of the binary. To leading order in $m\ll1$, $\text{sn}(\psi,m)=\sin(\psi)$ and Eq.~\eqref{eq:dphiz-1} now only depends on trigonometric functions.

Unfortunately, even after taking $m\rightarrow0$, there is still no closed form expression for the precession average of $d\phi_{z,-1}/dt_{\rm rr}$. To work around this, we make use of the fact that the $b_{n}$ and $d_{n}$ coefficients are functions of velocity (or $u$), and we PN expand the right-hand side of Eq.~\eqref{eq:dphiz-1}. There are two subtle aspects to doing this. First, we do not PN expand the overall factor of $J$ within Eq.~\eqref{eq:dphiz-1}. The reason for this was pointed out in~\cite{Chatziioannou:2017tdw}, wherein it was realized that expanding the analytic expression for $J$ in Eq.~\eqref{eq:J-of-L} actually constitutes a small mass ratio expansion since $L \sim \eta$. Thus, the expansion loses accuracy as one varies the mass ratio, and to avoid this, we factor out $J$ as was done in~\cite{Chatziioannou:2017tdw}. The second issue is with the PN expansion that follows. When PN expanding the $b_{n}$ coefficients, there are subtle cancellations that occur due to their structure. It is actually easier to recast the ${\cal{O}}(\bar{\zeta}_{2}^{0})$ part of Eq.~\eqref{eq:dphiz-1} in terms of the coefficients used in~\cite{Chatziioannou:2017tdw}. This mapping is given in Appendix~\ref{app:exp}.

As an example of how to properly proceed with the averaging, consider the calculation to leading PN order and within GR (i.e. $\bar{\zeta}_{2} = 0$). Equation~\eqref{eq:dphiz-1} becomes
\begin{equation}
    \label{eq:dphiz-temp}
    \frac{d\phi_{z,-1}}{dt_{\rm rr}} = J u^{6} \left[\frac{3+2\eta}{4\eta} + \frac{\kappa_{0}}{1 + \kappa_{1} \sin^{2}\psi} \right] + {\cal{O}}(J u^{7})\,.
\end{equation}
where $(\kappa_{0},\kappa_{1})$ are known constants, the latter of which is given in Eq.~\eqref{eq:k1}. By direct integration of the above expression with respect to $\psi$, one obtains
\begin{align}
    \int d\psi \frac{d\phi_{z,-1}}{dt_{\rm rr}} &= J u^{6} \left[\frac{3+2\eta}{4\eta} \psi
    \right.
    \nonumber \\
    &\left.
    + \frac{\kappa_{0}}{\sqrt{1+\kappa_{1}}} \tan^{-1}\left(\sqrt{1+\kappa_{1}} \tan\psi\right)\right]
\end{align}
The above expression has a branch cut within it at $\psi=\pi/2$. Evaluating this at the endpoints $\psi=0$ and $\psi=\pi$ is necessary to achieve the precession averaged evolution of $\phi_{z,-1}$, but due to the branch cut, we would obtain the incorrect average. To fix this behavior, we rely on a well-known technique in the modeling of eccentric binaries to remove the branch cut from the above expression. Specifically, we perform the replacement
\begin{align}
    \label{eq:atan-rep}
    \tan^{-1}\left(\sqrt{1+\kappa_{1}} \tan\psi\right) \rightarrow \psi + \tan^{-1}\left[\frac{\beta_{z} \sin(2\psi)}{1-\beta_{z} \cos(2\psi)}\right]\,,
\end{align}
with
\begin{align}
    \beta_{z} &= \frac{1}{e_{z}} \left(1 - \sqrt{1-e_{z}^{2}}\right)\,.
    \\
    e_{z} &= \frac{\kappa_{1}}{2+\kappa_{1}}\,.
\end{align}
The average can now be appropriately taken since the second term in Eq.~\eqref{eq:atan-rep} is purely oscillatory and possesses no branch cuts.

Returning to the calculation in dCS gravity, the procedure for averaging only acquires one extra step, namely an expansion about $\bar{\zeta}_{2} \ll 1$ before doing a PN expansion. The calculation is rather lengthy as all of the coefficients $(J, a, b_{n}, d_{n})$ in Eq.~\eqref{eq:dphiz-1} are shifted from their GR expressions. However, following the above procedure is straightforward and produces
\begin{align}
    \label{eq:dphiz-avg}
    \bigg\langle \frac{d\phi_{z,-1}}{dt_{\rm rr}}\bigg\rangle_{\psi} &= J_{0} \sum_{n=6} \Omega_{z,n}^{(-1)} u^{n} 
    \nonumber \\
    &+ \bar{\zeta}_{2} \left[\frac{\Omega_{z,6}^{(-1)}}{2 J_{0}} \left(\delta s_{+}^{(0)} + \delta s_{-}^{(0)}\right) + J_{0} \delta \Omega_{z,6}^{(-1)}\right]u^{6}\,.
\end{align}
In the above expression, $\Omega_{z,n}$ are the coefficients of the GR expression, $\delta \Omega_{z,6}$ is the leading PN order correction coming from the $(a,b_{n},d_{n}, A_{\phi}, B_{\phi})$ coefficients, the first term in the square brackets comes from the expansion of $J$ about $\bar{\zeta}_{2} \ll 1$, and $J_{0} = J(\bar{\zeta}_{2} \rightarrow 0)$ with $J$ given in Eq.~\eqref{eq:J-of-L}. Also, $J_{0}$ is a function of $u$ through $L$, which as discussed in Sec.~\ref{sec:const} takes its standard GR mapping. To integrate this, we divide by Eq.~\eqref{eq:dudt} and PN expand. After integrating with respect to $u$, we obtain
\begin{align}
    \label{eq:phiz-sec}
    \phi_{z,-1} &= \phi_{z,c} +  \sum_{n=-3} \Phi_{z,n}^{(-1)} \; \varphi_{n}(u) 
    \nonumber \\
    &+ \bar{\zeta}_{2} \left[\delta \Phi_{z,-3}^{(-1)} \; \varphi_{-3}(u) + \Phi_{z,-3}^{(-1)} \; \delta \varphi_{-3}(u)\right]\,,
\end{align}
where $\phi_{z,c}$ is an integration constant, and the constants $(\Phi_{z,n},\delta\Phi_{z,-3})$ and functions $[\phi_{n}(v),\delta\phi_{-3}(v)]$ are given in Appendix~\ref{app:rr-coeffs}. As we point out there, the coefficients $\Omega_{z,n}$, and as a result $\Phi_{z,n}$, found here are not the same as those in~\cite{Chatziioannou:2017tdw}. We discuss this more in Appendix~\ref{app:phiz-comp}.

As a last step, we complete the solution to Eq.~\eqref{eq:msa1} by finding the leading order oscillatory correction to $\phi_{z}$. Following the procedure of Eq.~\eqref{eq:u1-osc}, we obtain
\begin{align}
    \label{eq:phiz0}
    \phi_{z,0} &= \Omega_{\rm osc} J_{0} u^{6}\left\{\left(\Upsilon^{(0)}_{z,2} + \bar{\zeta}_{2} \delta \Upsilon_{z,2}^{(0)}\right) \tan^{-1}\left[\frac{\beta_{z} \tan(2\psi)}{1-\beta_{z} \cos(2\psi)}\right]
    \right.
    \nonumber \\
    &\left.
    + \bar{\zeta}_{2} \left[\frac{\Delta_{z,2}^{(0)} \sin(2\psi)}{1 + \kappa_{1} \sin^{2}\psi} + \Sigma^{(0)}_{z,2} \sin(2\psi)\right] \right\} + {\cal{O}}(J_{0} u^{7})\,,
\end{align}
where the coefficients $(\Upsilon_{z,2}^{(0)}, \delta \Upsilon_{z,2}^{(0)}, \Delta_{z,2}^{(0)}, \Sigma_{z,2}^{(0)})$ are given in Appendix~\ref{app:rr-coeffs}. The above result could be extended to higher PN order if one desires more accuracy, but we do not do so here. Further, recall from our discussion following Eq.~\eqref{eq:u1-osc} that this only fixes $\phi_{z,0}$ up to a purely secular correction. To find said secular corrections, one would have to carry the MSA to higher order. Lastly, this expression in the GR limit differs from the one presented in~\cite{Chatziioannou:2017tdw}, namely Eq.~(67) therein. While in principle they are equivalent solutions, the latter has known problems with branch cuts when $\psi=n\pi/2$ with $n$ an integer, whereas Eq.~\eqref{eq:phiz0} does not.

\begin{figure*}
    \centering
    \includegraphics[scale=0.37, trim=2cm 0.5cm 2cm 0.5cm, clip]{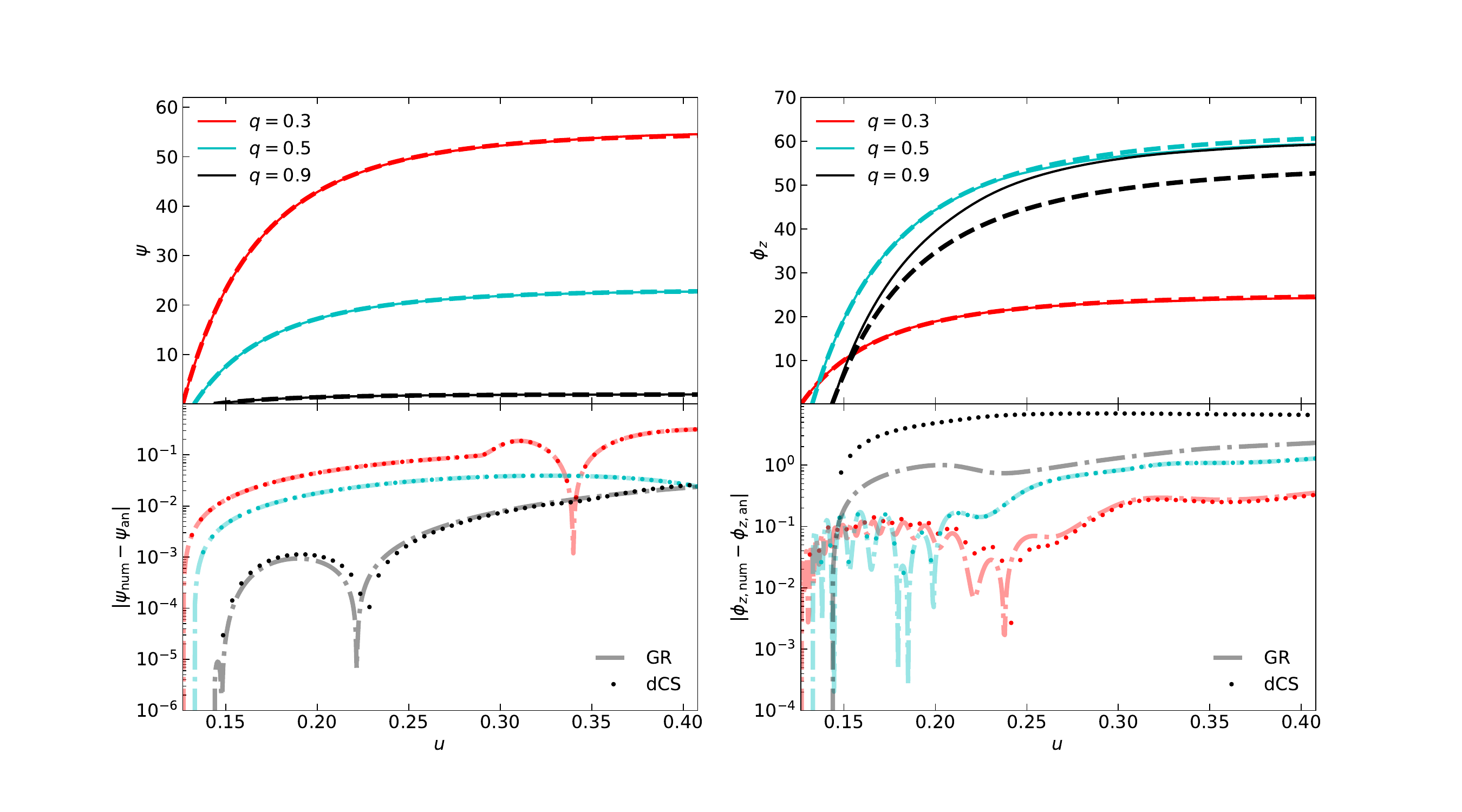}
    \includegraphics[scale=0.37, trim=2cm 0.5cm 2cm 0.5cm, clip]{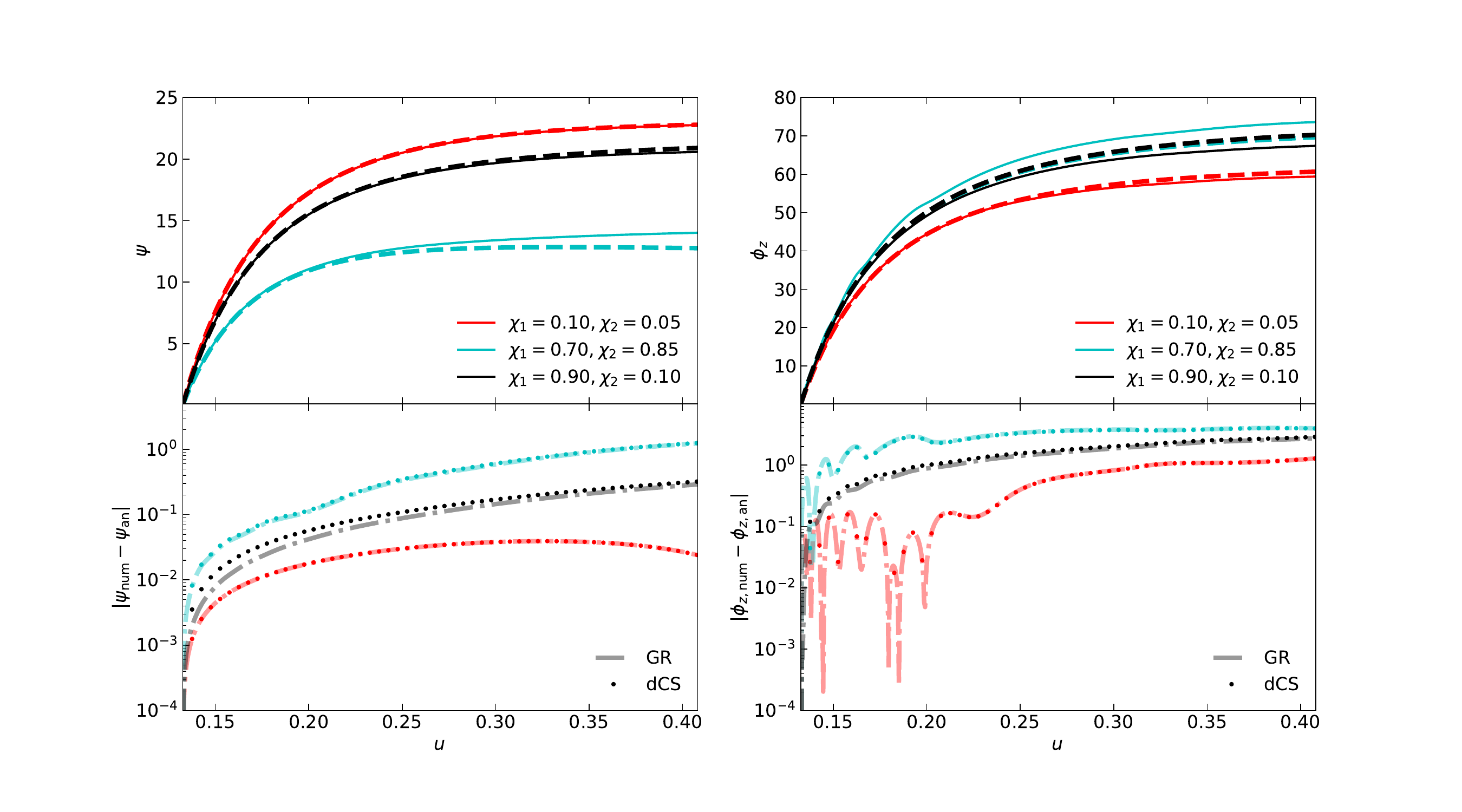}
    \caption{Left: Comparison of the analytic solution for $\psi(u)$ in Eq.~\eqref{eq:psi-pn} (dashed lines) to numerical evolutions of Eq.~\eqref{eq:dpsidt} (solid lines). The bottom panel provides the dephasing between the two solutions (dashed lines). The dephasing for the same systems in GR (i.e. $\bar{\zeta}_{2}=0$) is shown for reference (solid lines). Right: Same as the left but for $\phi_{z}(u)$. We neglect the oscillatory corrections from the MSA since they are PN suppressed.}
    \label{fig:prec}
\end{figure*}

Figure~\ref{fig:prec} provides a comparison between the analytic approximations of Eq.~\eqref{eq:psi-pn} for $\psi(u)$ and Eq.~\eqref{eq:phiz-sec} for $\phi_{z}(u)$, versus the numerical evolution of Eqs.~\eqref{eq:dpsidt} and~\eqref{eq:dphiz-dcs} for the systems in Table~\ref{tab1}. For $\phi_{z}(u)$ in the right panels, we only use the secular $\phi_{z,-1}$ as the analytic approximation,
and only take the leading PN order terms therein. The oscillatory effects are PN suppressed and do not provide significant improvement in the analytic approximation beyond removing oscillations in the dephasing. Once again, we also provide the dephasing for the GR solutions to compare against. Observe that there are no significant changes in the dephasing when allowing the dCS coupling parameter $\bar{\zeta}_{2}$ to be non-zero, with the exception of the $q=0.9$ case. This is a result of the unavoidable non-uniform expansion in the dCS coupling described below Eq.~\eqref{eq:cphi-temp}. Decreasing the dCS coupling parameter for this system results in better accuracy compared to the numerical integration. In spite of this, observe that the analytic approximation to the nutation phase evolution is accurate to better than a few radians for all systems considered.  

\subsection{Orbital Phases}
\label{sec:orb-rr}

To complete our discussion of radiation reaction effects, we need the evolution of orbital quantities under radiation reaction in order to compute the Fourier domain waveform. More specifically, we require $t(u)$ and $\phi(u)$. The former of these can be computed to leading order in the MSA by inverting Eq.~\eqref{eq:u0-avg}, which after integration becomes 
%
\begin{align}
    \label{eq:tofu}
    t(u) &= t_{c} - \frac{3 M}{8 a_{0}} u_{0}^{-8} \bigg\{1 + \sum_{n=2} \left(\langle t_{n} \rangle_{\psi} + \langle t_{n}^{l} \rangle_{\psi} \ln u_{0}\right) u_{0}^{n}
    \nonumber \\
    &+ \bar{\zeta}_{2} \left[\frac{8}{5} \langle \delta \beta_{3} \rangle_{\psi}^{\rm dCS} u_{0}^{3} + 2 \langle \delta \sigma_{4} \rangle_{\psi}^{\rm dCS} u_{0}^{4} 
    \right.
    \nonumber \\
    &\left.
    - 32 \frac{q^{4}(1-q)^{2}}{(1+q)^{4}} \langle \delta C\rangle_{\psi} u_{0}^{4}\right] \bigg\}
\end{align}
%
where $t_{c}$ is an integration constant, while $\langle \delta \beta_{3} \rangle_{\psi}^{\rm dCS}$ and $\langle \delta \sigma_{4}\rangle_{\psi}^{\rm dCS}$ are given in Eqs.~\eqref{eq:db3-avg} and~\eqref{eq:ds4-avg}, respectively.
These terms come from the averaging of $a_{3}$ and $a_{4}$, which contain the spin-orbit contribution $\beta_{3}$ and spin-spin contribution $\sigma_{4}$. These quantities depend on powers of dot products of the form $(\hat{L} \cdot \vec{S}_{A})$, which are modified from GR due to their dependence on $J$, $S^{2}$, and $\chi_{\rm eff}$. As we show in Appendix~\ref{app:avg}, the correction to these quantities enter at leading PN order, and thus enter Eq.~\eqref{eq:tofu} at relative 1.5PN order and 2PN order, respectively. Contrast this to the spin-aligned limit discussed in Sec.~\ref{sec:qc} where the leading-order correction enters at relative 2PN order due to dipole radiation, and is included in Eq.~\eqref{eq:tofu} through $\langle \delta C\rangle_{\psi}$ which is given in Eq.~\eqref{eq:dC-avg}. 

To find $\phi_{\rm orb}(u)$, we use the fact that $d\phi_{\rm orb}/dt = 2\pi F = u^{3}/M$. Working within the MSA, we may divide this expression by Eq.~\eqref{eq:u0-avg} and integrate to obtain 
%
\begin{align}
    \label{eq:phiofu}
    \phi_{\rm orb}(u) &= \phi_{c} - \frac{3M}{5a_{0}} u_{0}^{-5} \bigg\{1 + \sum_{n=2} \left(\langle \phi_{n} \rangle_{\psi} + \langle \phi_{n}^{l} \rangle_{\psi} \ln u_{0}\right) u_{0}^{n}
    \nonumber \\
    &+ \bar{\zeta}_{2} \left[\frac{5}{2} \langle \delta \beta_{3}\rangle_{\psi}^{\rm dCS} u_{0}^{3} + 5\langle \delta \sigma_{4}\rangle_{\psi}^{\rm dCS} u_{0}^{4}
    \right.
    \nonumber \\
    &\left.
    - 80 \frac{q^{4}(1-q)^{2}}{(1+q)^{4}} \langle \delta C\rangle_{\psi} u_{0}^{4}\right] \bigg\}
\end{align}
with $\phi_{c}$ an integration constant.

Lastly, for precessing binaries, the orbital phase $\phi$ is modulated by the Thomas phase $\phi_{T}$\footnote{Here we use $\phi_{T}$ for the Thomas phase, as opposed to~\cite{Chatziioannou:2017tdw} where $\zeta$ was used. We do so as to not create confusion with the dCS coupling parameters.}, which captures the Lens-Thirring effect. This quantity obeys the equation
\begin{align}
    \frac{d\phi_{T}}{dt} = \frac{d\phi_{z}}{dt} \cos\theta_{L}\,.
\end{align}
We solve this by once again working with MSA, whereby we treat $\phi_{T}$ in the same manner as $\phi_{z}$. There is one main difference, namely that the overall factor of $J$ that we factored out of Eq.~\eqref{eq:dphiz-1} actually cancels with a factor of $J$ in the denominator of $\cos\theta_{L}$. Thus, we can proceed with a standard PN expansion, without having to worry about possible issues with expanding $J$. The end result is
\begin{align}
    \label{eq:phiT-1}
    \phi_{T,-1}(u) &= \phi_{T,c} + u^{-3} \left[\sum_{n=0} \Phi_{T,n}^{(-1)} u_{0}^{n} + \bar{\zeta}_{2} \delta \Phi_{T,0}^{(-1)}\right]
\end{align}
where the coefficients $\Phi_{T,n}^{(-1)}$ and $\delta \Phi_{T,n}^{(-1)}$ are given in Appendix~\ref{app:rr-coeffs}. This completes the calculation of all necessary radiation reaction effects.

\section{\label{sec:gws}Gravitational Waves}

Having solved for the evolution of all relevant phase quantities under radiation reaction, we now move to obtaining the Fourier domain waveform in dCS gravity. As a matter of simplicity, we consider the construction of a TaylorF2-style approximate, where the waveform amplitudes are taken to be leading PN order, while the phases contain higher PN order corrections. The reason to do this is not merely to simplify the calculation. As we will show, the leading order dCS corrections enter at Newtonian order in the amplitudes. The calculation of the GR sector of the waveform amplitude can easily be extended to include higher PN effects and higher harmonics with the results of~\cite{Lundgren:2013jla,Boyle:2011,Arun:2008kb,Chatziioannou:2013dza,Chatziioannou:2017tdw}.

\subsection{\label{sec:wave} Fourier Domain Waveform}

Our starting point is the enumeration of the metric perturbation in the far zone, where the GW metric perturbation can be treated within the standard quadrupole approximation
\begin{equation}
    \label{eq:h-def}
    h_{ij} = \frac{2}{D_{L}} \ddot{I}_{<ij>}
\end{equation}
where $I_{ij} = \mu r_{i} r_{j}$ is the orbital quadrupole moment, $D_{L}$ is the luminosity distance to the source, and the far zone dCS corrections is given in Eq.~(118) in~\cite{Yagi:2011xp}. This correction scales as $u^{6}$ whereas Eq.~\eqref{eq:h-def} scales as $u^{2}$, and thus the dCS correction coming from the far zone scalar radiation is 2PN suppressed relative to standard quadrupole radiation in GR. One might expect that this is the dominant correction to the waveform amplitude. However, the orbital trajectory depends on the orientation of $\hat{L}$, which changes due to orbital precession, and is thus modified by dCS effects. Thus, the leading-order corrections actually come from the weak coupling expansion of Eq.~\eqref{eq:h-def}.

In a frame where $\hat{L}$ is aligned with the z-axis, the orbital plane is spanned by the vectors,
\begin{align}
    \hat{n}'' &= \left[\cos\phi_{C}, \sin\phi_{C}, 0\right]
    \\
    \hat{\lambda}'' &= \left[-\sin\phi_{C}, \cos\phi_{C}, 0\right]
\end{align}
where we have label them with a double prime superscript to distinguish the frame from those discussed in Sec.~\ref{sec:co-prec}. In the above, $\phi_{C} = \Phi+ \phi_{T}$ is the carrier phase, where $\Phi=\phi_{\rm orb} - 3v^{3}(2-\eta v^{2})\ln v$, with $\phi_{\rm orb}$ the orbital phase and the second term is the correction arising from GW tails~\cite{Arun:2008kb,Blanchet:1996pi}. The vectors in the non-precessing frame can then be found by performing two Euler rotations, first by $\theta_{L}$ about the y-axis, and then by $\phi_{z}$ about the new z-axis. Doing so, one obtains
\begin{widetext}
\begin{align}
    \hat{n} &= \left[\cos\theta_{L} \cos\phi_{z} \cos\phi_{C} + \sin\phi_{z} \sin\phi_{C}, -\cos\theta_{L} \sin\phi_{z} \cos\phi_{C} + \cos\phi_{z} \sin\phi_{C},\sin\theta_{L} \cos\phi_{C}\right]
    \\
    \hat{\lambda} &= \left[-\cos\theta_{L} \cos\phi_{z} \sin\phi_{C} + \sin\phi_{z} \cos\phi_{C}, \cos\theta_{L} \sin\phi_{z} \sin\phi_{C} + \cos\phi_{z}\cos\phi_{C}, -\sin\theta_{L} \sin\phi_{C}\right]
    \\
    \label{eq:Lhat}
    \hat{L} &= \left[-\cos\phi_{z} \sin\theta_{L}, \sin\phi_{z}\sin\theta_{L}, \cos\theta_{L}\right]
\end{align}
\end{widetext}
where Eq.~\eqref{eq:Lhat} is equivalent to Eq.~\eqref{eq:L-no-prec}. With this, one can evaluate the necessary derivative on $I_{<ij>}$ to obtain the metric pertrubations.

The observable waveform is not given by Eq.~\eqref{eq:h-def}, but by its transverse trace-less (TT) projection~\cite{PoissonWill}. The projection is performed by defining the line of sight vector $\vec{N}$ in a frame where $\vec{J}$ is aligned with the z-axis. The vector $\vec{N}$ is then determined by the angles $(\theta_{N}, \phi_{N})$ relative to $\vec{J}$. By performing the TT projection along $\vec{N}$, we obtain the two polarization states of the GW, which may be written as
\begin{align}
    h_{+} - i h_{\times} &=  h_{2} \sum_{m=-2}^{2} H_{2m}(\phi_{T}, \theta_{L}, \phi_{z}, \theta_{N}, \phi_{N}) e^{im\Phi}
\end{align}
where $h_{+,\times}$ are the plus and cross polarizations, and
\begin{align}
    \label{eq:Hm-def}
    H_{2m} = \sum_{m'=-2}^{2} (-1)^{m'+1} &{D^{2}}_{mm'}(\phi_{T}, \theta_{L}, -\phi_{z})  {_{-2}}Y_{2m'}(\theta_{N}, \phi_{N})
\end{align}
with $h_{2}$ a known functions of $u$ given in~\eqref{eq:h2}, ${D^{l}}_{mm'}$ the Wigner D matrices\footnote{Note that here we are using the phase convention provided explicitly in Eq.~\eqref{eq:D-to-P}. Different conventions are used throughout different pieces of the literature. As long as one is consistent in the phase conventions used, one obtains the same $h_{+} - i h_{\times}$ provided here.}, and ${_{s}}Y_{lm}$ the spin-weighted spherical harmonics. Note that, because we are working to leading PN order, we only have the $l=2$ modes in the above waveform. This can easily be extended by simply summing over all possible $l$ modes at the relevant PN order.

The amplitudes $H_{2m}$ are oscillatory functions in $\phi_{z}$ and $\phi_{T}$, but the dependence on $\theta_{L}$ is not purely oscillatory in the following sense. The expression for $\cos\theta_{L}$ is given in Eq.~\eqref{eq:thetaL}, which contains an average part that evolves monotonically on the radiation reaction timescale and an oscillatory part that evolves on the precession timescale. This can be explicitly seen by expanding out Eq.~\eqref{eq:thetaL}, specifically
\begin{align}
    \label{eq:thetaL-exp}
    \cos\theta_{L} &= \frac{J^{2} + L^{2} - S_{+}^{2}}{2 J L} - \frac{\left(S_{-}^{2} - S_{+}^{2}\right)}{2 J L} \sin^{2}\psi\,.
\end{align}
The standard technique of calculating the Fourier domain waveform is the use of the SPA, which was modified in~\cite{Klein:2014bua} to handle precessing waveforms through the introduction of SUA. In order for the SPA to be applicable, there must be a separation of scales between the variation of the amplitude and phase. In the inspiral phase, the amplitude evolves on the radiation-reaction timescale, while the phase evolves on the orbital timescale, and thus the condition of separation of scales is met. When applying SUA, there must be a further separation of scales between the standard phase variable and the phase characterizing the modulation of the amplitude. Once again, the precession timescale is intermediate to the orbital and radiation reaction timescales, and the conditions to apply SUA are met. Further, the SUA method implements a partial resummation procedure, so Eq.~\eqref{eq:thetaL-exp} does not need to be Fourier decomposed to extract the oscillatory modulation. Instead, this expression can simply be left as is without the need for expansions in GR. We will not review the ingredients of the SUA method here, but instead refer the reader to~\cite{Klein:2014bua}, and below we use the same notation as that introduced in that reference. 

The crux of the issue in dCS gravity is that $\theta_{L}$ given by Eq.~\eqref{eq:thetaL-exp} is not simply modified through the shifts in $S_{\pm}^{2}$, but also within $\psi$ through Eq.~\eqref{eq:psi-pn}. Within GR, the Wigner-D matrices can be left as they are, but a naive weak-coupling expansion within dCS gravity results in the expansion of trigonometric functions of $\psi$, which  loses accuracy as the binary evolves, thus spoiling the oscillatory amplitude modulation that is fundamental to precessing waveforms. To rectify this, we consider an expansion of relevant trigonometric functions of $\theta_{L}$ as a Fourier series on $\psi$. The starting point to this is the relationship between Wigner D-matrices and spin-weighted spherical harmonics,
\begin{align}
    \label{eq:D-to-P}
    {D^{l}}_{mm'}(\phi_{T}, \theta_{L}, -\phi_{z}) &= (-1)^{m'} N_{l} e^{im'\phi_{z}} {_{m'}}Y_{lm}(\theta_{L},\phi_{T}) \,,
    \nonumber \\
    &= (-1)^{m'} N_{lm} e^{i(m\phi_{T} + m' \phi_{z})} 
    \nonumber \\
    &\times {_{m'}}P_{lm}(\cos\theta_{L}) 
\end{align}
where $N_{l}$ and $N_{lm}$ are constants given in Appendix~\ref{app:amp-fourier}, and ${_{s}}P_{lm}$ are the spin-weighted associated Legendre polynomials (SALPs)~\cite{Breuer}. As we show in Appendix~\ref{app:amp-fourier}, any trigonometric function of $\theta_{L}$ can be written as a Fourier series in $\psi$, with the coefficients of the series only dependent on the radiation-reaction timescale. Thus, we can go one step further and write
\begin{align}
    \label{eq:D-decomp}
    {D^{l}}_{mm'}(\phi_{T}, \theta_{L}, -\phi_{z}) &= (-1)^{m'} e^{i(m\phi_{T} + m'\phi_{z})} 
    \nonumber \\
    &\times \sum_{n=-\infty}^{\infty} {P^{l}}_{mm'n}(u) e^{in\psi}\,,
\end{align}
where the ${P^{l}}_{mm'n}$ can generically be written in terms of Gegenbauer polynomials. Formally, $n$ in the above equation extends to infinitely many harmonics; however, as we detail in Appendix~\ref{app:amp-fourier}, there is a small parameter that can be used to truncate the sum at a finite number of harmonics. 

With the decomposition in Eq.~\eqref{eq:D-decomp}, we can consider the weak coupling expansion of the waveform amplitudes. Writing the phases as
\begin{align}
    \phi_{z}(u) &= \phi_{z,{\rm GR}}(u) + \bar{\zeta}_{2} \delta \phi_{z}(u)\,,
    \\
    \phi_{T}(u) &= \phi_{T,{\rm GR}}(u) + \bar{\zeta}_{2} \delta \phi_{T}(u)\,,
    \\
    \psi(u) &= \psi_{\rm GR}(u) + \bar{\zeta}_{2} \delta\psi(u)\,,
\end{align}
the Wigner D-matrices can be expanded as
\begin{align}
    {D^{l}}_{mm'}\left(\phi_{T}, \theta_{L}, -\phi_{z}\right) &= (-1)^{m'} N_{lm} \sum_{n} e^{i\Phi_{K}^{\rm p,GR}} e^{i\bar{\zeta}_{2}\delta\Phi^{\rm p,dCS}_{K}}
    \nonumber \\
    &\times
    P^{\rm GR}_{K}(u) \left[1 + \bar{\zeta}_{2} \delta P_{K}(u)\right]
\end{align}
where $K$ is the multi-spectral index $K=lmm'n$, and
\begin{align}
    \label{eq:PK-GR}
    P^{\rm GR}_{K} &= \underset{\bar{\zeta}_{2}\rightarrow0}{\lim} {P^{l}}_{mm'n}(u)
    \\
    \label{eq:PK-dCS}
    \delta P^{\rm dCS}_{K} &= \underset{\bar{\zeta}_{2}\rightarrow0}{\lim} \; \frac{\partial}{\partial \bar{\zeta}_{2}} \ln{P^{l}}_{mm'n}(u)
    \\
    \label{eq:PhiP-GR}
    \Phi_{K}^{\rm p,GR} &= m\phi_{T,{\rm GR}} + m'\phi_{z,{\rm GR}} + n\psi_{\rm GR}\,,
    \\
    \label{eq:PhiP-dCS}
    \delta \Phi^{\rm p,dCS}_{K} &= m\delta\phi_{T} + m'\delta\phi_{z} + n\delta \psi\,,
\end{align}
This completes the weak-coupling expansion of the waveform in dCS gravity. 

Before moving on to the calculation of the Fourier transform of the waveform, it is worth noting that $h_{2}$ depends of the orbital velocity, which is mapped to $u$ via the dCS modified Kepler's third law. We do not include this contribution here because is it higher PN order than the $\delta P_{K}$. Also note that this decomposition into harmonics of $\psi$ does not spoil the analytic behavior needed to apply the SUA method since the precession phase $\Phi^{p}_{K}$ is still oscillatory in $\psi$.

The Fourier phase of the waveform is $\Psi = 2\pi f t + m \Phi$, where recall that $\Phi$ is the orbital period plus the GW tail contribution. As is standard with integrals of this type, we can proceed with using the SPA. The stationary point is found by demanding $\dot{\Psi} = 0$, which gives
\begin{equation}
    \label{eq:spa}
    f = -m F(t_{\star})
\end{equation}
where $t_{\star}$ is the stationary point. Note that for $m>0$, this condition is only satisfied for negative Fourier frequencies $f$, while for $m<0$ it is satisfied for positive frequencies. Thus, only the $m<0$ terms survive. The SPA phase is then 
%
\begin{align}
    \label{eq:Psi-m}
    \Psi_{m}(f) &= 2\pi f t_{c} + m \phi_{c} - \frac{\pi}{4} - \frac{3m}{256\eta \tilde{u}^{5}} \bigg\{1 + \sum_{n=2} \langle \Psi_{n}^{\rm GR} \rangle_{\psi} \tilde{u}^{n} 
    \nonumber \\
    & \sum_{n=5} \langle \Psi_{n}^{l,\rm GR}\rangle_{\psi} \tilde{u}^{n} \ln\tilde{u}+ \bar{\zeta}_{2} \bigg[4 \langle \delta \beta_{3}\rangle_{\psi}^{\rm dCS} \tilde{u}^{3} + 10 \langle \delta \sigma_{4} \rangle_{\psi}^{\rm dCS} \tilde{u}^{4} 
    \nonumber \\
    &- 160 \frac{q^{4}(1-q)^{2}}{(1+q)^{4}}\langle \delta C\rangle_{\psi} \tilde{u}^{4}\bigg]\bigg\}
\end{align}
where $\tilde{u} = (2\pi M f/|m|)^{1/3}$, and recall that only the $m<0$ contribute. The higher PN coefficients $\langle \Psi_{n}^{\rm GR}\rangle_{\psi}$ are given in Appendix H of~\cite{Chatziioannou:2013dza}.

The SUA correction to the stationary point is obtained from 
\begin{equation}
    \label{eq:T-def}
    T_{m} = \left[-m \ddot\Phi\right]^{-1/2}
\end{equation}
To find the SUA corrections to the amplitude, one must evaluate the condition in Eq.~\eqref{eq:spa} at the shifted time $t_{\star} + k T_{m}$. Since $T_m$ is PN suppressed, we may Taylor expand the shifted version of Eq.~\eqref{eq:spa} to obtain 
%
\begin{align}
    \label{eq:sua}
    u_{k} &= \tilde{u} + \frac{k}{3} \sqrt{\frac{a_{0}}{|m|}} \tilde{u}^{7/2} \bigg\{1 + \sum_{n=2} \left(\upsilon_{n} + \upsilon_{n}^{l} \ln \tilde{u}\right)\tilde{u}^{n} 
    \nonumber \\
    &+\bar{\zeta}_{2} \bigg[- \frac{1}{2} \langle \delta \beta_{3}\rangle_{\psi}^{\rm dCS} \tilde{u}^{3} - \frac{1}{2} \langle \delta \sigma_{4} \rangle_{\psi}^{\rm dCS} \tilde{u}^{4} 
    \nonumber \\
    &+ 8 \frac{q^{4}(1-q)^{2}}{(1+q)^{4}} \langle \delta C\rangle_{\psi} \tilde{u}^{2}\bigg]\bigg\}\,,
\end{align}
where the PN coefficients $\upsilon_{n}$ are given in Appendix~\ref{app:amp-fourier}. The SUA amplitudes are then
\begin{align}
    \label{eq:amp}
    {\cal{A}}_{m}(f) &= \sum_{k=0}^{k_{\rm max}} \frac{a_{k,k_{\rm max}}}{2} \left[H_{2m}(u_{k}) + H_{2m}(u_{-k})\right]\,,
\end{align}
where $a_{k,k_{\rm max}}$ is given in Eq.~(79) in~\cite{Chatziioannou:2017tdw}, and the waveform can then simply be written as
\begin{align}
    \label{eq:h-fourier}
    \tilde{h}_{+} - i \tilde{h}_{\times} &= \sqrt{2\pi} \sum_{m<0} T_{m} {\cal{A}}_{m}(f) e^{i\Psi_{m}(f)}
    \nonumber \\
    &= \sqrt{\frac{2}{3}} \frac{{\cal{M}}^{5/6}}{D_{L} \pi^{1/6}} f^{-7/6} {\cal{A}}_{-2}(f) e^{i\Psi_{-2}(f)}\,.
\end{align}

\begin{figure*}
    \centering
    \includegraphics[scale=0.37, trim=2cm 0.5cm 2cm 0.5cm, clip]{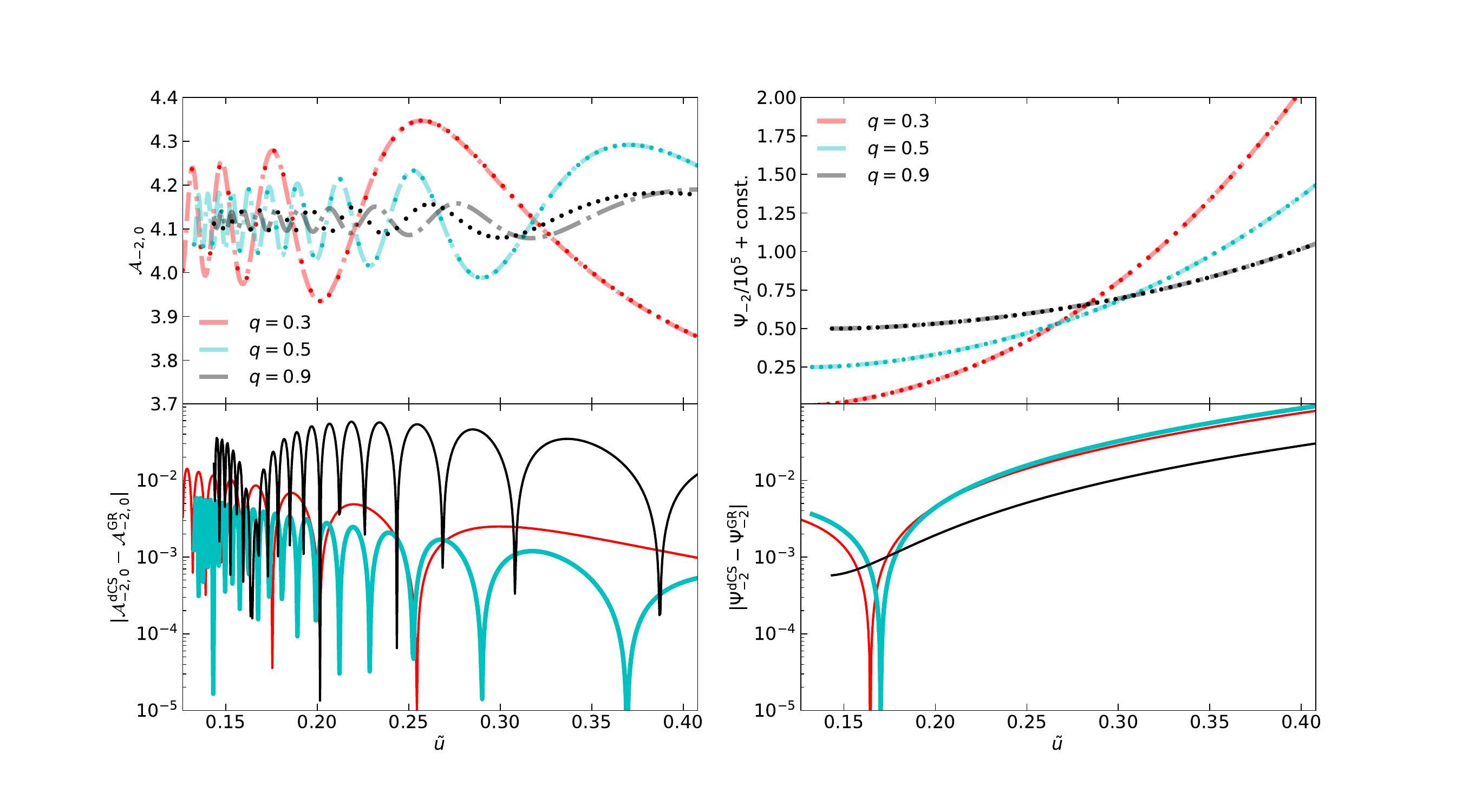}
    \includegraphics[scale=0.37, trim=2cm 0.5cm 2cm 0.5cm, clip]{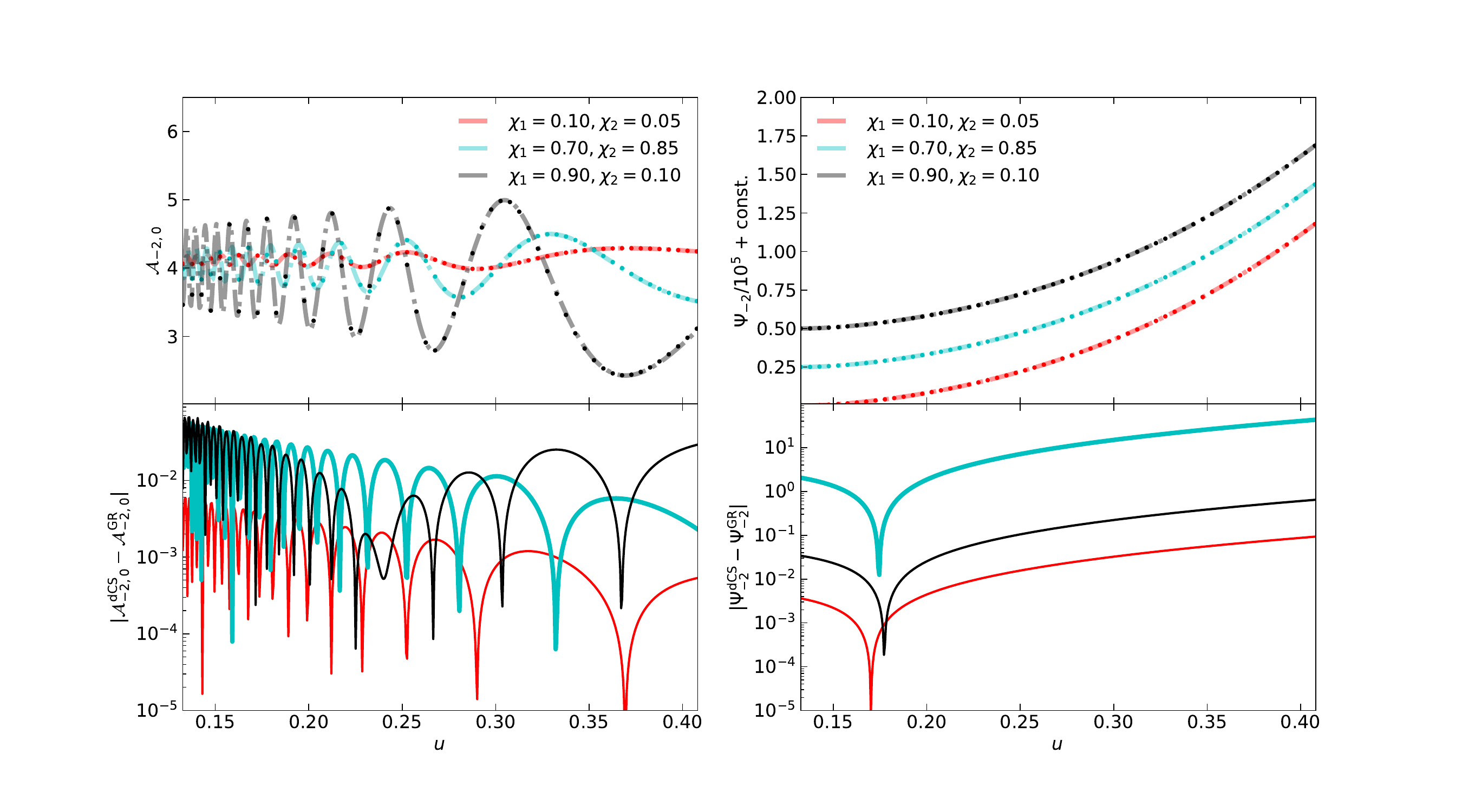}
    \caption{Left: Plot of the Fourier domain waveform amplitude ${\cal{A}}_{-2,0}$ for both GR (dot-dashed lines) and dCS gravity (dotted lines) for the systems in Table~\ref{tab1}, and with the line of sight angle $(\theta_{N},\phi_{N})=(\pi/4,0)$. Right: Plot of the SPA phase for both GR and dCS gravity. The bottom panels display the relative difference between dCS gravity and GR.}
    \label{fig:amp}
\end{figure*}

To highlight the difference between the waveform of Eq.~\eqref{eq:h-fourier} in GR and dCS gravity, we plot a part of the amplitude and the Fourier phase in Fig.~\ref{fig:amp} for the systems in Table~\ref{tab1}. The amplitude ${\cal{A}}_{-2}(f)$ can be decomposed into harmonics of $\psi$, specifically ${\cal{A}}_{-2} = \sum_{n} {\cal{A}}_{-2,n} e^{in\psi}$. The new amplitudes ${\cal{A}}_{-2,n}$ still contain dependence on $\phi_{z}$ and $\phi_{T}$, which we do not factor out since these generate most of the amplitude modulation. The harmonic with the largest amplitude is $n=0$, which we plot in the top panel of Fig.~\ref{fig:amp} for the same system parameters as Fig.~\ref{fig:J-chi}. Observe that the differences in the amplitude generally decrease with increasing frequency, as opposed to the waveform phase. From the SUA correction in Eq.~\eqref{eq:sua}, one might expect the opposite due to the dCS correction first appearing at 1.5PN order therein. However, the difference in the waveform amplitude is largely controlled by the correction in Eq.~\eqref{eq:PK-dCS}, which does not have a definite PN order. The difference between the waveform phases is typically small, but it increases with increasing spin. On the other hand, the difference in the waveform amplitudes is largest for the $q=0.9$ system, highlighting the nature of the non-uniform weak coupling expansion in dCS gravity. The large difference between precessing waveforms in GR and dCS gravity shown in the bottom panels of Fig.~\ref{fig:amp} qualitatively suggests that one should be able to place stringent constraints on the dCS coupling constant $\xi$ with precessing binaries. However, the ability to place stringent constraints is dependent on the existence of covariances and/or degeneracies among the physical parameters of the binary. Such effects need to be elucidated in a formal parameter estimation study, which we leave to future work.

\subsection{\label{sec:imr} Towards Inspiral-Merger-Ringdown Waveforms}

To review, the results of our analysis thus far have led to the development of an analytic, frequency-domain waveform for the inspiral phase of precessing BHs in dCS gravity. The TaylorF2-style waveform is given in Eq.~\eqref{eq:h-fourier}, with the Fourier phase $\Psi_{-2}(f)$ given in Eq.~\eqref{eq:Psi-m}. The Fourier amplitude ${\cal{A}}_{-2}(f)$ is given in Eq.~\eqref{eq:amp}, with the $u_{k}$ SUA variable given in Eq.~\eqref{eq:sua}. This constitutes the first fully analytic inspiral waveform calculated for spin precessing binaries in a modified theory of gravity.

The ultimate goal of waveform modelling is the development of a single waveform that captures the full coalescence of the binary, from the early inspiral, through the merger and ringdown when the remnant finally reaches a steady state. The so-called ``inspiral-merger-ringdown (IMR) waveforms'' seek to achieve this via different methods. Those of most relevance to the work at hand are the IMRPhenom waveform models~\cite{Hannam:2013oca,Husa:2016,Santamaria:2010,Khan:2015jqa,Khan:2018fmp}, which combine PN theory for the inspiral phase, black hole perturbation theory for the ringdown, and a fit to numerical relativity for the merger. The advent of analytic solutions to the PN spin precession problem has lead to the development of precessing IMRPhenom waveforms~\cite{Chatziioannou:2017tdw}, with IMRPhenomPv3~\cite{Khan:2018fmp} being the most up to date.

The IMRPhenomPv3 waveforms~\cite{Khan:2018fmp} make use of the analytic solutions to the PN precession equations derived in~\cite{Chatziioannou:2017tdw}. Since our work herein provides an extension of the calculations in~\cite{Chatziioannou:2017tdw} to include the leading-order, dCS corrections, it is straightforward to extend the inspiral phase of the IMRPhenomPv3 waveforms to include these. The mappings between the phases of the IMRPhenomPv3 waveforms and the phases herein (and within~\cite{Chatziioannou:2017tdw}) are $\alpha \leftrightarrow \phi_{z}$, $\beta \leftrightarrow \theta_{L}$, and $\epsilon \leftrightarrow \phi_{T}$. Then, the extension of these phases to dCS gravity is trivially
\begin{align}
&    \alpha(f) = \alpha_{\rm Pv3}(f) + \bar{\zeta}_{2} \left[\delta \Phi_{z,-3}^{(-1)} \; \varphi_{-3}(\tilde{u}_{k}) + \Phi_{z,-3}^{(-1)} \; \delta \varphi_{-3}(\tilde{u}_{k})\right]\,,
    \\
&    \cos[\beta(f)] = \cos[\beta_{\rm Pv3}(f)] + \bar{\zeta}_{2} \sum_{n} \delta {\cal{C}}_{L,n} \cos[n\psi(f)]\,,
    \\
&    \epsilon(f) = \epsilon_{\rm Pv3}(f) + \bar{\zeta}_{2} \delta \Phi_{T,0}^{(-1)} \tilde{u}_{k}^{-3}\,,
\end{align}
where $\psi(f)$ is given in Eq.~\eqref{eq:psi-pn}, $[\Phi_{z,-3}^{(-1)}, \delta\Phi_{z,-3}^{(-1)}, \delta\Phi_{T,0}^{(-1)}]$ are given in Eqs.~\eqref{eq:Phiz-3},\eqref{eq:dPhiz-3}, \&~\eqref{eq:PhiT-0}, $[\varphi_{-3},\delta\varphi_{-3}]$ are given in Eqs.~\eqref{eq:vphi-3} \&~\eqref{eq:dvphi-3}, $\delta C_{L,n}$ is defined as
\begin{equation}
    \delta C_{L,n} = \underset{\bar{\zeta}_{2}\rightarrow0}{\lim} \frac{\partial C_{L,n}}{\partial \bar{\zeta}_{2}}\,,
\end{equation}
with the $C_{L,n}$ given in Eq.~\eqref{eq:CL0}-\eqref{eq:CL2}, and $[\alpha_{\rm Pv3}, \beta_{\rm Pv3},\epsilon_{\rm Pv3}]$ are the standard GR expression appearing in the IMRPhenomPv3 waveforms. The IMRPhenomPv3 waveforms are also specified by two effective spin parameters, the first being $\chi_{\rm eff}$, which we have computed in Sec.~\ref{sec:const} and is given in Eq.~\eqref{eq:chieff-of-t}. The second is the effective precession spin, given by
\begin{align}
    \label{eq:chip}
    \chi_{p} &= \frac{\max\left[(2+\frac{3}{2}q) \left|\hat{L}\times \vec{S}_{1}\right|, (2+\frac{3}{2q}) \left|\hat{L}\times\vec{S}_{2}\right|\right]}{(2+\frac{3}{2}q) m_{1}^{2}}\,.
\end{align}
This quantity can be computed in the precession averaged scheme detailed in Appendix~\ref{app:avg}, with the dCS corrections to the cross products given in Eqs.~\eqref{eq:dP-1}-\eqref{eq:dP-2}. 

The above considerations provide all of the information necessary to extend the inspiral phase of the IMRPhenomPv3 waveforms to dCS gravity. While we have not considered the merger-ringdown phase of the coalescence, the first numerical relativity simulations of black hole mergers in dCS gravity have been computed within the effective field theory framework in~\cite{Okounkova:2019dfo,Okounkova:2019zjf}. One may be tempted to then use those results to re-fit the intermediate and merger-ringdown phases of IMRPhenomPv3 against those numerical simulations. Unfortunately, to do so would require a large bank of such simulations, which does not currently exist. The recent work in~\cite{Perkins:2022fhr}, however, suggests that modifications to the intermediate and the merger-ringdown phase of IMRPhenomPv3 are not required to carry out the first tests of dCS gravity. Indeed, tests that only modify the inspiral phase have been shown to be \textit{conservative}~\cite{Perkins:2022fhr}. That is, if we had a full dCS IMRPhenomPv3 model, then tests with this IMR model would not invalidate tests carried out with modifications in the inspiral only, buy rather the former would merely strengthen the latter.

\section{\label{sec:disc}Discussion}

The calculations performed here have resulted in the first analytic Fourier domain waveforms for the quasi-circular inspiral of spin-precessing BH binaries in a modified theory of gravity. Our theory of choice to perform this calculation, dCS gravity, is a prime target for study with these types of systems due to parameter degeneracies that otherwise prevent constraints on the dCS coupling parameter in the spin-aligned case~\cite{Nair:2019,Perkins:2021mhb}. We have shown here that the methods developed within GR for analytically studying spin-precessing binaries also apply to dCS gravity, with one caveat. The equal mass limit of the precession problem possesses unique properties in GR due to the no-hair theorems, which are violated in dCS gravity. Physically, this causes nutation of the BHs' spins to be present in dCS gravity in the equal-mass case, and this mathematically forces us to work with a non-uniform expansion in the coupling parameter. However, from the context of data analysis and tests of GR, this behavior is intriguing since it presents the possibility of a smoking gun test. Indeed, as can be seen from the results of Fig.~\ref{fig:amp}, the difference between the dCS and GR waveforms is largest for approximately equal mass and high spin systems, the former producing the largest difference in the waveform amplitude and the latter producing the largest difference in waveform phase.

Another aspect of these waveforms that is intriguing is the fact that the phase modification is parameterized by not one beyond-GR correction, but three. The reason for this is that modifications to the spin-precession equations will alter the evolution of the relative orientation of the spins and orbital angular momentum, thus modifying the spin-coupling terms in the GR part of the Fourier phase. For dCS gravity, these effects enter at relative Newtonian order, and thus produce corrections to the phase at 1.5PN order for spin-orbit corrections and 2PN order for spin-spin corrections. On the other hand, these effects are suppressed in the spin-aligned limit, and the correction to the GW phase is dominated by the emission of dipole radiation, which enters at 2PN order in dCS gravity. This behavior can be seen from the last three columns of Table~\ref{tab1} as a function of the orientation angle $\theta_{L}$. 

There are at least two issues that we have not addressed in this paper and that could serve as the basis of future investigations. The first is how general are the results of dCS gravity to other modified theories. So far, the only modified spin precession equations that have been derived are in dCS gravity. One could consider other  modified theories and perform the same calculation carried out here to study the effects on the emission of GWs. Alternatively, one could posit a set of theory agnostic modifications to Eq.~\eqref{eq:prec} and study the effect of these on the precession dynamics. A perhaps more suitable avenue would be to develop a \textit{parameterized post-Einsteinian} (ppE)~\cite{Yunes:2009ke} waveform for spin precessing binaries. The ppE framework posits that modified gravity effects in the waveform may be captured by a set of theory agnostic parameters that scale as particular PN orders. The results found here can be used to motivate such a framework. In addition,~\cite{Loutrel:2022ant} studied the corrections to binary dynamics due to non-axisymmetric mass quadrupole effects, which generically modify the precession of the binary. While this study did not include spin effects, the end result is a waveform analogous to those derived here, with the main difference being the overall analytic structure of the precession dynamics. Using both the non-axisymmetric case and the dCS gravity case to motivate a precessing ppE framework will allow us to capture a wide variety of beyond-GR and beyond-vacuum scenarios.

The second issue is how the waveforms derived here aid in our ability to perform tests of GR. As we have already pointed out, dCS gravity has known parameter degeneracies that makes placing constraint on the coupling constants of the theory currently impossible solely with GW observations~\cite{Nair:2019,Perkins:2021mhb}. It is also well known that spin precession can break degeneracies when performing parameter estimation of certain systems~\cite{Chatziioannou:2014coa,Stavridis:2009mb}. Therefore, we may expect that dCS gravity can be constrained stringently with observations of GWs from spin precessing binaries, as predicted in~\cite{Alexander:2017jmt}. Bayesian inference provides the most informative approach for performing parameter estimation studies of GW observations, but it is time consuming to fully sample the posterior distribution of the waveform parameters. We leave an in-depth parameter estimation study, as well as the development of a precessing ppE framework, to future work. 

\acknowledgements
We would like to thank Katerina Chatziioannou for useful discussions. N.L. acknowledges financial support provided under the European Union's H2020 ERC, Starting
Grant agreement no.~DarkGRA--757480. We also acknowledge support under the MIUR PRIN (Grant
2020KR4KN2) and FARE programmes (GW- NEXT, CUP: B84I20000100001).
N.Y. acknowledges support from the Simons Foundation through Award number 896696. 

\appendix

\section{\label{app:exp}Precession Coefficients}

We here provide various expressions related to the conservative precession equations. The polynomials $(f_{3},f_{5})$ given in Eq.~\eqref{eq:chi-eff} are
\begin{align}
    f_{3} &= 1 + q + \frac{700}{603} q^{2} + \frac{350}{603} q^{3}\,,
    \\
    \addlinespace
    f_{5} &= 1 + q + \frac{350}{603} q^{2} - \frac{350}{603} q^{3} - q^{4} - q^{5}\,.
\end{align}
The coefficients of $dS^{2}/dt$ appearing in Eq.~\eqref{eq:dSdt-gr} are
\begin{align}
\label{eq:A-coeff}
A^{2} &= \frac{9 q v^{10} (-1 + v \chi_{\rm eff})^2}{4 L^2 M^2 (1 + q)^2}\,,
\\
B &= -2 J^{2} + L^{2} \left(\frac{1}{q} + q\right) - \left(\frac{1 - q}{q}\right) \left(q S_{1}^{2} - S_{2}^{2}\right) 
\nonumber \\
&+ 2 L M^{2} \chi_{\rm eff}\,,
\\
C_{0} &= J^{4} + J^{2} \left[-2 L^{2} - \frac{2}{q} (-1 + q)\left(q S_{1}^{2} - S_{2}^{2}\right) - 2 L M^{2} \chi_{\rm eff}\right] 
\nonumber \\
&+ L \Bigg[ L^{3} - \frac{2}{q} (1 - q) L S_{1}^{2} + 2 L (1 - q) S_{2}^{2} + 2 L^{2} M^{2} \chi_{\rm eff} 
\nonumber \\
&+ 2 M^{2} \left(\frac{-1 + q}{1+q}\right) \left(S_{1}^{2} - S_{2}^{2}\right) \chi_{\rm eff} + \frac{4 L M^{4} q \chi_{\rm eff}^{2}}{(1+q)^{2}}\Bigg]\,,
\\
\label{eq:D0-coeff}
D_{0} &= -\frac{(1 - q)}{q (1 + q)} \Big\{J^4 [q^2 S_{1}^2 - S_{2}^2 + q (S_{1}^2 - S_{2}^2)]
\nonumber \\
&- 2 J^2 L \left[L (q S_{1}^2 + q^2 S_{1}^2 - S_{2}^2 - q S_{2}^2) 
\right.
\nonumber \\
&\left.
+ M^2 q (S_{1}^2 - S_{2}^2) \chi_{\rm eff}\right]+ L^2 \left[(-1 + q^2) (S_{1}^2 - S_{2}^2)^2 
\right.
\nonumber \\
&\left.
+ L^2 (q S_{1}^2 + q^2 S_{1}^2 - S_{2}^2 - q S_{2}^2)
\right.
\nonumber \\
&\left.
+ 2 L M^2 q (S_{1}^2 - S_{2}^2) \chi_{\rm eff}\right]\Big\}\,.
\end{align}
The dCS corrections to $dS^{2}/dt$ in Eqs.~\eqref{eq:Seq-dcs}-\eqref{eq:dD} are
\begin{widetext}
\begin{align}
\label{eq:dC_2}
\delta C_{2} &= - \frac{(J - L)^{2} (J + L)^{2}}{1344 L^{3} M^{2} (1 - v \chi_{\rm eff})^{2}} \left[ \delta C_{2,0} + \delta C_{2,2} J^{2} + \delta C_{2,4} J^{4} \right]\,,
\\
\delta C_{3} &= - \frac{J^{2} - L^{2}}{1344 L^{4} M^{2} (q+1)^{3} q (1 - v \chi_{\rm eff})^{2}} \left[\delta C_{3,0} + \delta C_{3,2} J^{2} + \delta C_{3,4} J^{4}\right]\,,
\\
\delta D_{2} &= -\frac{(1-q) (J-L)^{2} (J+L)^{2}}{1344 L^{3} M^{2} (q + 1) q (1 - v \chi_{\rm eff})^{2}} \left[\delta D_{2,0} + \delta D_{2,2} J^{2} + \delta D_{2,4} J^{4}\right]\,,
\\
\delta D_{3} &= -\frac{J^{2} - L^{2}}{1344 L^{4} M^{2} (q + 1)^{3} q (1 - v \chi_{\rm eff})^{2}} \left[\delta D_{3,0} + \delta D_{3,2} J^{2} + \delta D_{3,4} J^{4} \right]\,,
\end{align}
where
\allowdisplaybreaks[4]
\begin{align}
\delta C_{2,0} &= \frac{q L^{3} M^{2}}{(1 + q)^{2}} \left(1809 + 1109 q + 1147 q^2 + 3150 q^3 + 1147 q^4 - 97 q^5 + 603 q^6\right)\,,
\\
\delta C_{2,2} &= - \frac{q L M^{2}}{(1 + q)^{2}} \left(1809 + 2412 q + 1147 q^2 + 2800 q^3 + 1147 q^4 + 1206 q^5 + 603 q^6\right)\,,
\\
\delta C_{2,4} &= 3 q v \left(201 + 350 q^2 + 201 q^4\right)\,,
\\
\delta C_{3,0} &= - L^{5} M^{4} q^{2} \left(7236 + 9045 q + 1809 q^2 + 15750 q^3 + 14350 q^4 + 2197 q^5 
\right.
\nonumber \\
&\left.
+ 1615 q^6 + 2412 q^7\right)\,,
\\
\delta C_{3,2} &= 2 L^{3} M^{4} q^{2} \left(3618 + 5427 q + 3909 q^2 + 8050 q^3 + 5950 q^4 + 3403 q^5 
\right.
\nonumber \\
&\left.
+ 3015 q^6 + 1206 q^7\right)\,,
\\
\delta C_{3,4} &= - 9 L M^{4} q^{3} \left(201 + 201 q + 350 q^2 + 350 q^3 + 201 q^4 + 335 q^5\right)\,,
\\
\delta D_{2,0} &= \frac{L^{3} M^{2} q}{1+q} \left[ q S_{1}^{2} \left( 1809 + 1712 q + 2450 q^2 + 3500 q^3 + 1497 q^4 + 1206 q^5 + 1206 q^6\right)
\right.
\nonumber \\
&\left.
- S_{2}^{2} \left(2412 + 2412 q + 1497 q^2 + 3500 q^3 + 2450 q^4 + 506 q^5 + 603 q^6\right)\right]\,,
\\
\delta D_{2,2} &= \frac{L M^{2} q}{1 + q} \left[-q S_{1}^{2} \left(1809 + 3015 q + 1750 q^2 + 3850 q^3 + 2197 q^4 + 1809 q^5 + 1206 q^6\right) 
\right.
\nonumber \\
&\left.
+ S_{2}^{2} \left(2412 + 3015 q + 2197 q^2 + 3850 q^3 + 1750 q^4 + 1809 q^5 + 603 q^6 \right) \right]\,,
\\
\delta D_{2,4} &= 3 v q (1 + q) \left(201 + 350 q^2 + 201 q^4\right) \left(q S_{1}^{2} - S_{2}^{2}\right)\,,
\\
\delta D_{3,0} &= L^{5} M^{4} (q-1) q \big[ -q S_{1}^{2} \left(7236 + 9045 q + 2800 q^2 + 14350 q^3 + 11200 q^4 
\right.
\nonumber \\
&\left.
+ 1809 q^5 + 3618 q^6 + 2412 q^7 \right) + S_{2}^{2} \left( 4824 + 7236 q + 4221 q^2 + 12600 q^3 
\right.
\nonumber \\
&\left.
+ 10150 q^4 + 7000 q^5 + 4027 q^6 + 2412 q^7\right) \big]\,,
\\
\delta D_{3,2} &= 2 L^{3} M^{4} (q - 1) q \big[ q S_{1}^{2} \left( 3618 + 5427 q + 2100 q^2 + 8050 q^3 + 4900 q^4 + 2509 q^5 
\right.
\nonumber \\
&\left.
+ 2412 q^6 + 1206 q^7\right) - S_{2}^{2} \left(2412 + 3618 q + 3015 q^2 + 7000 q^3 + 5950 q^4 + 2800 q^5 
\right.
\nonumber \\
&\left.
+ 4221 q^6 + 1206 q^7 \right)\big]\,,
\\
\label{eq:dD_3,4}
\delta D_{3,4} &= - 9 L M^{4} (q - 1) q^{3} \big[ S_{1}^{2} \left(201 + 350 q^2 + 201 q^4 + 134 q^5\right) - S_{2}^{2} \left( 201 + 350 q^2 + 335 q^4\right)\big]\,,
\end{align}
\end{widetext}

To find the roots of $dS^{2}/dt$ in either Eq.~\eqref{eq:dSdt-gr} or Eq.~\eqref{eq:Seq-dcs}, the following procedure can be used. Consider a polynomial of the form $P(x) = x^{3} + B x^{2} + C x + D$, and let us assume we desire to find its roots, ie.~solve for $x$ in $P(x) = 0$, with all solutions assumed to be real. We begin by defining $t = x - B/3$, such that the polynomial becomes $P(t) = t^{3} - p t + q$, where
\begin{align}
p = \frac{B^{2}}{3} - C\,, \qquad q = D - \frac{B C}{3} + \frac{2 B^{3}}{27}\,.
\end{align}
We now apply the transformation $t = z \cos\theta$, which converts the polynomial to $P(\theta) = 4 \cos^{3}\theta - (4p/z^{2}) \cos\theta + (4q/z^{3})$. The goal is to now write this so that we can exploit the trigonometric identity $4\cos^{3}\theta - 3 \cos\theta - \cos(3\theta) = 0$. To do this, we demand $(4p/z^{2}) = 3$, which fixes $z = (4p/3)^{1/2}$. With this, the problem of finding the roots of the polynomial reduces to solving $\cos(3\theta) = - (q/2) (3/p)^{3/2}$. Transforming back to the original variable, the roots are given by
\begin{equation}
\label{eq:xn}
x_{n} = - \frac{B}{3} + 2 \sqrt{\frac{p}{3}} \cos\left(\frac{1}{3} \left\{\arccos\left[-\frac{q}{2} \left(\frac{3}{p}\right)^{3/2}\right] - 2 \pi n\right\}\right)\,,
\end{equation}
which satisfy $x_{0} > x_{1} > x_{2}$. If we then apply this to the problem of solving Eq.~\eqref{eq:dSdt-gr} (or Eq.~\eqref{eq:Seq-dcs}), we have $S_{+}^{2} = x_{0}$, $S_{-}^{2} = x_{1}$, and $S_{3}^{2} = x_{2}$.  

The angular momenta vectors in the non-precessing frame are given by
\begin{widetext}
\begin{align}
\label{eq:L-no-prec}
\hat{L} &= \left[\frac{A_{1} A_{2}}{2J} \cos \phi_{z}, \frac{A_{1} A_{2}}{2J} \sin \phi_{z}, \frac{J^{2} + L^{2} - S^{2}}{2J}\right]\,,
\\
\label{eq:S1-no-prec}
\vec{S}_{1} &= \left[\frac{\cos\phi_{z}}{4 J S^{2}} \left(- A_{1} A_{2} (S^{2} + S_{1}^{2} - S_{2}^{2}) + A_{3} A_{4} (J^{2} - L^{2} + S^{2}) \cos\phi'\right) - \frac{A_{3} A_{4}}{2 S} \sin\phi' \sin \phi_{z},
\right.
\nonumber \\
& \left.
\qquad \frac{\sin\phi_{z}}{4 J S^{2}} \left(- A_{1} A_{2} (S^{2} + S_{1}^{2} - S_{2}^{2}) + A_{3} A_{4} (J^{2} - L^{2} + S^{2}) \cos\phi'\right) + \frac{A_{3} A_{4}}{2 S} \sin\phi' \cos\phi_{z},
\right.
\nonumber \\
&\left.
\qquad \frac{1}{4 J S^{2}} \left((J^{2} - L^{2} + S^{2}) (S^{2} + S_{1}^{2} - S_{2}^{2}) + A_{1} A_{2} A_{3} A_{4} \cos\phi'\right)\right]\,,
\\
\label{eq:S2-no-prec}
\vec{S}_{2} &= \left[-\frac{\cos\phi_{z}}{4 J S^{2}} \left(A_{1} A_{2} (S^{2} + S_{1}^{2} - S_{2}^{2}) + A_{3} A_{4} (J^{2} - L^{2} + S^{2}) \cos\phi'\right) + \frac{A_{3} A_{4}}{2 S} \sin\phi' \sin \phi_{z},
\right.
\nonumber \\
& \left.
\qquad -\frac{\sin\phi_{z}}{4 J S^{2}} \left(- A_{1} A_{2} (S^{2} + S_{1}^{2} - S_{2}^{2}) + A_{3} A_{4} (J^{2} - L^{2} + S^{2}) \cos\phi'\right) - \frac{A_{3} A_{4}}{2 S} \sin\phi' \cos\phi_{z},
\right.
\nonumber \\
&\left.
\qquad \frac{1}{4 J S^{2}} \left((J^{2} - L^{2} + S^{2}) (S^{2} + S_{1}^{2} - S_{2}^{2}) - A_{1} A_{2} A_{3} A_{4} \cos\phi'\right)\right]\,.
\end{align}

The coefficients of $d\phi_{z}/dt$ in Eqs.~\eqref{eq:dphizdt-GR} \& Eq.~\eqref{eq:dphizdt-dcs} are
\allowdisplaybreaks[4]
\begin{align}
\label{eq:b0}
b_{0} &= v^6 (J^4 q + L^4 q + q S_{+}^4 + 6 L^3 M^2 q \chi_{\rm eff} (-1 + v \chi_{\rm eff}) 
+ 6 L M^2 q S_{+}^2 \chi_{\rm eff} (-1 + v \chi_{\rm eff}) 
\nonumber \\
&- 2 J^2 q [L^2 + S_{+}^2 + 3 L M^2 \chi_{\rm eff} (-1 + v \chi_{\rm eff})] 
+ 2 L^2 \{3 (-1 + q^2) S_{1}^2 (-1 + v \chi_{\rm eff}) 
\nonumber \\
&- 3 (-1 + q^2) S_{2}^2 (-1 + v \chi_{\rm eff}) 
+ S_{+}^2 [-3 - 7 q - 3 q^2 + 3 (1 + q)^2 v \chi_{\rm eff}]\})\,,
\\
b_{2} &= 2 (S_{-} - S_{+}) (S_{-} + S_{+}) v^6 \{q (-J^2 + S_{+}^2) + 3 L M^2 q \chi_{\rm eff} (-1 + v \chi_{\rm eff}) 
\nonumber \\
&+ L^2 [-3 - 7 q - 3 q^2 + 3 (1 + q)^2 v \chi_{\rm eff}]\}\,,
\\
b_{4} &= q (S_{-}^2 - S_{+}^2)^2 v^6\,,
\\
d_{0} &= 2 M^3 q (J - L - S_{+}) (J + L - S_{+}) (J - L + S_{+}) (J + L + S_{+})\,,
\\
d_{2} &= 4 M^3 q (J^2 + L^2 - S_{+}^2) (-S_{-}^2 + S_{+}^2)\,,
\\
\label{eq:c4}
d_{4} &= 2 M^3 q (S_{-}^2 - S_{+}^2)^2\,.
\end{align}
\end{widetext}
As we pointed out below Eq.~\eqref{eq:dphizdt-GR}, our expression for $d\phi_{z}/dt$ is equivalent to Eq.~(30) of~\cite{Chatziioannou:2017tdw}. The mapping between the $c_{n}$ coefficients therein and the $b_{n}$ coefficients found in Eq.~\eqref{eq:dphizdt-GR} is
\begin{equation}
    b_{n} = c_{n} + a d_{n}
\end{equation}
where
\begin{equation}
    a = \frac{v^{6}}{2M^{3}} \left[1 + \frac{3}{2\eta}(1-v\chi_{\rm eff})\right]\,.
\end{equation}
The coefficients $(\bar{A}_{\phi}, \bar{C}_{\pm}, \bar{n}_{\pm})$ in Eq.~\eqref{eq:phiz-GR} are functions of these, specifically
\begin{align}
\label{eq:A-phi}
\bar{A}_{\phi} &= \frac{J b_{4}}{d_{0} \bar{n}_{+} \bar{n}_{-}}\,,
\\
\bar{C}_{\pm} &= \frac{J\left(b_{4} + b_{2} \bar{n}_{\pm} + b_{0} \bar{n}_{\pm}^2\right)}{d_{0} \bar{n}_{\pm} (\bar{n}_{\pm} - \bar{n}_{\mp})}\,,
\\
\label{eq:n-d}
\bar{n}_{\pm} &= - \frac{1}{2 d_{0}} \left(d_{2} \pm \sqrt{d_{2}^{2} - 4 d_{0} d_{4}}\right)\,,
\end{align}
The full expressions for $(\delta A_{\phi}', \delta B_{\phi}')$ are too lengthy to provide here but can be readily derived from Eq.~\eqref{eq:dphizdt-dcs}. Instead, we provide their leading order PN expansion since these are of most relevance to the present calculations, specifically
\begin{align}
    \delta A_{\phi}' &= \frac{25}{32}\frac{q^{2} u^{6}}{c_{1}^{3} M^{3}} \frac{(1-q)^{2}}{(1+q)} \left(c_{1}^{2} + s_{+}^{(0)}\right) \left[c_{1} (1+q) - M^{2} q \chi_{c}\right]\,,
    \\
    \delta B_{\phi}' &= \frac{25}{32} \frac{q^{2} u^{6}}{c_{1}^{3} M^{3}} \frac{(1-q)^{2}}{(1+q)} \left(s_{-}^{(0)} - s_{+}^{(0)}\right) \left[c_{1} (1+q) - M^{2} q \chi_{c}\right]\,.
\end{align}

\section{\label{app:spn}PN Expansion of $(S_{+}^{2}, S_{-}^{2}, S_{3}^{2})$}

We here provide the PN coefficients of the roots $(S_{+}^{2}, S_{-}^{2}, S_{3}^{2})$ as described in Sec.~\ref{sec:const}. To achieve this, we must PN expand Eq.~\eqref{eq:xn} for $n=0,1,2$, respectively. It is convenient to consider the PN expansion of the quantity within the $\arccos$ of Eq.~\eqref{eq:xn}, specifically
\begin{align}
    \label{eq:yn}
    -\frac{q}{2} \left(\frac{3}{p}\right)^{3/2} &= -1 + v^{4} \sum_{n=0} y_{n} v^{n} + \bar{\zeta}_{2} \delta y_{0} v^{4}
\end{align}
where
\begin{align}
    y_{0} &= \frac{54 q^{2} \Delta_{1} \Delta_{2}}{M^{8}(1-q)^{8}(1+q)^{4} \eta^{4}}\,,
    \\
    \delta y_{0} &= \frac{225c_{1}q^{5}}{8M^{8}(1-q)^{6}(1+q)^{5}\eta^{4}} 
    \nonumber \\
    &\times
    \Big\{(1-q)^{3}(1+q)^{2}\left(S_{2}^{2}-S_{1}^{2}\right)\left[c_{1}(1+q) - q M^{2} \chi_{\rm eff}\right] 
    \nonumber \\
    &- \left[2 c_{1}^{2} q (1+q)^{2} - (1-q^{2})^{2} \left(S_{1}^{2} + S_{2}^{2}\right) 
    \right.
    \nonumber \\
    &\left.
    - 2 c_{1} M^{2} q (1+q)^{2} \chi_{\rm eff} + 2 M^{4} q^{2} \chi_{\rm eff}^{2}\right]
    \nonumber \\
    &\times \left[c_{1}(1+q)^{2} - q (3+q) M^{2} \chi_{\rm eff}\right] \Big\}
\end{align}
with
\begin{align}
    \Delta_{1} &= c_{1}^2 (1 + q)^2 - (-1 + q^2)^2 S_{1}^2 
    \nonumber \\
    &- 2 c_{1} M^2 q (1 + q) \chi_{\rm eff} +M^4 q^2 \chi_{\rm eff}^2\,,
    \\
    \Delta_{2} &= c_{1}^2 q^2 (1 + q)^2 - (-1 + q^2)^2 S_{2}^2 
    \nonumber \\
    &- 2 c_{1} M^2 q^2 (1 + q) \chi_{\rm eff} + M^4 q^2 \chi_{\rm eff}^2\,.
\end{align}
The remaining $y_{n}$ are too lengthy to provide here, but can be derived from Eq.~\eqref{eq:yn}. From here, we may compute the PN expansions for the roots, the explicit form of which are given in Eqs.~\eqref{eq:Spm-pn}-\eqref{eq:S3-pn}. For $S_{\pm}^{2}$ in Eq.~\eqref{eq:Spm-pn}, the coefficients of these equations can generically be written as polynomials in $\chi_{c}$, specifically
\begin{align}
    s_{\pm}^{(n)} &= \sum_{k=0}^{n+2} \sigma_{\pm,k}^{(n)} (M^{2} \chi_{c})^{k}\,.
\end{align}
Up to relative 2PN order, the coefficients of these polynomials are
\begin{align}
    \sigma_{\pm,0}^{(0)} &= -\frac{2c_{1}\eta}{\Delta_{m}^{2}} - \Delta_{m}^{2} Y_{0} + Z_{0}
    \\
    \sigma_{\pm,1}^{(0)} &= \frac{2c_{1}\eta}{\Delta_{m}^{2}}
    \\
    \sigma_{\pm,2}^{(0)} &= \frac{2\eta^{2}}{\Delta_{m}^{2}}
    \\
    \sigma_{\pm,0}^{(1)} &= -\frac{8c_{1}^{3}q(1+q)^{2}}{M^{2}(1-q)^{4}} \pm Y_{1} 
    \nonumber \\
    &+ \frac{2c_{1}\eta}{M^{2} q^{2} \Delta_{m}^{2}} \left(q^{2} Z_{2} + q^{3} X_{1}^{\mp} + q X_{2}^{\mp} + q^{2} X_{d}^{\pm}\right)\,,
    \\
    \sigma_{\pm,1}^{(1)} &= \frac{12c_{1}^{2} q (1+q)^{2} + (1-q)^{2} X_{q}^{\pm}}{M^{2}(1-q)^{4}}\,,
    \\
    \sigma_{\pm,2}^{(1)} &= -\frac{4 c_{1} q(1+4q+q^{2})}{M^{2}(1-q)^{4}}\,,
    \\
    \sigma_{\pm,3}^{(1)} &= \frac{4q^{2}}{M^{2}(1-q)^{4}}\,,
    \\
    \sigma_{\pm,0}^{(2)} &= \sum_{n=0}^{8} \left(U_{2,n}^{\pm} q^{n} + U_{1,n}^{\pm} q^{16-n}\right)
    \\
    \sigma_{\pm,1}^{(2)} &= \frac{2\eta^{3/2}}{M^{4} q^{9/2} \Delta_{m}^{6}} \left\{38 c_{1}^{3} \frac{q^{9/2}}{\eta^{1/2}} \pm \frac{M^{2} \Delta_{m}^{4} q^{9/2}}{\eta^{3/2}} Y_{1} 
    \right.
    \nonumber \\
    &\left.
    - \frac{c_{1}q^{2} \Delta_{m}^{2}}{\eta} \left[q^{5/2}\left(V_{s}^{\mp} - 3 W_{s}^{\mp}\right) \mp 3 Y_{0} \eta^{2} (1+q^{5})\right]\right\}
    \\
    \sigma_{\pm,2}^{(2)} &= \frac{\eta^{4}}{M^{4} q^{6} \Delta_{m}^{6}} \left[2 c_{1}^{3} q^{3} \left(23+68q+23q^{2}\right) 
    \right.
    \nonumber \\
    &\left.
    + 2(1-q)^{2} q W_{q}^{\mp}\right]
    \\
    \sigma_{\pm,3}^{(2)} &= \frac{4c_{1} q (1+q)^{2} (2+15q+2q^{2})}{M^{4}(1-q)^{6}}
    \\
    \sigma_{\pm,4}^{(2)} &= -\frac{2q^{2}(4+11q+4q^{2})}{M^{4}(1-q)^{6}}
\end{align}
where we have defined
\allowdisplaybreaks[4]
\begin{align}
    \Delta_{m} &= \frac{m_{1}-m_{2}}{M}\,,
    \\
    \langle S^{2} \rangle_{\psi,0} &= \frac{1}{2} \left(s_{+}^{(0)} + s_{-}^{(0)}\right)\,,
    \\
    Z_{n} &= q^{n} S_{1}^{2} + q^{-n} S_{2}^{2}\,,
    \\
    Z_{d} &= Z_{0} - \langle S^{2}\rangle_{\psi,0}\,,
    \\
    Z_{q} &= (1 + 4 q + 3 q^{2})S_{1}^{2} + (3 + 4q + q^{2}) S_{2}^{2}\,,
    \\
    Y_{0} &= \sqrt{\frac{2y_{0}}{3}} \frac{M^{4} q}{3(1+q)^{2}}\,,
    \\
    Y_{1} &= \frac{y_{1} M^{4} (-1+q)^{2}q}{3 \sqrt{6y_{0}} (1+q)^{4}}\,,
    \\
    Y_{2} &= \frac{y_{2} M^{4} (1-q)^{2} \eta^{2}}{3\sqrt{6y_{0}} q}
    \\
    X_{A}^{\pm} &= 2S_{A}^{2} + Z_{d} \pm 2Y_{0}\,,
    \\
    X_{d}^{\pm} &= Z_{0} + 2 Z_{d} \pm 4 Y_{0}\,,
    \\
    X_{q}^{\pm} &= (1+q)^{2} \langle S^{2} \rangle_{\psi,0} - Z_{q} \pm (1-q)^{2} Y_{0}\,,
    \\
    W_{q}^{\pm} &= q Z_{d} + q^{3} Z_{d} \pm 3 Y_{0} \eta^{2} \pm 3 q^{4} Y_{0} \eta^{2} 
    \nonumber \\
    &+ q^{2} \left(4 Z_{d} \mp 6Y_{0} \eta^{2} + 2Z_{0} + 3Z_{1}\right)\,,
    \\
    W_{A}^{\pm} &= S_{A}^{2} \pm Y_{0} \eta^{2}\,,
    \\
    W_{s}^{\pm} &= q^{3/2} W_{1}^{\pm} + q^{-3/2} W_{2}^{\pm}\,,
    \\
    V_{A}^{\pm} &= 13 S_{A}^{2} + 6 Z_{d} - 6 W_{A}^{\pm}\,,
    \\
    V_{s}^{\pm} &= q^{1/2} U_{1}^{\pm} + q^{-1/2} U_{2}^{\pm}\,,
    \\
    U_{A,0}^{\pm} &= Y_{0}^{2} \eta^{4}\,,
    \\
    U_{A,1}^{\pm} &= \mp 2 S_{2}^{2} Y_{0} \eta^{2}\,,
    \\
    U_{A,2}^{\pm} &= \mp 4 Z_{d} Y_{0} \eta^{2} \pm 8 c_{1} M^{2} Y_{1} \eta^{3} - 8 Y_{0}^{2} \eta^{4}\,,
    \\
    U_{A,3}^{\pm} &= 2 \eta^{2} \left(\pm 6 c_{1}^{2} Y_{0} \mp S_{B}^{2} Y_{0} \mp M^{4} Y_{2} \pm 16 c_{1} M^{2} Y_{1} \eta\right) 
    \nonumber \\
    &+ 2 S_{A}^{2} \left(Z_{d} \pm 6 Y_{0} \eta^{2}\right)
    \\
    U_{A,4}^{\pm} &= Z_{d}^{2} \pm 16 Z_{d} Y_{0} \eta^{2} \pm 16 c_{1} M^{2} Y_{1} \eta^{3} 
    \nonumber \\
    &- 4 c_{1}^{2} (5 S_{A}^{2} \mp 6 Y_{0} \eta^{2}) \pm 4 \eta^{2} (2 S_{B}^{2} Y_{0} + 2 S_{A}^{2} Y_{0} 
    \nonumber \\
    &- 2 \langle S^{2} \rangle_{\psi,0} Y_{0} + M^{4} Y_{2} \pm 7 Y_{0}^{2} \eta^{2})
    \\
    U_{A,5}^{\pm} &= -2 \Big[-S_{B}^{4} + 4 S_{A}^{4} - 4 S_{A}^{2} \langle S^{2} \rangle_{\psi,0} \pm 15 S_{A}^{2} Y_{0} \eta^{2} 
    \nonumber \\
    &\mp 3 M^{4} Y_{2} \eta^{2} \pm 48 c_{1} M^{2} Y_{1} \eta^{3} + S_{B}^{2} \left(3 S_{A}^{2} + \langle S^{2} \rangle_{\psi,0} 
    \right.
    \nonumber \\
    &\left.
    \mp 6 Y_{0} \eta^{2}\right) + 2 c_{1}^{2} \left(10 S_{A}^{2} - 4 \langle S^{2} \rangle_{\psi,0} \pm 15 Y_{0} \eta^{2}\right) \Big]
    \\
    U_{A,6}^{\pm} &= 4 \Big[19 c_{1}^{4} - S_{B}^{4} - S_{A}^{4} + 2 S_{A}^{2} \langle S^{2} \rangle_{\psi,0} - \langle S^{2} \rangle_{\psi,0}^{2} 
    \nonumber \\
    &\mp 8 S_{A}^{2} Y_{0} \eta^{2} \pm 8 \langle S^{2} \rangle_{\psi,0} Y_{0} \eta^{2} \mp 7 Z_{d} Y_{0} \eta^{2} \mp 4 M^{4} Y_{2} \eta^{2} 
    \nonumber \\
    &\mp 34 c_{1} M^{2} Y_{1} \eta^{3} - 14 Y_{0}^{2} \eta^{4} - 2 S_{B}^{2} \left(S_{A}^{2} - \langle S^{2} \rangle_{\psi,0} 
    \right.
    \nonumber \\
    &\left.
    \pm 4 Y_{0}\eta^{2}\right) - c_{1}^{2} \left(5 S_{B}^{2} - 5 S_{A}^{2} - 12Z_{d} \mp 24 Y_{0} \eta^{2}\right) \Big]
    \\
    U_{A,7}^{\pm} &= 2 \Big[ 152 c_{1}^{4} - 4 S_{B}^{2} + 6 S_{A}^{2} - 6 S_{A}^{2} \langle S^{2} \rangle_{\psi,0} \pm 20 S_{A}^{2} Y_{0} \eta^{2} 
    \nonumber \\
    &\mp 2 M^{4} Y_{2} \eta^{2} \pm 32 c_{1} M^{2} Y_{1} \eta^{3} + S_{B}^{2} \left(2 S_{A}^{2} + 4 \langle S^{2} \rangle_{\psi,0} 
    \right.
    \nonumber \\
    &\left.
    \mp 15 Y_{0} \eta^{2}\right) + 4 c_{1}^{2} \left(S_{B}^{2} + 16 S_{A}^{2} - 6 \langle S^{2} \rangle_{\psi,0} 
    \right.
    \nonumber \\
    &\left.
    - 3 Z_{d} \pm 3 Y_{0} \eta^{2}\right)  \Big]
    \\
    U_{A,8}^{\pm} &= 228 c_{1}^{4} + 3 S_{B}^{4} + 3 S_{A}^{4} - 6 S_{A}^{2} \langle S^{2} \rangle_{\psi,0} + 3 \ssqav ^{2} 
    \nonumber \\
    &\pm 24 S_{A}^{2} Y_{0} \eta^{2} \mp 24 \ssqav Y_{0} \eta^{2} \pm 16 Z_{d} Y_{0} \eta^{2} 
    \nonumber \\
    &\pm 12 M^{4} Y_{2} \eta^{2} \pm 112 c_{1} M^{2} Y_{1} \eta^{3} + 35 Y_{0}^{2} \eta^{4} 
    \nonumber \\
    &+ 6 S_{B}^{2} \left(S_{A}^{2} - \ssqav \pm 4 Y_{0}\eta^{2}\right) 
    \nonumber \\
    &
    + c_{1}^{2}\left(58 S_{B}^{2} + 58 S_{A}^{2} - 48 \ssqav \pm 72 Y_{0} \eta^{2}\right)
\end{align}
The $(A,B)$ in the coefficients $U_{A,n}^{\pm}$ belong to the set $\{1,2;A\neq B\}$. Note that $U_{A,8}^{\pm}$ is symmetric under the exchange $A\leftrightarrow B$, i.e. $U_{1,8}^{\pm} = U_{2,8}^{\pm}$. Also, in general the ``$-$" coefficients can be found by making the replacement $Y_{n} \rightarrow -Y_{n}$ in the ``$+$" coefficients. For $S_{3}^{2}$ in Eq.~\eqref{eq:S3-pn}, the coefficient can also be written as polynomials in $\chi_{c}$, but the maximum power changes, specifically
\begin{align}
    s_{3}^{(n)} &= \sum_{k=0}^{n} \sigma_{3,k}^{(n)} (M^{2} \chi_{c})^{k}\,.
\end{align}
Once again, to relative 2PN order, the $\sigma_{3,k}^{(n)}$ are
\begin{align}
    \sigma_{3,0}^{(0)} &= \frac{M^{4} \eta^{2}}{q} (1-q)^{2}\,,
    \\
    \sigma_{3,0}^{(1)} &= 4 c_{1} M^{2} \eta\,,
    \\
    \sigma_{3,1}^{(1)} &= -2 M^{2} \eta\,,
    \\
    \sigma_{3,0}^{(2)} &= \frac{4c_{1}^{2}q}{(1-q)^{2}} + \ssqav - Z_{d} - Z_{1}\,,
    \\
    \sigma_{3,1}^{(2)} &= -\frac{4c_{1}q}{(1-q)^{2}}\,,
    \\
    \sigma_{3,2}^{(2)} &= \frac{4q^{2}}{(1-q^{2})^{2}}\,.
\end{align}
As a final note, these expression are supposed to be GR quantities, i.e. they are independent of the dCS coupling parameter $\xi$ (or $\bar{\zeta}_{2}$), yet they depend on $\chi_{c}$. However, within the GR limit, $\chi_{c}$ becomes the standard constant $\chi_{\rm eff}$ from the PN precession equations in GR.

\section{\label{app:avg}Precession Averages}

When considering the evolution of various quantities under radiation reaction within a MSA, we require the precession averages of dot products between different angular momentum vectors. Specifically, the $(a_{n}, b_{n})$ coefficients of Eq.~\eqref{eq:dudt} explicitly given in Appendix A of~\cite{Chatziioannou:2013dza} depend on spin-orbit, spin-spin, and higher PN spin coupling effects. 

As an example, the spin-orbit correction to $a_{3}$ is
\begin{equation}
    \beta_{3} = \frac{1}{M^{2}} \left(\frac{113}{12} + \frac{25}{4} q\right) \left(\vec{S}_{1} \cdot \hat{L}\right) + \left(1\leftrightarrow2\right)\,.
\end{equation}
We thus require the average of $\vec{S}_{A}\cdot\hat{L}$. To achieve this, we use Eqs.~\eqref{eq:L-no-prec}-\eqref{eq:S2-no-prec}, and explicitly expand all relevant precession quantities in terms of $\psi$. The full average is then a generic function of $u$, which we may PN expand. For our purposes here, we have found it suffices to truncate these expressions at leading PN order. For example, the linear in spin averages are
\begin{align}
    \langle \vec{S}_{1} \cdot \hat{L} \rangle_{\psi} &= \frac{c_{1}(1+q) - q M^{2} \chi_{c}}{(1-q^{2})} \nonumber \\
    &- \frac{25}{48}\bar{\zeta}_{2} \frac{q^{3}(1-q)}{(1+q)^{2}} \left[c_{1}(1+q) - q M^{2} \chi_{c}\right] + {\cal{O}}(u)\,,
    \\
    \langle \vec{S}_{2} \cdot \hat{L} \rangle_{\psi} &= -\frac{q\left[c_{1}(1+q) - M^{2}\chi_{c}\right]}{(1-q^{2})} 
    \nonumber \\
    &+ \frac{25}{48} \frac{q^{3}(1-q)}{(1+q)^{2}} \bar{\zeta}_{2} \left[c_{1}(1+q) - M^{2} q \chi_{c}\right] + {\cal{O}}(u)\,,
\end{align}
where we have performed the weak coupling expansion. When considering the averages in GR, the final expression would be given by the above equation with $\bar{\zeta}_{2} \rightarrow 0$. However, the average is shifted from its GR value in dCS gravity, and in general this is true for all precession averages. Utilizing Eq.~\eqref{eq:chieff-of-t}, we find the spin-orbit correction becomes
\begin{align}
    \langle \beta_{3} \rangle_{\psi} &= \langle \beta_{3} \rangle_{\psi}^{\rm GR} + \bar{\zeta}_{2} \langle \delta \beta_{3} \rangle_{\psi}^{\rm dCS}\,,
    \\
    \langle \beta_{3} \rangle_{\psi}^{\rm GR} &= \frac{19}{6} \frac{c_{1}}{M^{2}} + \frac{25}{4} \chi_{c}\,.
    \\
    \label{eq:db3-avg}
    \langle \delta \beta_{3} \rangle_{\psi}^{\rm dCS} &= \frac{625}{192} \frac{q^{2}}{M^{2}} \frac{(1-q)^{2}}{(1+q)} \left[c_{1}(1+q) - q M^{2} \chi_{c}\right]
\end{align}

For the remaining spin couplings, the necessary averages to leading PN order are
\begin{widetext}
\begin{align}
\Big\langle \vec{S}_{1} \cdot \vec{S}_{2} \Big\rangle_{\psi} &= \frac{1}{4}\left(s_{+}^{(0)} + s_{-}^{(0)} - 2 S_{1}^{2} - 2 S_{2}^{2}\right) + \frac{1}{4} \bar{\zeta}_{2} \left(\delta s_{+}^{(0)} + \delta s_{-}^{(0)}\right) {\cal{O}}(u)\,,
\\
\Big\langle \left(\hat{L} \cdot \vec{S}_{1}\right)^{2}\Big\rangle_{\psi} &= \left[\frac{c_{1} (1 + q) - M^{2} q \chi_{c}}{(1-q^{2})}\right]^{2} - \frac{25}{24} \frac{q^{3}}{(1+q)^{3}} \bar{\zeta}_{2} \left[c_{1}(1+q) - M^{2} q \chi_{c}\right]^{2} +  {\cal{O}}(u)\,,
\\
\Big\langle \left(\hat{L} \cdot \vec{S}_{2}\right)^{2} \Big\rangle_{\psi} &= \left[q \frac{c_{1} (1 + q) - M^{2} \chi_{c}}{(1-q^{2})}\right]^{2} - \frac{25}{24} \frac{q^{4}}{(1+q)^{3}} \bar{\zeta}_{2} \left[c_{1}(1+q) - M^{2} \chi_{c}\right] \left[c_{1}(1+q) - M^{2} q \chi_{c}\right] + {\cal{O}}(u) \,,
\\
\Big\langle \left(\hat{L} \cdot \vec{S}_{1}\right) \left(\hat{L} \cdot \vec{S}_{2}\right) \Big\rangle_{\psi} &= - \frac{q}{(1-q^{2})^{2}}\left[c_{1}(1+q) - M^{2} q \chi_{c}\right]\left[c_{1}(1+q) - M^{2} \chi_{c}\right] 
\nonumber \\
& 
+ \frac{25}{48} \bar{\zeta}_{2} \left(\frac{q}{1+q}\right)^{3} \left[c_{1}(1+q) - M^{2} q\chi_{c}\right] \left[c_{1}(1+q)^{2} - 2 M^{2} q \chi_{c}\right] + {\cal{O}}(u) \,.
\end{align}
The averages of the higher PN order spin coupling are then
\allowdisplaybreaks[4]
\begin{align}
    \langle \sigma_{4} \rangle_{\psi} &= \langle \sigma_{4} \rangle_{\psi}^{\rm GR} + \bar{\zeta}_{2} \langle \delta \sigma_{4} \rangle_{\psi}^{\rm dCS}\,,
    \\
    \langle \sigma_{4} \rangle_{\psi}^{\rm GR} &= \frac{-719+1442q-719q^{2}}{96(1-q)^{2}}\chi_{c}^{2} - \frac{c_{1} (1+q)^{2}}{24 M^{2} (1-q)^{2}} \chi_{c} + \frac{(1+q)^{2}}{192M^{4} q^{2}(1-q)^{2}} \bigg\{ 8 c_{1}^{2} q^{2} + (1-q)^{2}\left[466 q^{2} S_{1}^{2} 
    \right.
    \nonumber \\
    &\left.
    + 466 S_{2}^{2} - 247q \left(2 S_{1}^{2} + 2 S_{2}^{2} - s_{+}^{(0)} - s_{-}^{(0)}\right) \right] \bigg\} \,,
    \\
    \label{eq:ds4-avg}
    \langle \delta \sigma_{4} \rangle_{\psi}^{\rm dCS} &= -\frac{25}{2304} \frac{q^{2}}{M^{4}(1+q)} \left[c_{1}(1+q) - M^{2} q \chi_{c}\right]\left[2c_{1}(1+q)^{2} + M^{2} \left(719 - 1442q + 719q^{2}\right) \chi_{c}\right]\,,
    \\
    \langle \beta_{5} \rangle_{\psi}^{\rm GR} &= \left(\frac{21611}{1008} - \frac{79}{6}\eta\right)\frac{c_{1}}{M^{2}} + \left(\frac{809}{84}-\frac{281}{8}\eta\right)\chi_{c}\,,
    \\
    \langle \beta_{6} \rangle_{\psi}^{\rm GR} &= \frac{\pi}{6} \left(74 \frac{c_{1}}{M^{2}} + 151 \chi_{c}\right)\,,
    \\
    \langle \beta_{7} \rangle_{\psi}^{\rm GR} &= \left(\frac{1932041}{18144} - \frac{40289}{288} \eta + \frac{10819}{432}\eta^{2}\right)\frac{c_{1}}{M^{2}} + \left(\frac{1195759}{18144} - \frac{257023}{1008}\eta + \frac{2903}{32}\eta^{2}\right)\chi_{c}\,,
\end{align}
and the 2PN dCS dipole radiation term in Eq.~\eqref{eq:dC} becomes
\begin{align}
    \label{eq:dC-avg}
    \langle \delta C\rangle_{\psi} &= \frac{5}{344064} \chi_{1} \frac{(1+q)^{2}}{q^{3}} \left(17815+20311q^{3}\right) + \frac{5}{344064} \chi_{2} \frac{(1+q)^{2}}{q^{3}} \left(20311 + 17815 q^{3}\right) 
    \nonumber \\
    &-\frac{5(1+q)^{6}}{688128 M^{2}(1-q)^{2}q^{4}}\left[18 c_{1}^{2} \left(6537+11410 q^{2} + 6537 q^{4}\right) + 17815 (1-q)^{2} q \left(s_{+}^{(0)} + s_{-}^{(0)}\right)\right]
    \nonumber \\
    &+ \frac{15c_{1}\chi_{c}(1+q)^{6}}{57343M^{2}(1-q)^{2}q^{4}}\left(6537-6537q+12242q^{2}-6537q^{3}+6537q^{4}\right)
    \nonumber \\
    &-\frac{15\chi_{c}^{2}(1+q)^{4}}{114688(1-q)^{2}q^{4}}\left(6537+11410q^{3}+6537q^{6}\right)\,.
\end{align}
\end{widetext}
The precession averaged coeffiencts in Eq.~\eqref{eq:u0-avg} are then found by replacing the spin couplings $\beta_{n}$ and $\sigma_{4}$ in Eqs.~(A2)-(A8) in~\cite{Chatziioannou:2013dza} with the above averaged expresssions.

Technically, Eq.~\eqref{eq:db3-avg} constitutes the lowest PN order correction to the GW phase for dCS gravity, entering at relative 1.5PN order. However, for spin aligned binaries, it is well known that the dCS correction enters at 2PN order due to dipole radiation~\cite{Yagi:2012vf}. It then must be that $\langle \delta\beta_{3}\rangle_{\psi}^{\rm dCS}$ is suppressed in the aligned limit, and the same is true for $\langle\delta\sigma_{4}\rangle_{\psi}^{\rm dCS}$. This implies, however, that we have multiple dCS effects at different PN orders that compete against one another for which is the dominant correction. From the arguments regarding the spin aligned limit, we can expect that the dipole radiation term $\langle \delta C\rangle_{\psi}^{\rm dCS}$ will dominate when $\theta_{L}$ is small. However, this need not be true for arbitrary alignments. Numerical values for these terms are provided in Table~\ref{tab1} for the systems therein. Indeed, we find the expected behavior as $\theta_{L}\rightarrow0$, but for more significant misalignments, we find that the dipole radiation term can be 2-3 orders of magnitude smaller than both the spin-orbit and spin-spin corrections. In order to achieve a waveform that is uniform across possible values of the angle $\theta_{L}$, we keep all three terms in the waveform's phase in Eq.~\eqref{eq:spa}.

The last averages we need to consider are those of the cross products used to define the effective precession spin in Eq.~\eqref{eq:chip}. This quantity depends on the magnitudes $|\hat{L}\times \vec{S}_{A}|$, which are frame independent and, as such, we compute them in the co-precessing frame to simplify the calculation. In general, the precession average is difficult to compute due to the fact that it depends on the square root of a trigonmetric polynomial. Instead, we compute the root-mean-square of these quantities to obtain them analytically. Following the average procedures above, we find
\begin{align}
    \langle \hat{L}\times \vec{S}_{A}\rangle_{\rm rms} &= \sqrt{\Big\langle |\hat{L}\times\vec{S}_{A}|^{2} 
    \Big\rangle_{\psi}}
    \nonumber \\
    &= {\cal{P}}_{A} + \frac{\bar{\zeta}_{2}}{2} \frac{\delta {\cal{P}}_{A}}{{\cal{P}}_{A}}
\end{align}
where 
\begin{align}
    {\cal{P}}_{1} &= \frac{1}{(1-q)(1+q)} \left[(1-q^{2})^{2}S_{1} - c_{1}^{2}(1+q)^{2} 
    \right.
    \nonumber \\
    &\left.
    + 2 c_{1} M^{2} q (1+q) \chi_{c} - M^{4} q^{2} \chi_{c}^{2}\right]^{1/2}
    \\
    {\cal{P}}_{2} &= \frac{1}{(1-q)(1+q)} \left[(1-q^{2})^{2} S_{2} - c_{1}^{2} q^{2}(1+q)^{2} \right.
    \nonumber \\
    &\left.
    + 2 c_{1} M^{2} q^{2} (1+q) \chi_{c} - M^{4} q^{2} \chi_{c}^{2}\right]^{1/2}
    \\
    \label{eq:dP-1}
    \delta {\cal{P}}_{1} &= \frac{25}{24} \frac{q^{3}}{(1+q)^{3}} \left[c_{1}(1+q) - M^{2} q \chi_{c}\right]^{2}
    \\
    \label{eq:dP-2}
    \delta {\cal{P}}_{2} &= \frac{25}{24} \frac{q^{4}}{(1+q)^{3}} \left[c_{1}^{2}(1+q)^{2} - c_{1} (1+q)^{2} M^{2} \chi_{c} + M^{4} q \chi_{c}^{2}\right]
\end{align}
\section{\label{app:rr-coeffs}Radiation Reaction Coefficients}

We here provide the PN coefficients of various quantities when considering radiation reaction. The coefficients $\Sigma_{n}$ appearing in Eq.~\eqref{eq:dchi-eff-dL} for $\chi_{\rm eff}$ are
\begin{align}
    \Sigma_{0} &= \frac{c_{1}^{2} (f_{5}  - q f_{3} )}{(1-q)^{2}}\,,
    \\
    \Sigma_{1} &= \frac{c_{1} M^{2} q \left[q(1+q)f_{3} - 2 f_{5} \right]}{(1-q)^{2}(1+q)}\,,
    \\
    \Sigma_{2} &= \frac{M^{4}q^{2}\left(f_{5} - q f_{3} \right)}{(1-q^{2})^{2}}\,.
\end{align}

For $\psi(u)$ in Eq.~\eqref{eq:psi-pn}, the coefficients up to 2PN order are
\begin{widetext}
\begin{align}
    \label{eq:psi-1}
    \psi_{1} &= \frac{3}{4 g_{0} s_{3}^{(0)}} \left[2 g_{1} s_{3}^{(0)} - g_{0} \left(s_{3}^{(1)} + 2 s_{3}^{(0)} \chi_{c}\right)\right]\,,
    \\
    \psi_{2} &= \frac{3}{8g_{0} \left(s_{3}^{(0)}\right)^{2}} \left\{8 g_{2} \left(s_{3}^{(0)}\right)^{2} - 4 g_{1} s_{3}^{(0)} \left(s_{3}^{(1)} + 2 s_{3}^{(0)} \chi_{c}\right) - g_{0} \left[\left(s_{3}^{(1)}\right)^{2} + 4 s_{3}^{(0)} s_{3}^{(2)} - 4 s_{3}^{(0)} s_{+}^{(0)} - 4 s_{3}^{(0)} s_{3}^{(1)} \chi_{c}\right]\right\}\,,
    \\
    \psi_{3} &= -\frac{3}{16 g_{0} \left(s_{3}^{(0)}\right)^{3}} \left[16 g_{3} \left(s_{3}^{(0)}\right)^{3} - 2 g_{1} s_{3}^{(0)} \left(s_{3}^{(1)}\right)^{2} - g_{0} \left(s_{3}^{(1)}\right)^{3} - 8 g_{1} \left(s_{3}^{(0)}\right)^{2} s_{3}^{(2)} - 4 g_{0} s_{3}^{(0)} s_{3}^{(1)} s_{3}^{(2)} - 8 g_{0} \left(s_{3}^{(0)}\right)^{2} s_{3}^{(3)} 
    \right.
    \nonumber \\
    &\left.
    + 8 g_{1} \left(s_{3}^{(0)}\right)^{2} s_{+}^{(0)} + 4 g_{0} s_{3}^{(0)} s_{3}^{(1)} s_{+}^{(0)} + 8 g_{0} \left(s_{3}^{(0)}\right)^{2} s_{+}^{(1)} + 8 g_{1} \left(s_{3}^{(0)}\right)^{2} s_{3}^{(1)} \chi_{c} + 2g_{0} s_{3}^{(0)} \left(s_{3}^{(1)}\right)^{2}\chi_{c} + 8 g_{0} \left(s_{3}^{(0)}\right)^{2} s_{3}^{(2)} \chi_{c} 
    \right.
    \nonumber \\
    &\left.
    - 8 g_{0} \left(s_{3}^{(0)}\right)^{2} s_{+}^{(0)}\chi_{c} - 8 g_{2} \left(s_{3}^{(0)}\right)^{2} \left(s_{3}^{(1)} + 2 s_{3}^{(0)} \chi_{c}\right)\right]\,,
    \\
    \label{eq:psi-4}
    \psi_{4} &= \frac{3}{128 g_{0} \left(s_{3}^{(0)}\right)^{4}} \Bigg\{16 g_{2} \left(s_{3}^{(0)}\right)^2 \left(s_{3}^{(1)}\right)^2 + 8 g_1 s_{3}^{(0)} \left(s_{3}^{(1)}\right)^3 + 5 g_0 \left(s_{3}^{(1)}\right)^4 + 
 16 g_0 \left(s_{3}^{(0)}\right)^2 \left(s_{3}^{(0)} - s_{+}^{(0)}\right)^2 
 \nonumber \\
 &+ 64 g_2 \left(s_{3}^{(0)}\right)^3 \left(s_{3}^{(2)} - s_{+}^{(0)} - s_{3}^{(1)} \chi_{c}\right) - 
 16 g_1 \left(s_{3}^{(0)}\right)^2 s_{3}^{(1)} \left(-2 s_{3}^{(2)} + 2 s_{+}^{(0)} + s_{3}^{(1)} \chi_{c}\right) - 
 8 g_0 s_{3}^{(0)} \left(s_{3}^{(1)}\right)^2 \left(-3 s_{3}^{(2)} 
 \right.
 \nonumber \\
 &\left.
 + 3 s_{+}^{(0)} + s_{3}^{(1)} \chi_{c}\right) + 
 64 g_1 \left(s_{3}^{(0)}\right)^3 \left[s_{3}^{(3)} - s_{+}^{(1)} + \left(-s_{3}^{(2)} + s_{+}^{(0)}\right) \chi_{c}\right] + 
 32 g_0 \left(s_{3}^{(0)}\right)^2 s_{3}^{(1)} \left[s_{3}^{(3)} - s_{+}^{(1)} + \left(-s_{3}^{(2)} + s_{+}^{(0)}\right) \chi_{c}\right] 
 \nonumber \\
 &- 64 g_0 \left(s_{3}^{(0)}\right)^3 \left[s_{+}^{(2)} + \left(s_{3}^{(3)} - s_{+}^{(1)}\right) \chi_{c}\right] + 
 64 \left(s_{3}^{(0)}\right)^3 \left[-2 g_4 s_{3}^{(0)} + g_3 \left(s_{3}^{(1)} + 2 s_{3}^{(0)} \chi_{c}\right)\right]    \Bigg\}\,,
\end{align}
\end{widetext}
We have verified that $(\psi_{1},\psi_{2})$ given above are equivalent to the expressions given in Eqs.~(C1)-(C2) in Appendix C of~\cite{Chatziioannou:2017tdw}. The dCS correction coefficient is
\begin{align}
    \label{eq:dpsi2}
    \delta \psi_{2} &= \frac{\delta y_{0} M^{4} (1-q)^{2} \eta^{2}}{2\sqrt{6 y_{0}} q s_{3}^{(0)}} 
    \nonumber \\
    &+ \frac{3}{s_{3}^{(0)}} \left[\frac{25}{16} c_{1}^{2} q^{3} - \frac{25}{16} \frac{c_{1} M^{2} q^{4} (3+q) \chi_{c}}{(1+q)^{2}} 
    \right.
    \nonumber \\
    &\left.
    - \frac{1}{2} \left(\delta s_{+}^{(0)} + \delta s_{-}^{(0)}\right) - \left(s_{3}^{(0)} - M^{4} \eta\right) \chi_{e,1}\right]\,.
\end{align}

For $\phi_{z}(u)$ given in Eq.~\eqref{eq:phiz-sec}, the functions $\varphi_{n}(u)$ are defined by
\begin{align}
    \label{eq:phiz-funcs}
    \varphi_{n}(u) &= \int du J_{0}(u) u^{n}\,,
\end{align}
where recall that $J_{0}(u) = J(u, \bar{\zeta}_{2}\rightarrow0)$, with $J$ given in Eq.~\eqref{eq:J-of-L}.  Note that these functions are formally equivalent to those in Eqs.~(D22)-(D29) in Appendix D of~\cite{Chatziioannou:2017tdw}. The only difference is whether one computes the integral in terms of $u$ as above, or whether one converts the integral to be over $L$ through the Newtonian mapping $L=\eta M^{2}/u$. We leave the integral in terms of $u$, and writing 
\begin{equation}
    \label{eq:J0}
    J_{0} = u^{-1} (j_{0} + j_{1} u + j_{2} u^{2})^{1/2}\,,
\end{equation}
we find
\begin{align}
    \label{eq:vphi-3}
    \varphi_{-3}(u) &= \frac{J_{0}}{24 j_{0}^{2} u^{2}} \left[-8 j_{0}^{2} + 3 j_{1}^{2} u^{2} - 2 j_{0} u \left(j_{1} + 4 j_{2} u\right)\right] 
    \nonumber \\
    &+ \frac{j_{1}}{8 j_{0}^{5/2}} \left(j_{1}^{2} - 4 j_{0} j_{2}\right) \tau_{1}\,,
    \\
    \varphi_{-2}(u) &= - \frac{J_{0}}{4j_{0}u} \left(2j_{0} + j_{1} u\right) - \frac{\left(j_{1}^{2} - 4 j_{0} j_{2}\right)}{4j_{0}^{3/2}} \tau_{1}\,,
    \\
    \varphi_{-1}(u) &= -J_{0} - \sqrt{j_{2}} \ell_{1} + \frac{j_{1}}{\sqrt{j_{0}}} \tau_{1}\,,
    \\
    \varphi_{0}(u) &= J_{0} u + 2 \sqrt{j_{0}} \tau_{1} - \frac{j_{1}}{2 \sqrt{j_{2}}} \ell_{1}\,,
    \\
    \varphi_{1}(u) &= \frac{J_{0}}{4j_{2}} \left(j_{1} + 2 j_{2} u\right) + \frac{\left(j_{1}^{2} - 4 j_{0} j_{2}\right)}{8 j_{2}^{3/2}} \ell_{2}\,,
\end{align}
with
\begin{align}
    j_{0} &= M^{4} \eta^{2}\,,
    \\
    j_{1} &= 2 c_{1} M^{2} \eta\,,
    \\
    j_{2} &= \langle S^{2}\rangle_{\psi,0}\,,
    \\
    \tau_{1} &= \tanh^{-1}\left[\frac{u}{\sqrt{j_{0}}} \left(\sqrt{j_{2}} - J_{0}\right)\right]\,,
    \\
    \ell_{1} &= \ln\left[j_{1} +2u\left(j_{2} - J_{0} \sqrt{j_{2}}\right)\right]\,,
    \\
    \ell_{2} &= \ln\left(j_{2}\right) + \ell_{1}\,.
\end{align}
The function $\delta \varphi_{-3}(u)$ appearing in the dCS correction to Eq.~\eqref{eq:phiz-sec} is defined as follows
\begin{align}
    \delta \varphi_{-3}(u) &= \int du \left(\frac{\partial J}{\partial \bar{\zeta}_{2}}\right)_{\bar{\zeta}_{2} = 0} u^{-3} 
    \nonumber \\
    &= \frac{1}{4} \left(\delta s_{+}^{(0)} + \delta s_{-}^{(0)}\right) \int \frac{du}{J_{0} u^{3}}\,.
\end{align}
Evaluating, we find
\begin{align}
    \label{eq:dvphi-3}
    \delta \varphi_{-3}(u) &= -\frac{1}{4} \left(\delta s_{+}^{(0)} + \delta s_{-}^{(0)}\right) \left(\frac{J_{0}}{j_{0}} - \frac{j_{1}}{j_{0}^{3/2}} \tau_{1}\right)\,.
\end{align}

For the coefficients of Eq.~\eqref{eq:dphiz-avg}, up to relative 0.5 PN order we find
\begin{align}
    \Omega_{z,n}^{(-1)} &= \sum_{j,k} \rho_{n,j,k} \left(c_{+} c_{-}\right)^{-j} \left(M^{2} \chi_{c}\right)^{k}\,,
\end{align}
where we have defined
\begin{align}
    c_{\pm} = \sqrt{c_{1}^{2} - s_{\pm}^{(0)}}\,,
\end{align}
and the non-zero $\rho_{n,j,k}$ are
\begin{align}
    \rho_{6,0,0} &= \frac{1}{4M^{3}} \left(2 + \frac{3}{\eta}\right)\,,
    \\
    \rho_{6,1,0} &= -\frac{3}{8M^{3}q^{2}} \left[c_{1}^{2} (1+q)^{2} - \left(1-q^{2}\right) \left(S_{1}^{2} - S_{2}^{2}\right)\right]\,,
    \\
    \rho_{6,1,1} &= \frac{3c_{1}}{4M^{3}q}\,,
    \\
    \rho_{7,0,0} &= -\frac{3c_{1}(1+q)^{4}}{8M^{5}q^{3}}\,,
    \\
    \rho_{7,0,1} &= \frac{3}{8M^{5}q^{2}} \left(1+q\right)^{2} \left(1-2q\right)\,,
    \\
    \rho_{7,1,0} &= \frac{3c_{1}(1+q)^{4}}{16M^{5}q^{3}}\left(2c_{1}^{2} - s_{-}^{(0)} - s_{+}^{(0)}\right)\,,
    \\
    \rho_{7,1,1} &= \frac{3(1+q)}{16M^{5}q^{2}}\Bigg[2(1-q)\left(S_{2}^{2} - S_{1}^{2}\right) 
    \nonumber \\
    &+ (1+q) \left(s_{-}^{(0)} + s_{+}^{(0)}\right)\Bigg]\,,
    \\
    \rho_{7,1,2} &= -\frac{3c_{1}}{4M^{5}q}\,,
    \\
    \rho_{7,3,0} &= -\frac{3(1+q)}{32 M^{5}q^{3}} \left[c_{1}^{2} (1+q) - (1-q) \left(S_{1}^{2} - S_{2}^{2}\right)\right] \,,
    \nonumber \\
    &
    \times \left[c_{1} \left(1+q\right)^{2} \left(s_{-}^{(0)} - s_{+}^{(0)}\right)^{2} + 2 c_{1}^{2} M^{2} q \left(s_{-}^{(1)} + s_{+}^{(1)}\right) 
    \right.
    \nonumber \\
    &\left.
    - 2 M^{2} q \left(s_{-}^{(1)} s_{+}^{(0)} + s_{-}^{(0)} s_{+}^{(1)}\right)\right]\,,
    \\
    \rho_{7,3,1} &= \frac{3c_{1}}{16 M^{5} q^{2}} \left[c_{1} (1+q)^{2} \left(s_{-}^{(0)} - s_{+}^{(0)}\right)^{2} 
    \right.
    \nonumber \\
    &\left.
    + 2 c_{1}^{2} M^{2} q \left(s_{-}^{(1)} + s_{+}^{(1)}\right) - 2 M^{2} q \left(s_{-}^{(1)} s_{+}^{(0)} + s_{-}^{(0)} s_{+}^{(1)}\right)\right]\,.
\end{align}
We stop the calculations at this PN order due to the fact that the higher PN order expression for $\Omega_{z,n}^{(-1)}$ becomes increasing complicated and too lengthy to provide here. There is no difficulty with extending the calculation to higher PN order if one desires more accuracy. The dCS correction coefficients is
\begin{align}
    \delta \Omega_{z,6}^{(-1)} &= \sum_{j,k} \delta \rho_{n,j,k} (c_{+} c_{-})^{-j} \left(M^{2} \chi_{c}\right)^{k}\,,
    \\
    \delta \rho_{6,0,0} &= \frac{25(1-q)^{2}q^{2}}{64 c_{1}^{2} M^{3}}\left(2c_{1}^{2} + s_{-}^{(0)} + s_{+}^{(0)}\right)\,,
    \\
    \delta \rho_{6,0,1} &= -\frac{25(1-q)^{2} q^{3}}{64c_{1}^{3} M^{3}(1+q)} \left(2c_{1}^{2} + s_{-}^{(0)} + s_{+}^{(0)}\right)\,,
    \\
    \delta \rho_{6,3,0} &= -\frac{3}{16 M^{3} q^{2}} \left[c_{1}^{2}(1+q)^{2} - \left(1-q^{2}\right)\left(S_{1}^{2} - S_{2}^{2}\right)\right] 
    \nonumber \\
    &\times \left(c_{+}^{2} \delta s_{-}^{(0)} + c_{-}^{2} \delta s_{+}^{(0)}\right)\,,
    \\
    \delta \rho_{6,3,1} &= \frac{3 c_{1}}{8 M^{3} q} \left(c_{+}^{2} \delta s_{-}^{(0)} + c_{-}^{2} \delta s_{+}^{(0)}\right)\,.
\end{align}

The coefficients of $\phi_{z,-1}$ in Eq.~\eqref{eq:phiz-sec} are found by dividing Eq.~\eqref{eq:dphiz-1} by Eq.~\eqref{eq:u0-avg} and PN expanding. The first few of these are
\begin{align}
    \label{eq:Phiz-3}
    \Phi_{z,-3}^{(-1)} &= \frac{3M \Omega_{z,6}^{(-1)}}{a_{0}}\,,
    \\
    \Phi_{z,-2}^{(-1)} &= \frac{3M \Omega_{z,7}^{(-1)}}{a_{0}}\,,
    \\
    \label{eq:dPhiz-3}
    \delta \Phi_{z,-3}^{(-1)} &= \frac{3M \delta \Omega_{z,6}^{(-1)}}{a_{0}}\,.
\end{align}
The coefficients of the oscillatory correction to $\phi_{z}$ in Eq.~\eqref{eq:phiz0} are
\begin{align}
    \label{eq:om-osc}
    \Omega_{\rm osc} &= \frac{(3+2\eta)\sqrt{s_{3}^{(0)}}}{8M}
    \\
    \label{eq:k1}
    \kappa_{1} &= \frac{c_{-}^{2}}{c_{+}^{2}} - 1\,,
    \\
    \Upsilon_{z,2}^{(0)} &= \frac{3(1+q)^{2}}{M^{2} q^{2} c_{+} c_{-} (3 + 8 q + 3 q^{2}) \sqrt{s_{3}^{(0)}}} 
    \nonumber \\
    &\times \left[c_{1}^{2}(1+q)^{2} - (1-q^{2}) \left(S_{1}^{2} - S_{2}^{2}\right) - 2 c_{1} M^{2} q \chi_{c}\right]\,,
    \\
    \delta \Upsilon_{z,2}^{(0)} &= - \frac{3(1+q)^{2} \left(c_{+}^{2} \delta s_{-}^{(0)} + c_{-}^{2} \delta s_{+}^{(0)}\right)}{2 c_{+}^{3} c_{-}^{3} M^{2} q^{2} (3+8q + 3q^{2}) \sqrt{s_{3}^{(0)}}} \nonumber \\
    &\times \left[c_{1}^{2} (1+q)^{2} - (1-q^{2}) \left(S_{1}^{2} - S_{2}^{2}\right) - 2 c_{1} M^{2} q \chi_{c}\right]\,,
    \\
    \Delta_{z,2}^{(0)} &= \frac{3 (1+q)^{2} \left(c_{+}^{2} \delta s_{-}^{(0)} - c_{-}^{2} s_{+}^{(0)}\right)}{4 c_{-}^{2} c_{+}^{4} M^{2} q^{2} \left(3 + 8 q + 3 q^{2}\right) \sqrt{s_{3}^{(0)}}} 
    \nonumber \\
    &\times \left[c_{1}^{2} (1+q)^{2} - (1-q^{2}) \left(S_{1}^{2} - S_{2}^{2}\right) - 2 c_{1} M^{2} q \chi_{c}\right]\,,
    \\
    \Sigma_{z,2}^{(0)} &= \frac{25 (1-q)^{2} q^{2} (1+q)}{16 c_{1}^{3} M^{2} (3 + 8q + 3q^{2}) \sqrt{s_{3}^{(0)}}} 
    \nonumber \\
    &\times \left(s_{+}^{(0)} - s_{-}^{(0)}\right) \left[c_{1} (1+q) - M^{2} q \chi_{c}\right]\,.
\end{align}

To obtain the coefficients of $\phi_{T,-1}$ in Eq.~\eqref{eq:phiT-1}, it is convenient to write
\begin{equation}
    \label{eq:dphiT-1}
    \frac{d\phi_{T,-1}}{dt} = \sum_{n=5} \Omega_{T,n}^{(-1)} u_{0}^{n} + \bar{\zeta}_{2} \delta \Omega_{T,5}^{(-1)} u_{0}^{5}\,,
\end{equation}
where
\begin{align}
    \Omega_{T,5}^{(-1)} &= M^{2} \eta \Omega_{z,6}^{(-1)}\,,
    \\
    \Omega_{T,6}^{(-1)} &= c_{1} \Omega_{z,6}^{(-1)} + M^{2} \eta \Omega_{z,7}^{(-1)}\,,
    \\
    \delta \Omega_{T,5}^{(-1)} &= M^{2} \eta \delta \Omega_{z,6}^{(-1)}\,,
\end{align}
By dividing Eq.~\eqref{eq:dphiT-1} by Eq.~\eqref{eq:dudt}, PN expanding, and then integrating, we may find the $\Phi_{T,n}^{(-1)}$ coefficients, specifically
\begin{align}
    \label{eq:PhiT-0}
    \Phi_{T,0}^{(-1)} &= -\frac{M}{a_{0}} \Omega_{T,5}^{(-1)}\,,
    \\
    \Phi_{T,1}^{(-1)} &= - \frac{3M}{2a_{0}} \Omega_{T,6}^{(-1)}\,,
    \\
    \delta \Phi_{T,0}^{(-1)} &= - \frac{M}{a_{0}} \delta \Omega_{T,5}^{(-1)}\,.
\end{align}

For $t(u)$ given in Eq.~\eqref{eq:tofu}, the coefficents $\langle t_{n} \rangle_{\psi}$ and $\langle t_{n}^{l}\rangle_{\psi}$ are found in Appendix G of~\cite{Chatziioannou:2013dza}, with the replacements $a_{n} \rightarrow \langle a_{n} \rangle_{\psi}$ and $b_{n} \rightarrow \langle b_{n} \rangle_{\psi}$. Similarly, for $\phi(u)$ given in Eq.~\eqref{eq:phiofu}, the coefficients can be found in Appendix D of~\cite{Chatziioannou:2013dza} with the same replacements.

\section{\label{app:phiz-comp}Comparison of $\phi_{z}$ Results}

As we pointed out in the main text below Eq.~\eqref{eq:phiz-sec}, the $\Omega_{z,n}^{(-1)}$ coefficients given in the previous appendix are not in general the same as those found in Appendix D of~\cite{Chatziioannou:2017tdw} (hereafter CKYC). It turns out that the leading order coefficient $\Omega_{z,6}^{(-1)}$ is the same as $\Omega_{z,0}$ therein, but the higher PN order terms are not. There are some subtle differences between the calculation of CKYC and those herein that shed some light on the reason for this difference. First, CKYC Taylor expand $S_{\pm}^{2}$ about their initial values, which appears to provide better convergence to an exact (numerical integration) answer under radiation reaction. Here, we do not do this and explicitly PN expand $S_{\pm}^{2}$ as in Eq.~\eqref{eq:Spm-pn} and in Appendix~\ref{app:spn}, since these quantities are shifted from their GR values in a non-trivial way when considering dCS modification. Second, CKYC defines the $\varphi(u)$ functions in Eq.~\eqref{eq:phiz-funcs} differently than we do here. CKYC uses $L$ as the integration variable to obtain these functions, whereas we use our PN expansion variable $u=(2\pi M F)^{1/3}$. 

\begin{figure}
    \centering
    \includegraphics[width=\columnwidth, trim=0cm 1.5cm 0cm 1.5cm, clip]{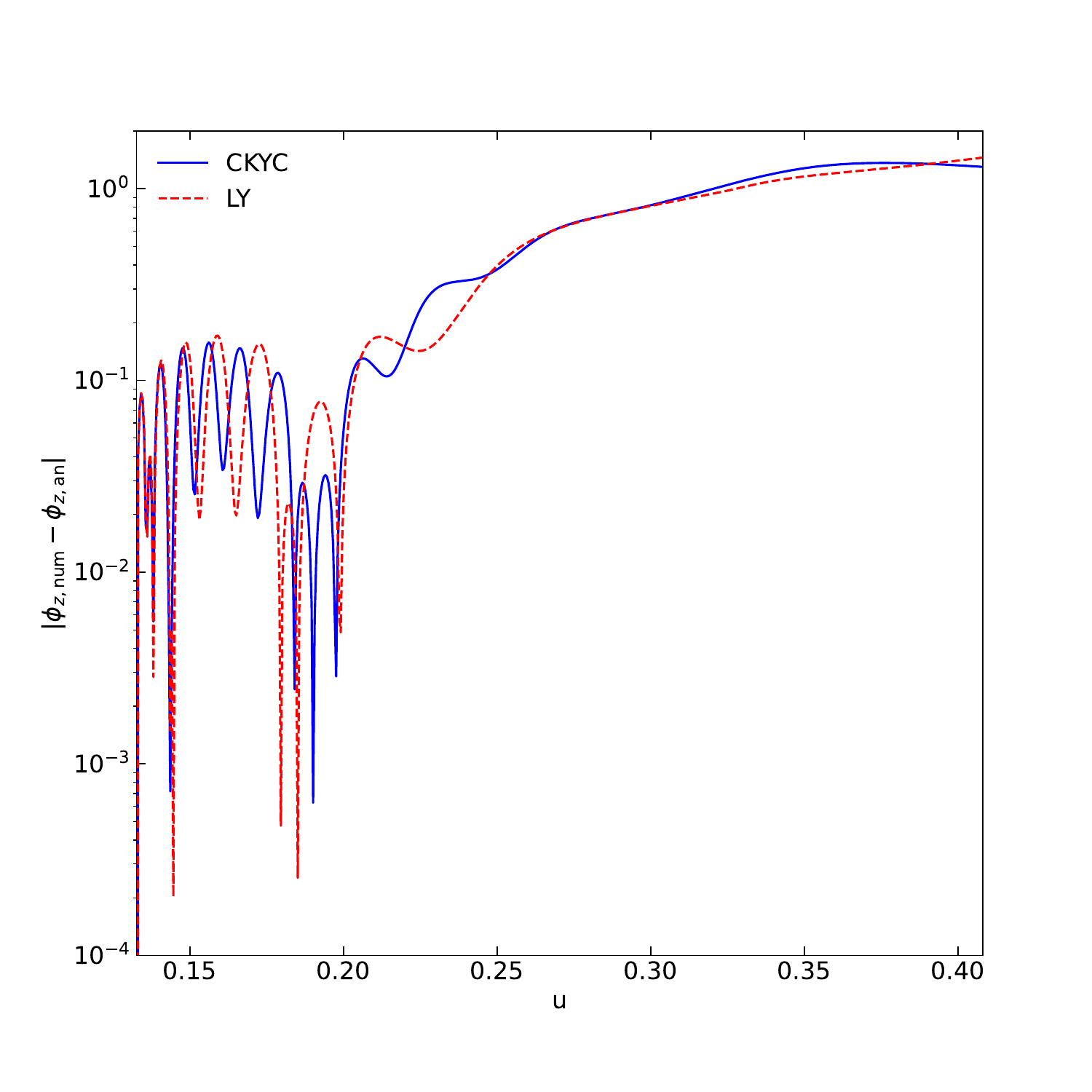}
    \caption{Comparison of the analytic $\phi_{z,-1}$ found in~\cite{Chatziioannou:2017tdw} to our result, namely Eq.~\eqref{eq:phiz-sec}. The error betweent these analytic approximations and numerical integration of the precession and radiation reaction equations does not change significantly, despite us not obtaining the same expression.}
    \label{fig:phiz-comp}
\end{figure}

These differences between our work and that of CKYC are actually minor, as can be seen in Fig.~\ref{fig:phiz-comp}, where we compare the error in the analytic representations of $\phi_{z,-1}$ to the numerical integration described in Sec.~\ref{sec:const}. For comparison, in this figure we use the same PN order in CKYC's expression, namely 0.5PN. Going to higher PN order does not significantly improve the error. The error in our result (labeled LY) and CKYC's result are comparable, with the primary difference being because of oscillatory effects that we have not included in the analytic $\phi_{z}$ for this comparison.

Regardless of the difference between CKYC's and our results, the largest source of error in $\phi_{z,-1}$ is due to the PN expansion of the GR part, which as CKYC points out, is not well behaved and typically gives a relatively large error. One can improve this by creating a hybrid model, wherein the PN expansion of $\phi_{z}$ is replaced with a numerical integration, while all other quantities are analytic. This was considered in CKYC, where it was found that doing so produces better faithfulness to numerical waveforms. Here we are not working in the context of GR anymore. Since the largest source of error in $\phi_{z}$ is in the GR part, one could replace this with the numerical version, while keeping the dCS correction to be the analytic results found here. 

\section{\label{app:amp-fourier}Fourier Decomposition of Waveform Amplitudes}

When computing the waveform in Sec.~\ref{sec:wave}, we performed a Fourier decomposition of the Wigner D-matrices in order to properly expand the non-GR modifications to the amplitude. We here provide more details on how this is done.

The mapping of the Wigner D-matrices to SALPs in Eq.~\eqref{eq:D-to-P} is merely a result of their properties. The SALPS depend on trigonometric combinations of $\theta_{L}$, or more specifically, powers of $\cos\theta_{L}$, given in Eq.~\eqref{eq:thetaL-exp}, and
\begin{align}
    \label{eq:sinL}
    \sin\theta_{L} &= \sqrt{1 - \frac{\left(J^{2} + L^{2} - S^{2}\right)^{2}}{4 J^{2} L^{2}}}
    \nonumber \\
    &= \sqrt{{\cal{S}}_{0}(u) - 2 {\cal{S}}_{2}(u) \sin^{2}\psi + {\cal{S}}_{4}(u) \sin^{4} \psi}\,,
\end{align}
where
\begin{align}
    {\cal{S}}_{0}(u) &= \frac{2J^{2} (L^{2} + S_{+}^{2}) - J^{4} - (L^{2} - S_{+}^{2})^{2}}{4 J^{2} L^{2}}\,,
    \\
    {\cal{S}}_{2}(u) &= \frac{(J^{2} + L^{2} - S_{+}^{2})(S_{+}^{2} - S_{-}^{2})}{4 J^{2} L^{2}}\,,
    \\
    {\cal{S}}_{4}(u) &= - \frac{(S_{-}^{2} - S_{+}^{2})^{2}}{4 J^{2} L^{2}}\,.
\end{align}
The Fourier decomposition of $\cos\theta_{L}$ can be read off directly from Eq.~\eqref{eq:thetaL-exp}, namely
\begin{align}
    \label{eq:cos-fourier}
    \cos\theta_{L} &= \sum_{n=0}^{1} {\cal{C}}_{L,n} \left(e^{in\psi} + e^{-in\psi}\right)\,,
    \\
    \label{eq:CL0}
    {\cal{C}}_{L,0} &= \frac{2J^{2} + 2L^{2} - S_{+}^{2} - S_{-}^{2}}{2 J L}\,,
    \\
    \label{eq:CL2}
    {\cal{C}}_{L,2} &= \frac{S_{-}^{2} - S_{+}^{2}}{8 J L}\,.
\end{align}
On the other hand, the decomposition of $\sin\theta_{L}$ is not so straightforward.

To properly perform the decomposition, consider the generating function of Gegenbauer polynomials,
\begin{equation}
    \frac{1}{\left(1 -2xt + t^{2}\right)^{\alpha}} = \sum_{n=0}^{\infty} C_{n}^{(\alpha)}(x) t^{n}\,.
\end{equation}
The expression in Eq.~\eqref{eq:sinL} can be written in a similar form by defining $t=({\cal{S}}_{4}/{\cal{S}}_{0})^{1/2}\sin^{2}\psi$ and $x={\cal{S}}_{2}/\sqrt{{\cal{S}}_{0}{\cal{S}}_{4}}$. We then recognize that $\sin\theta_{L}$ can be written in terms of Gegenbauer polynomials of order $\alpha = -1/2$, specifically
\begin{equation}
    \label{eq:sin-gegen}
    \sin\theta_{L} = \sqrt{{\cal{S}}_{0}} \sum_{n=0}^{\infty} \left(\frac{{\cal{S}}_{4}}{{\cal{S}}_{0}}\right)^{n/2} C^{(-1/2)}_{n}\left(\frac{{\cal{S}}_{2}}{\sqrt{{\cal{S}}_{0} {\cal{S}}_{4}}}\right) \sin^{2n}\psi\,.
\end{equation}
While the above sum formally extends to $n=\infty$, it is worth noting that ${\cal{S}}_{4}/{\cal{S}}_{0} \ll 1$ since it scales like $v^{4}$ and is further suppressed by an effective factor of $\chi_{A}^{4}$. In practice, the above sum can then be truncated at a finite number of terms. Furthermore, the same holds true for the $n\neq0$ terms in Eq.~\eqref{eq:cos-fourier} since these are also PN and spin suppressed. For generality, the remainder of the discussion here keeps Eq.~\eqref{eq:sin-gegen} as an infinite sum, but the analyses carried out in Sec.~\ref{sec:wave} truncates at the appropriate order. 

We can go one step further and write
\begin{align}
    \label{eq:sin-reduce}
    \sin^{2n}\psi &=  \sum_{k=0}^{n} (-1)^{n-k} \left(\frac{1+\delta_{k,0}}{2^{2n}}\right) \binom{2n}{k} \cos[2(n-k)\psi]\,,
\end{align}
which allows us to write $\sin\theta_{L}$ as a formal Fourier series, specifically
\begin{align}
    \sin\theta_{L} &= \sum_{n=0}^{\infty} {\cal{S}}_{L,n} \left(e^{in\psi} + e^{-in\psi}\right)\,,
\end{align}
where $S_{L,n}$ are found by combining Eqs.~\eqref{eq:sin-gegen} \&~\eqref{eq:sin-reduce}. Note that while the ${\cal{C}}_{L,n}$ and ${\cal{S}}_{L,n}$ are functions of orbital velocity, we do not formally PN expand them for the reasons explained before Eq.~\eqref{eq:dphiz-temp}.

The Wigner D-matrices are related to the spin-weighted spherical harmonics and SALPs through Eq.~\eqref{eq:D-to-P}, where~\cite{Breuer}
\begin{align}
    N_{l} &= \sqrt{\frac{4\pi}{2l+1}}\,, \qquad 
    N_{lm} = \sqrt{\frac{(l-m)!}{(l+m)!}}\,.
\end{align}
The SALPs are related to spin-weighted spherical harmonics through Eq.~(2.6) in~\cite{Breuer}, and those of relevance to the Fourier domain waveform in Sec.~\ref{sec:wave} are
\begin{align}
    {_{0}}P_{2,-2}(\theta_{L}) &= \frac{1}{8} \sin^{2}\theta_{L}\,,
    \\
    {_{\pm1}}P_{2,-2}(\theta_{L}) &= -\frac{1}{4\sqrt{6}} \sin\theta_{L} \left(1 \pm \cos\theta_{L}\right)\,,
    \\
    {_{\pm2}}P_{2,-2}(\theta_{L}) &= \frac{1}{16\sqrt{6}} \left[3 \pm 4 \cos\theta_{L} + \cos(2\theta_{L})\right]\,.
\end{align}

In Eq.~\eqref{eq:D-decomp}, we Fourier decompose the SALPs into harmonics of $\psi$. As an example of this, consider ${_{0}}P_{2,-2}$ above, which is decompsed as
\begin{align}
    {_{0}}P_{2,-2}(u) &= \sum_{n=-4}^{4} {P^{2}}_{-2,0,n}(u) e^{in\psi(u)}\,,
    \\
    {P^{2}}_{-2,0,0}(u) &= \frac{1}{32J^{2} L^{2}} \left[-8 J^{4} - 8 L^{4} - 3 S_{-}^{4} 
    \right.
    \nonumber \\
    &\left.
    - 2S_{-}^{2} S_{+}^{2} - 3 S_{+}^{4} + 8 L^{2}(S_{-}^{2} + S_{+}^{2}) 
    \right.
    \nonumber \\
    &\left.
    + 8 J^{2} (2 L^{2} + S_{-}^{2} + S_{+}^{2})\right]\,,
    \\
    {P^{2}}_{-2,0,\pm2}(u) &= \frac{1}{16 J^{2} L^{2}} \left(S_{-}^{2} - S_{+}^{2}\right) \left(S_{-}^{2} + S_{+}^{2} - 2 J^{2} - 2 L^{2}\right)\,,
    \\
    {P^{2}}_{-2,0,\pm4}(u) &= -\frac{1}{64 J^{2} L^{2}} \left(S_{-}^{2} - S_{+}^{2}\right)\,,
\end{align}
The Fourier coefficients for the remaining SALPs can be found by combining Eqs.~\eqref{eq:cos-fourier},~\eqref{eq:sin-gegen}-\eqref{eq:sin-reduce}, and are generically written in terms of the Gegenbauer polynomials.

The above Fourier coefficients ${P^{l}}_{mm'n}$ are exact in GR, but in dCS gravity, we must perform a weak coupling expansion due to the fact that $J^{2}$ and $S_{\pm}^{2}$ are shifted from their GR values. The corrections $\delta P_{K}$ are found by taking the replacements
\begin{align}
    J^{2} &= J_{0}^{2} + \frac{\bar{\zeta}_{2}}{2} \left(\delta s_{+}^{(0)} + \delta s_{-}^{(0)}\right)\,,
    \\
    S_{\pm}^{2} &= s_{\pm}^{(0)} + \bar{\zeta}_{2} \delta s_{\pm}^{(0)} + {\cal{O}}(u)\,,
\end{align}
where $J_{0}$ is given in Eq.~\eqref{eq:J0}, and then applying Eq.~\eqref{eq:PK-dCS}. As an example,
\begin{align}
    \delta {P^{2}}_{-2,0,\pm4}(u) &= 2\left(\delta s_{-}^{(0)} - \delta s_{+}^{(0)}\right) 
    \nonumber \\
    &- \frac{1}{2J_{0}^{2}} \left(s_{-}^{(0)} - s_{+}^{(0)}\right) \left(\delta s_{-}^{(0)} + \delta s_{+}^{(0)}\right)\,,
\end{align}
with $L=\eta M^{2}/u$. 

The coefficient $h_{2}$ is
\begin{equation}
    \label{eq:h2}
    h_{2} = \frac{8\eta M u^{2}}{D_{L}} \sqrt{\frac{\pi}{5}}\,.
\end{equation}
The PN corrections to the SUA frequency mapping in Eq.~\eqref{eq:sua} up to 3PN order are
\begin{align}
    \upsilon_{2} &= \frac{\langle a_{2}\rangle_{\psi}}{2}\,,
    \\
    \upsilon_{3} &= \frac{\langle a_{3} \rangle_{\psi}}{2}\,,
    \\
    \upsilon_{4} &= \frac{1}{8}\left(4 \langle a_{4} \rangle_{\psi} - \langle a_{2}\rangle_{\psi}^{2}\right)\,,
    \\
    \upsilon_{5} &= \frac{1}{4} \left(2 \langle a_{5}\rangle_{\psi} - \langle a_{2}\rangle_{\psi} \langle a_{3} \rangle_{\psi}\right)\,,
    \\
    \upsilon_{6} &= \frac{1}{16} \left(\langle a_{2}\rangle_{\psi}^{3} - 2 \langle a_{3}\rangle_{\psi}^{2} - 4 \langle a_{2}\rangle_{\psi} \langle a_{4}\rangle_{\psi} + 8 \langle a_{6}\rangle_{\psi}\right)
    \\
    \upsilon_{6}^{l} &= \frac{3}{2} \langle b_{6} \rangle_{\psi}\,.
\end{align}

\bibliographystyle{apsrev4-1}
\bibliography{refs}
\end{document}